\documentclass[12pt, draftclsnofoot, onecolumn]{IEEEtran}

\usepackage{blindtext}

\usepackage{latexsym}
\usepackage[utf8]{inputenc}
\usepackage{graphicx}
\usepackage{amsmath,amssymb,amsthm,mathtools}
\usepackage{color}
\usepackage{float}
\usepackage[linesnumbered,ruled,algo2e]{algorithm2e}
\usepackage{algorithm,algpseudocode,float}
\usepackage{upgreek}
\usepackage{psfrag}\usepackage{bm}
\usepackage{bbm}
\usepackage{cite}
\usepackage{multido}   
\usepackage{arydshln} 
\usepackage{blindtext}
\usepackage{soul}
\usepackage{mathrsfs}
\usepackage{setspace}
\usepackage{caption}
\usepackage{subcaption}

\usepackage[acronym,shortcuts]{glossaries}
\newacronym{3G}{3G}{third generation}
\newacronym{4G}{4G}{fourth generation}
\newacronym{LTE}{LTE}{Long-term Evolution}
\newacronym{5G}{5G}{fifth generation}
\newacronym{6G}{6G}{sixth generation}
\newacronym{IoT}{IoT}{Internet of Things}
\newacronym{SotA}{SotA}{state-of-the-art}
\newacronym{SNR}{SNR}{signal-to-noise-ratio}
\newacronym{SINR}{SINR}{signal-to-interference-plus-noise ratio}
\newacronym{CA}{CA}{carrier aggregation}
\newacronym{eMBB}{eMBB}{enhanced mobile broadband}
\newacronym{URLLC}{URLLC}{ultra reliable low latency communications}
\newacronym{mMTC}{mMTC}{massive machine type communications}
\newacronym{DoF}{DoF}{degrees of freedom}
\newacronym{QoS}{QoS}{Quality of Service}
\newacronym{i.i.d.}{i.i.d.}{independent and identically distributed}
\newacronym{PDF}{PDF}{probability density function}
\newacronym{CDF}{CDF}{cumulative density function}
\newacronym{ML}{ML}{maximum likelihood}
\newacronym{SE}{SE}{spectral efficiency}
\newacronym{MS}{MS}{multidimensional signal}
\newacronym{MMSE}{MMSE}{minimum mean square error}
\newacronym{MUI}{MUI}{multiuser interference}
\newacronym{ISI}{ISI}{intra-symbol interference}
\newacronym{RHS}{RHS}{right-hand side}
\newacronym{LHS}{LHS}{left-hand side}
\newacronym{RAM}{RAM}{random-access memory}
\newacronym{DA}{DA}{discreteness-awareness}
\newacronym{IA}{IA}{imperfection-awareness}
\newacronym{SCA}{SCA}{successive convex approximation}
\newacronym{flops}{flops}{floating point operations}
\newacronym{COTS}{COTS}{commercial off-the-shelf}
\newacronym{LS}{LS}{least square}
\newacronym{FPGA}{FPGA}{field-programmable gate array}
\newacronym{NR}{NR}{new radio}

\newacronym{CSI}{CSI}{channel state information}
\newacronym{OMA}{OMA}{orthogonal multiple access}
\newacronym{TDMA}{TDMA}{time division multiple access}
\newacronym{CDMA}{CDMA}{code division multiple access}
\newacronym{OFDMA}{OFDMA}{orthogonal frequency division multiple access}
\newacronym{BS}{BS}{base station}
\newacronym{RX}{RX}{receiver}
\newacronym{TX}{TX}{transmitter}
\newacronym{BF}{BF}{beamforming}
\newacronym{MIMO}{MIMO}{multiple-input multiple-output}
\newacronym{MISO}{MISO}{Multiple-Input Single-Output}
\newacronym{PA}{PA}{power amplifier}
\newacronym{DAC}{DAC}{digital-to-analog converter}
\newacronym{ADC}{ADC}{analog-to-digital converter}
\newacronym{RF}{RF}{radio frequency}
\newacronym{FD}{FD}{full-duplex}
\newacronym{ZF}{ZF}{zero-forcing}
\newacronym{LMMSE}{LMMSE}{linear minimum mean square error}
\newacronym{BER}{BER}{bit error rate}
\newacronym{V-BLAST}{V-BLAST}{Vertical Bell Labs Layered Space-Time}
\newacronym{QAM}{QAM}{quadrature amplitude modulation}
\newacronym{PAM}{PAM}{pulse amplitude modulation}
\newacronym{SOAV}{SOAV}{sum-of-absolute-values}
\newacronym{SCSR}{SCSR}{sum of complex sparse regularizers}
\newacronym{NOMA}{NOMA}{non-orthogonal multiple access}
\newacronym{PD-NOMA}{PD-NOMA}{Power Domain NOMA}
\newacronym{AWGN}{AWGN}{additive white Gaussian noise}
\newacronym{IBFD}{IBFD}{In-Band Full-Duplex}
\newacronym{PIC}{PIC}{parallel interference cancellation}
\newacronym{QPSK}{QPSK}{quadrature phase shift keying}

\newacronym{LLR}{LLR}{log-likelihood ratio}
\newacronym{DAPSD}{DAPSD}{discreteness-aware probabilistic soft-quantizion detector}
\newacronym{DAPZF}{DAPZF}{discreteness-aware penalized zero-forcing}
\newacronym{DAGED}{DAGED}{discreteness-aware generalized eigenvalue detector}
\newacronym{SISO}{SISO}{single-input single-output} 
\newacronym{CGMT}{CGMT}{convex Gaussian min-max theorem} 
\newacronym{SBR}{SBR}{simplicity-based recovery} 
\newacronym{GIGD}{GIGD}{graph-based
iterative Gaussian detector} 
\newacronym{ERTS}{ERTS}{enhanced reactive tabu search}
\newacronym{IDAGLS}{IDAGLS}{imperfection- and discreteness-aware generalized least square}
\newacronym{DALMMSE}{DALMMSE}{discreteness-aware linear MMSE}
\newacronym{DAMP}{DAMP}{discreteness-aware approximate message passing}
\newacronym{AMP}{AMP}{approximate message passing}
\newacronym{$E_b/N_0$}{$E_b/N_0$}{energy per bit to noise power spectral density ratio}
 \newacronym{LDPC}{LDPC}{low-density parity-check code}
 \newacronym{GLS}{GLS}{generalized least square}
 \newacronym{UE}{UE}{user equipment}
 \newacronym{ULA}{ULA}{uniform linear array}
 \newacronym{AoA}{AoA}{angle of arrival}
 \newacronym{SVD}{SVD}{singular value decomposition}
 \newacronym{IDLS}{IDLS}{Iterative Discrete Least Square}

\newacronym{FP}{FP}{fractional programming}
\newacronym{SDR}{SDR}{semidefinite relaxation}
\newacronym{SDP}{SDP}{semidefinite programming}
\newacronym{ADMM}{ADMM}{alternating direction method of multipliers}
\newacronym{GEP}{GEP}{generalized eigenvalue problem}
\newacronym{GTRS}{GTRS}{generalized trust region subproblem}
\newacronym{QCQP}{QCQP}{quadratically constrained quadratic problem}
\newacronym{KKT}{KKT}{Karush Kuhn Tucker}
\newacronym{QT}{QT}{quadratic transform}
\newacronym{QCQP-1}{QCQP-1}{quadratically constrained quadratic program with one convex constraint}
\newacronym{DCP}{DCP}{disciplined convex programming}
\newacronym{SCCR}{SCCR}{sum of concave-over-convex ratios}
\newacronym{NP}{NP}{non-deterministic polynomial-time}
\newacronym{PG}{PG}{proximal gradient}
\newacronym{LASSO}{LASSO}{least absolute shrinkage and selection operator}

\newacronym{CS}{CS}{compressed sensing}
\newacronym{BP}{BP}{basis pursuit}
\newacronym{OMP}{OMP}{orthogonal matching pursuit}
\newacronym{BPDN}{BPDN}{basis pursuit denoising}
\newacronym{RIP}{RIP}{restricted isometry property}
\newacronym{RIC}{RIC}{restricted isometric constant}
\newacronym{MDV}{MDV}{mixed-norm discrete vector}

\newcommand{\norm}[1]{\left\lVert#1\right\rVert}
\DeclareMathAlphabet{\bbmsl}{U}{bbm}{m}{sl}			

\newcommand{\abs}[1]{\left|{#1}\right|}
\newcommand{\E}[1]{\mathbb{E}\!\left[{#1}\right]}
\newcommand{\Ev}[2]{\mathbb{E}_{#1}\!\!\left[{#2}\right]}
\newcommand{\Tr}[1]{\mathrm{Tr}\left({#1}\right)}

\renewcommand{\smallskip}{\vspace{0.25cm}}

\makeatletter
\newcommand{\nosemic}{\renewcommand{\@endalgocfline}{\relax}}
\newcommand{\dosemic}{\renewcommand{\@endalgocfline}{\algocf@endline}}
\let\oldnl\nl
\newcommand{\nonl}{\renewcommand{\nl}{\let\nl\oldnl}}
\makeatother

\newcommand{\referencesrootdir}{./}
\newcommand{\myreferences}{\referencesrootdir/listofpublications.bib}

\def\BibTeX{{\rm B\kern-.05em{\sc i\kern-.025em b}\kern-.08em T\kern-.1667em\lower.7ex\hbox{E}\kern-.125emX}}
\begin{document}

\title{Robust Symbol Detection in Overloaded NOMA Systems}

\author{Hiroki Iimori,~\IEEEmembership{Graduate Student Member,~IEEE,} Giuseppe~Thadeu~Freitas~de~Abreu,~\IEEEmembership{Senior Member,~IEEE,}\\ Takanori Hara, \IEEEmembership{Student Member, IEEE}, Koji Ishibashi, \IEEEmembership{Senior Member, IEEE},\\  Razvan-Andrei~Stoica,~\IEEEmembership{Member,~IEEE,} David~Gonz{\'a}lez~G.,~\IEEEmembership{Senior Member,~IEEE,} and Osvaldo~Gonsa\\
\thanks{H.~Iimori and G.~Abreu are with the Department of Electrical and Computer Engineering, Jacobs University Bremen, Campus Ring 1, 28759 Bremen, Germany (e-mail: h.iimori@ieee.org, g.abreu@jacobs-university.de).}
 \thanks{T. Hara and K. Ishibashi are with the Advanced Wireless \& Communication Research Center (AWCC), The University of Electro-Communications, 1-5-1 Chofugaoka, Chofu-shi, Tokyo 182-8585, Japan, (emails: hara@awcc.uec.ac.jp, koji@ieee.org).}
 \thanks{A.~Stoica is with WIOsense GmbH \& Co. KG, Lesumer Heerstra{\ss}e 42-44, 28717 Bremen (e-mail: rstoica@ieee.org)}
  \thanks{D.~Gonz{\'a}lez~G. and O.~Gonsa are with Advanced Connectivity Technologies Group, Continental AG, Wilhelm-Fay Strasse 30, 65936 Frankfurt/Main, Germany (e-mail: [david.gonzalez.gonzalez, osvaldo.gonsa]@continental-corporation.com).}}

  \maketitle

\begin{abstract}
We present a framework for the design of low-complexity and high-performance receivers for multidimensional overloaded \ac{NOMA} systems.  
The framework is built upon a novel \acf{CS} regularized \acf{ML} formulation of the discrete-input detection problem, in which the $\ell_0$-norm is introduced to enforce adherence of the solution to the prescribed discrete symbol constellation.
Unlike much of preceding literature \textit{e.g.,} \cite{TasneemSys20, JeongTVT19, NagaharaSPL15, MasoudIET20, HayakawaTWC2017, HayakawaAccess2018,MingICC20}, the method is not relaxed into the $\ell_1$-norm, but rather approximated with a continuous and asymptotically exact expression without resorting to \acf{PIC}.
The objective function of the resulting formulation is thus a sum of concave-over-convex ratios, which is then tightly convexized via the \acf{QT}, such that its solution can be obtained via the iteration of a simple closed-form expression that closely resembles that of the classic \acf{ZF} receiver. 
By further transforming the aforementioned problem into a \acf{QCQP-1}, the optimal regularization parameter to be used at each step of the iterative algorithm is then shown to be the largest generalized eigenvalue of a pair of matrices which are given in closed-form.
The method so obtained, referred to as the \acf{IDLS}, is then extended to address several factors of practical relevance, such as noisy conditions, imperfect \acf{CSI}, and hardware impairments, thus yielding the Robust \ac{IDLS} algorithm.
Simulation results show that the proposed art significantly outperforms both classic receivers, such as the \ac{LMMSE}, and recent \ac{CS}-based \acf{SotA} alternatives, such as the \acf{SOAV} and the \acf{SCSR} detectors.
\end{abstract}

\glsresetall

\section{Introduction}
\label{sect:Intro}

The rapid growth in demand for rate and user capacity led the \ac{5G} wireless communications systems to widely adopt \ac{MIMO} technologies and incorporate higher frequency bands of operation \cite{Popovski2018}.
Although these trends are expected to continue, it is also a consensus that the mere expansion of spatial and bandwidth resources is not sufficient for future systems, such as \ac{6G} wireless networks, to cope with looming wireless-driven applications, examples of which are augmented reality and fully connected autonomous driving that require not only extremely high data rates and system capacity, but also  stringent latency constraints \cite{GiordaniComMag2020}.

To meet these future challenges, greater efficiency must be achieved in exploiting spatial, temporal and spectral \ac{DoF}, via the elimination of access control overhead; thus motivating the proposal of \ac{NOMA}, cell-free \ac{MIMO}, and grant-free systems \cite{AndreiAccessMCNOMA2019, CaireAsilomar2019, GanesanSPAWC2020}  in which resources are not exclusively allocated to users, such that the total number of data streams may be larger than the dimension of received signals \cite{MyListOfPapers:LiuTWC2018}.

Receivers suitable to such overloaded \ac{NOMA} conditions differ fundamentally from the conventional and well-known linear \ac{LS} and \ac{LMMSE} detectors widely employed within previous and current generation wireless devices, which own their popularity to the relative ease of implementation and reduced complexity, but which are known to result in high error floors when employed in overloaded scenarios. 

In the past, \ac{MUI} and \ac{ISI} were largely dealt with by means of \ac{PIC} exploiting sphere detection and lattice-reduction methods \cite{DattaERTS12, ChenTWC2013}.
But although those earlier contributions illustrate the feasibility of asymptotically approaching the optimal \ac{ML} detection performance in overloaded conditions, despite the exponential computational complexity in scaling to massive dimensions, the approach suffers additionally from the fundamental limitation of operating under continuous multidimensional spaces.
However, detection under massively overloaded conditions requires the maximization of reconstruction likelihoods subject to constraints imposed by the discreteness of constellations.

Taking advantage of recent advances in \ac{CS} \cite{RauhutBook:2013, das2013finite}, lower complexity signal detectors based on discrete constellation sparse regularizers have emerged \cite{NagaharaSPL15, HayakawaTWC2017, HayakawaAccess2018}, in which the search of likely signals is biased towards the discrete constellation set, while maintaining the continuity of the search space, so that optimum solutions can efficiently be obtained.
To describe a couple of such alternatives, a non-closed-form signal detector based on the Douglas-Rachford algorithm for large overloaded \ac{MIMO} systems was proposed in \cite{HayakawaTWC2017}, which was shown to yield significant \ac{BER} gains over the classic \ac{LMMSE}.

That technique, referred to as \ac{SOAV}, was later generalized into the \ac{SCSR} method proposed in \cite{HayakawaAccess2018}, in which the \ac{ADMM} algorithm is leveraged to enable the detection of complex-valued discrete signals.
The \ac{SOAV} and \ac{SCSR} detectors were also shown to outperform previous \ac{SotA} schemes of both a more classic construct, such as the one proposed in \cite{DattaERTS12}, and of a more modern sparse processing-based design, such as the method of \cite{NagaharaSPL15}, in terms of not only detection accuracy but also computational complexity.
As a result, we have considered \ac{SOAV} of \cite{HayakawaTWC2017} and \ac{SCSR} of \cite{HayakawaAccess2018} as the primary benchmarking schemes for our contribution.

Despite their simplicity and excellent performance, a drawback of the \ac{SOAV} and \ac{SCSR} detectors is that the $\ell_0$-norm, which is employed in their formulation in the form of a regularization function with the crucial role of enforcing the discreteness of the solution, is ultimately replaced by the $\ell_1$-norm.
Although typical in the compressive sensing literature \cite{RauhutBook:2013} due to its convexity which helps reduce complexity of resulting algorithms, the replacement of the $\ell_0$-norm by the $\ell_1$-norm adopted in \ac{SOAV} and \ac{SCSR} is known to be rather loose and lead to significant sacrifice in performance \cite{DonohoTIT06}.

In addition to the above, another general limitation of modern \ac{CS}-based receivers for overloaded systems is the fact that all methods proposed so far -- including those aforementioned -- require numerical solutions obtained by means of implicit optimization problems (\textit{e.g.,} via the interior point method), which do not yield much insight on how the optimal solutions under the state conditions are achieved.

\vspace{-1ex}
\subsection{Contributions}
\label{sect:Intro.Contributions}

The contribution of this article is manyfold and can be summarized as follows:

\begin{itemize}
\item In Subsection \ref{sec:IDLSFormulation}, a \textbf{new tightly-convexized \ac{ML}-regularized formulation} of the  detection problem of overloaded \ac{NOMA}/\ac{MIMO} systems is proposed, in which the $\ell_0$-norm is not replaced, but instead tightly approximated by a novel asymptotically-exact expression.
A recently-proposed \ac{FP} technique \cite{ShenTSP2018} is then utilized to further transform the latter into a tractable quadratic problem, leading to an original constellation-constrained discrete generalization of the classic \ac{LS} algorithm that preserves the optimality potential approaching that of the \ac{ML} detector.

\item Building on this new \ac{ML}-like formulation of \ac{NOMA}/\ac{MIMO} receivers, a \textbf{novel iterative discrete least square framework} of the overloaded \ac{NOMA}/\ac{MIMO} detection problem is offered in Subsection \ref{sec:IDLSSolution}, which is presented in the form of a simple iteration of a closed-form expression closely resembling that of the classic \ac{ZF} receiver.
This new detector, referred to as the \ac{IDLS} receiver, is shown to outperform the classic \ac{MMSE} as well as the recently-proposed \ac{SOAV} and \ac{SCSR} \ac{SotA} schemes, without need for channel statistics, unlike belief propagation methods.
%
%

\item In contrast to the vast majority of related literature, where the  parameterization of regularized optimization problems is typically dealt with as separate problem (see $e.g.$ \cite{WenAccess2018}), in Subsection \ref{sec:auto_parameterization} a \textbf{novel method for the optimal auto-parameterization of the \ac{ML}-derived regularized formulation}.
Based on this formulation, the \ac{IDLS} detector is described and shown to be given by the largest generalized eigenvalue of a matrix pencil constructed only with knowledge of the received signal, the channel estimate, and noise variance.

\item Finally, in Section \ref{sec:practical}, several variations of the \ac{IDLS} receiver for overloaded \ac{NOMA} systems are given, which enable the \textbf{mitigation of impairment factors}, such as excessive noise, inexact knowledge of \ac{CSI}, and hardware imperfection.
These practical aspects are incorporated in the form of simple generalizations of the closed-form expression at the core of the \ac{IDLS} detector, effectively extending the latter to a Robust \ac{IDLS} scheme.
\end{itemize}

All in all, alluding to the fact that the article goes beyond proposing a single scheme, offering instead a broader and flexible approach to the design of effective detectors for \ac{NOMA} systems, enabling the incorporation of multiple robustness features, and including also the corresponding optimization of regularization parameters, we refer to the overall contribution as a \emph{framework} for symbol detection in \ac{NOMA} systems.

\vspace{-1ex}
\subsection{Notation}
\label{sect:Intro.Notation}
Throughout the text hereafter, the following notation is applied.
The sets of real and complex numbers are respectively denoted by $\mathbb{R}$ and $\mathbb{C}$.
Real-valued matrices and vectors are denoted as in $\mathbf{X}$ and $\mathbf{x}$, respectively, so as to distinguish  from complex-valued matrices and vectors which are respectively denoted as in $\bm{X}$ and $\bm{x}$.
Scalars are denoted as in $x$, irrespective of their belonging to $\mathbb{R}$ or $\mathbb{C}$.
The operators $\Re\{\bm{X}\}$ and $\Im\{\bm{X}\}$ denote the real and imaginary parts of $\bm{X}$, respectively.
The $\ell_p$-norm is denoted by $\|\mathbf{x}\|_{p}$, where $p \geq 0$.
The transpose, conjugate, and conjugate transpose (Hermitian) and of a matrix $\bm{X}$ are denoted as in $\bm{X}^{\text{T}}, \bm{X}^{\text{*}}$ and $\bm{X}^{\text{H}}$, respectively.
The $N$-sized identity, all-one and all-zero matrices are respectively represented as $\mathbf{I}_N$, $\mathbf{1}_N$ and $\mathbf{0}_N$. 
Finally, the null space of the linear mapping $\bm{V}$ is denoted by $\mathcal{N}(\bm{V})$

\vspace{-1ex}
\section{Preliminaries}
\label{sect:SystemModel_and_ProblemSetup}

\subsection{System Model}
\label{sect:SystemModel}

Consider a possibly underdetermined/overloaded wireless communication system with $N_t$ transmit and $N_r$ receive wireless resources ($i.e.$ antennas, subcarriers or time slots), such that the overloading ratio of the system can be defined as $\gamma \triangleq N_t / N_r$.
Assuming perfect channel knowledge at the receiver\footnote{Hardware and \ac{CSI} imperfection are considered later in Section \ref{sec:practical}.}, the received signal can then be modeled as
\begin{equation}
\label{eqn:receivedsignal}
{\bm{y}} = {\bm{H}} {\bm{s}} + {\bm{n}},\; {\bm{y}}\in\mathbb{C}^{N_r\times 1},
\end{equation}
where transmit symbols are normalized to a unit average power per symbol, $i.e.$ $\E{\bm{s}\bm{s}^\text{H}} = \mathbf{I}_{N_t}$, and such that each element of $\bm{s}$ is sampled from the same discrete and regular\footnotemark\, \ac{QAM} constellation set $\mathcal{C}\!=\!\{c_1,\cdots\!,c_{2^b}\}$ of cardinality $2^b$, where $b$ denotes the number of bits per symbol; while ${\bm{n}}\in\mathbb{C}^{N_r\times 1}$ is an \ac{i.i.d.} circularly symmetric complex \ac{AWGN} vector with zero mean and covariance matrix $\sigma^2_{{\bm{n}}}\mathbf{I}_{N_r}$, and ${\bm{H}}\in\mathbb{C}^{N_r\times N_t}$ describes the flat fading channel matrix between transmitter and receiver.

It will prove convenient hereafter to also express the complex-valued quantities in equation \eqref{eqn:receivedsignal} in terms of their real and imaginary parts, by defining
\begin{eqnarray}
\label{eq:Complex2RealMapping}
\mathbf{y} \triangleq \left[{\Re\{\bm{y}\}\atop \Im\{\bm{y}\}}\right],&&
\mathbf {H} \triangleq \left[{{\scriptstyle \mathrm \Re\{\bm{H}\} \; - \mathrm \Im\{\bm{H}\}}\atop {\scriptstyle \mathrm \Im\{\bm{H}\} \quad \scriptstyle \Re\{\bm{H}\}}}\right],\nonumber\\
\mathbf{s} \triangleq \left[{\Re\{\bm{s}\}\atop \Im\{\bm{s}\}}\right],&&
\mathbf{n} \triangleq \left[{\Re\{\bm{n}\}\atop \Im\{\bm{n}\}}\right],
\end{eqnarray}
such that we may write
\begin{equation}
\label{eqn:receivedsignalRealImag}
\mathbf{y}=\mathbf{H}\,\mathbf{s} + \mathbf{n}, \; \mathbf{y}\in\mathbb{R}^{2N_r\times 1}.
\end{equation}

\vspace{-3ex}
\subsection{State-of-the-Art Problem Formulation}
\label{sect:ProblemSetup}

Given the above, the \ac{ML} detection of the complex transmit signal vector $\bm{s}$ in equation \eqref{eqn:receivedsignal} can be expressed as the following constellation-constrained $\ell_2$-norm minimization problem
\begin{subequations}
\label{eqn:ML_complex}
\begin{eqnarray}
{\mathop {\mathrm{minimize}} \limits_{\bm{s} \in \mathbb {C}^{N_t}}} &&\!\!\!\!\|{\bm{y}} - \bm{H}\bm{s}\|^2_2\\
\label{eqn:MLConstraint_complex}
\mathrm{subject\,to} &&\!\!\!\! \bm{s}\in\mathcal{C}^{N_t},
\end{eqnarray}
\end{subequations} 
or equivalently
\begin{subequations}
  \label{eqn:ML_real}
  \begin{eqnarray}
  {\mathop {\mathrm{minimize}} \limits_{\mathbf{s} \in \mathbb {R}^{2N_t}}} &&\!\!\!\!\|\mathbf{y} - \mathbf{H}\mathbf{s}\|^2_2\\
  \label{eqn:MLConstraint_real}
  \mathrm{subject\,to} &&\!\!\!\! \mathbf{s}\in\mathcal{P}^{2N_t},
  \end{eqnarray}
\end{subequations} 
where $\mathcal{P} \triangleq \Re\{\mathcal{C}\} = \{p_1,\cdots,p_{2^{b/2}}\}$ is a \ac{PAM} constellation consisting of the $\abs{\mathcal{P}} = 2^{b/2}$ real/imaginary parts of the symbols in $\mathcal{C}$.

It is evident that due to the disjoint constraints \eqref{eqn:MLConstraint_complex} and \eqref{eqn:MLConstraint_real}  the optimization problems formulated as in equations \eqref{eqn:ML_complex} and \eqref{eqn:ML_real} are non-convex, such that their exact solution require exhaustive searches among all possible combinations of the elements of $\mathcal{C}$ and $\mathcal{P}$, respectively, resulting in a prohibitive complexity of order $2^{b N_t}$. 
Consequently, a continuous-space reformulation of the latter problem is desired, which allows for convexification methods to be applied, enabling the subsequent design of efficient algorithms to solve the problem at much lower complexities.
To that end, one approach that emerged relatively recent is to employ ideas of \ac{CS} \cite{das2013finite, RauhutBook:2013} to rewrite the constraints  \eqref{eqn:MLConstraint_complex} and \eqref{eqn:MLConstraint_real} in terms of equivalent expressions utilizing the $\ell_0$-norm, leading to
\begin{subequations}
\label{eqn:L0ML_complex}
\begin{eqnarray}
{\mathop {\mathrm{minimize}} \limits_{\bm{s} \in \mathbb {C}^{N_t}}} &&\!\!\!\!\|\bm{y} - \bm{H}\bm{s}\|^2_2\\[-2ex]
\label{eqn:L0ML_const_complex}
\mathrm{subject\,to} &&\!\!\!\! \sum^{2^{b}}_{i=1} \|\bm{s} - c_i\mathbf{1}\|_0 = N_t\cdot(2^{b}-1),
\end{eqnarray}
\end{subequations}
\text{and}
\begin{subequations}
\label{eqn:L0ML_real}
\begin{eqnarray}
\label{eqn:L0ML_obj_real}
{\mathop {\mathrm{minimize}} \limits_{\mathbf{s} \in \mathbb {R}^{2N_t}}} &&\!\!\!\!\|\mathbf{y} - \mathbf{H}\mathbf{s}\|^2_2\\[-2ex]
\label{eqn:L0ML_const_real}
\mathrm{subject\,to} &&\!\!\!\! \sum^{2^{b/2}}_{i=1} \|\mathbf{s} - p_i\mathbf{1}\|_0 = 2N_t\cdot(2^{\frac{b}{2}}-1).
\end{eqnarray}
\end{subequations} 

In order to understand the equivalence between the constraints \eqref{eqn:MLConstraint_complex} and \eqref{eqn:L0ML_const_complex}, notice that indeed each $t$-th element of the vector $\bm{s}$ can be only one symbol in $\mathcal{C}$, such that the differences $(s_t - c_i)$ are non-zero for all but one index $i$, for each entry $t$, the same being true for constraints \eqref{eqn:MLConstraint_real} and \eqref{eqn:L0ML_const_real}.

Unlike constraints \eqref{eqn:MLConstraint_complex} and \eqref{eqn:MLConstraint_real}, however, the constraints \eqref{eqn:L0ML_const_complex} and \eqref{eqn:L0ML_const_real} are continuous functions of $\bm{s}$ and $\mathbf{s}$, respectively.
In other words, no relaxation penalty results from the substitution of the disjoint constraint \eqref{eqn:MLConstraint_complex} by  \eqref{eqn:L0ML_const_complex} or 
\eqref{eqn:MLConstraint_real} by  \eqref{eqn:L0ML_const_real}, respectively, such that exact solutions of equations \eqref{eqn:L0ML_complex} and \eqref{eqn:L0ML_real} yield in fact the \ac{ML} symbol estimates corresponding to the system models described by equations \eqref{eqn:receivedsignal} and \eqref{eqn:receivedsignalRealImag}, respectively.

The advantage of the formulations given in equations \eqref{eqn:L0ML_complex} and \eqref{eqn:L0ML_real} over those of equations \eqref{eqn:ML_complex} and \eqref{eqn:ML_real} is, however, that the problems in equations \eqref{eqn:L0ML_complex} and \eqref{eqn:L0ML_real} can be reformulated to allow for low complexity solutions by applying convex relaxation techniques, with the corresponding performance determined by the tightness of the particular relaxation method employed. 

An example of a possible procedure to convexize the formulations given in equations \eqref{eqn:L0ML_complex} and \eqref{eqn:L0ML_real} is as follows.
First, consider the  regularized mixed-norm  minimization problems
\vspace{-1ex}
\begin{subequations}
\label{eqn:L0L2OP}
\begin{eqnarray}
\label{eqn:L0L2OP_complex} 
{\mathop {\mathrm{minimize}} \limits _{\bm{s} \in \mathbb {C}^{N_t}}} \:\:{\lambda\sum _{i=1}^{2^{b}} \| \bm{s}-c_{i}\mathbf{1}\|_{0} + \|\bm{y} - \bm{H} \bm{s} \|_{2}^{2}},\\
\label{eqn:L0L2OP_real} 
{\mathop {\mathrm{minimize}} \limits _{\mathbf{s} \in \mathbb {R}^{2N_t}}} \:\:{\lambda\sum _{i=1}^{2^{b/2}} \| \mathbf{s}-p_{i}\mathbf{1}\|_{0} + \|\mathbf{y} - \mathbf{H} \mathbf{s} \|_{2}^{2}},
\end{eqnarray}
\end{subequations}
in which the constraints \eqref{eqn:L0ML_const_complex} and \eqref{eqn:L0ML_const_real} were respectively relaxed by their insertion into the objective functions with the introduction of the penalization parameter $\lambda$.

Then, consider the convex relaxation of the latter problems via the classical replacement of the non-convex $\ell_0$-norm by the convex $\ell_1$-norm, which yields
\vspace{-1ex}
\begin{subequations}
\label{eqn:L1L2OP}
\begin{eqnarray}
\label{eqn:L1L2OP_complex} 
{\mathop {\mathrm{minimize}} \limits _{\bm{s} \in \mathbb {C}^{N_t}}} \:\:{\lambda\sum _{i=1}^{2^{b}} \| \bm{s}-c_{i}\mathbf{1}\|_{1} + \|\bm{y} - \bm{H} \bm{s} \|_{2}^{2}},\\
\label{eqn:L1L2OP_real} 
{\mathop {\mathrm{minimize}} \limits _{\mathbf{s} \in \mathbb {R}^{2N_t}}} \:\:{\lambda\sum _{i=1}^{2^{b/2}} \| \mathbf{s}-p_{i}\mathbf{1}\|_{1} + \|\mathbf{y} - \mathbf{H} \mathbf{s} \|_{2}^{2}}.
\end{eqnarray}
\end{subequations}

The latter two formulations are precisely those proposed in \cite{HayakawaAccess2018} and \cite{HayakawaTWC2017}, respectively, whose solutions yield the \ac{SCSR} and \ac{SOAV} receivers, and their compact and relatable description indicate that, at their core, both methods are equivalent, provided that the penalization parameter $\lambda$ is properly chosen for each method.
And due to their combination of good performance and low computational complexity, we will hereafter take this prior art as \ac{SotA} alternatives against which the improvements achieved by the receivers  designed in subsequent sections will be compared.

A qualitative inspection of the aforementioned \ac{SCSR} and \ac{SOAV} formulation reveals, however, the following limitations.
Firstly, the replacement of the $\ell_0$-norm originally appearing in equations \eqref{eqn:L0ML_complex} to \eqref{eqn:L0L2OP} by the $\ell_1$-norm employed in equation \eqref{eqn:L1L2OP} is known to be loose \cite{RauhutBook:2013}, such that a significant performance loss compared to the \ac{ML} detector can be expected.

\vspace{-1.5ex}
\footnotetext{By \emph{regularity}, it is meant that the sets $\Re\{\mathcal{C}\}$ and $\Im\{\mathcal{C}\}$ are identical. The assumption is without loss of generality and adopted  to simplify the exposition in alignment with the constellation sets used in practical \ac{5G} systems.}

Secondly, the penalization parameter $\lambda$ introduced in the reformulation of equations \eqref{eqn:L0ML_complex} and \eqref{eqn:L0ML_real} into equations \eqref{eqn:L0L2OP_complex} and \eqref{eqn:L0L2OP_real} is left unoptimized.
And thirdly, various issues of practical relevance such as the influence of noise, imperfect \ac{CSI}, channel correlation and hardware imperfection are left unaddressed, significantly compromising the robustness of both schemes.

In what follows, we will first introduce in Section \ref{sect:sect3} an improved method for the design of low-complexity and high-performing symbol detectors for overloaded systems, which not only resolves the first two of the aforementioned limitations, but in fact leads to a powerful framework that is subsequently extended in Section \ref{sec:practical} to incorporate robustness to all the aforementioned impairment factors, namely noise, CSI errors, channel correlation and hardware imperfection.
Better still, the resulting framework yields receivers whose algorithms resume to iterations of simple closed-form expressions, which greatly resemble, and thus in fact generalize, the classic \ac{ZF} and \ac{LMMSE} detectors.

\vspace{-1ex}
\section{Fundamentals of New Detection Framework}
\label{sect:sect3}
We seek to improve over the state of the art on large-scale overloaded multidimensional signal detection schemes.
To this end, in this section we propose a new framework for the symbol detection in overloaded \ac{NOMA} systems.

\vspace{-1ex}
\subsection[]{New Tightly Convexized \ac{ML}-derived Formulation}
\label{sec:IDLSFormulation}
  
Given the equivalence between the complex- and real-valued formulations of equations \eqref{eqn:L0L2OP_complex} and \eqref{eqn:L0L2OP_real}, we shall, for simplicity of exposition, focus hereafter on the complex-valued variation of equation \eqref{eqn:L0L2OP_complex}, without loss of generality.
Our objective is therefore to obtain a tight relaxation of the \ac{ML}-derived detection problem described by equation \eqref{eqn:L0L2OP_complex} which, unlike that of equation \eqref{eqn:L1L2OP_complex}, does \emph{not} resort to the convex $\ell_1$-norm, while still circumventing the non-convexity of the $\ell_0$-norm in a manner that allows for a low-complexity solution in the form of iterations of a simple closed-form expression.

\begin{figure}[H]
\centering
\includegraphics[width=0.45\columnwidth]{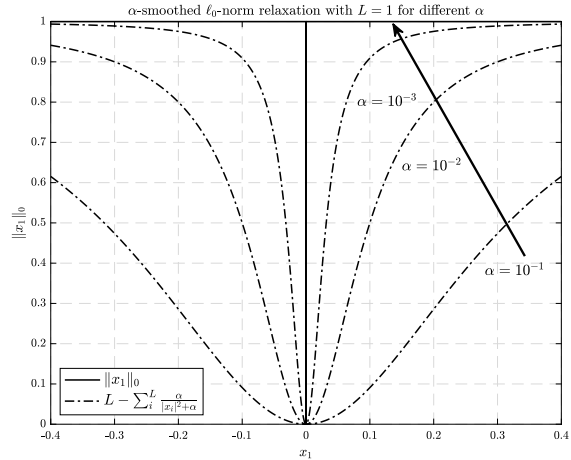}
\caption{Accuracy of $\ell_0$-norm approximation via equation \eqref{eqn:L0Approx}, for different values of $\alpha$. It is visible that the smooth approximation asymptotically approaches the $\ell_0$-norm as $\alpha\to 0$.}
\label{fig:l0-normApproximation}
\vspace{-2ex}
\end{figure}

To that end, consider the following asymptotically tight and smooth approximation of the $\ell_0$-norm
\begin{equation}
\label{eqn:L0Approx}
\|\mathbf{x}\|_{0} \approx \sum^{L}_{i=1} \frac{|x_{i}|^2}{|x_{i}|^2 + \alpha} = L - \sum^{L}_{i=1}\frac{\alpha}{|x_{i}|^2 + \alpha},
\end{equation}
where $\mathbf{x}$ denotes an arbitrary sparse vector of length $L$, with $0 < \alpha \ll 1$, such that for $\alpha\to 0$ the approximation becomes exact, as illustrated in Figure \ref{fig:l0-normApproximation}.

Substituting equation \eqref{eqn:L0Approx} into equation \eqref{eqn:L0L2OP_complex} yields
\begin{equation}
\label{eqn:DAZF_form1}
{\mathop {\mathrm{minimize}} \limits _{\bm{s} \in \mathbb {C}^{N_t}}} \:\:{ - \lambda\sum _{i=1}^{2^{b}}\sum _{j=1}^{N_t} \frac{\alpha}{|s_j-c_{i}|^2 + \alpha} + \|\bm{y} - \bm{H} \bm{s} \|_{2}^{2}}.
\end{equation}

Notice that the objective in equation \eqref{eqn:DAZF_form1} is, unlike that of equation \eqref{eqn:L0L2OP_complex}, a smooth and differentiable function which, despite non-convexity with respect to $\bm{s}$, is characterized by a \ac{SCCR}.
And although it is known that optimization problems with \ac{SCCR} objectives can be convexified via Taylor series approximations \cite{Bjornson2013}, a more effective technique to that end, referred to as the \ac{QT}, has been recently proposed \cite{ShenTSP2018}.

The advantage of \ac{QT}-based convexification of \acp{SCCR} objective functions is that it leads to \ac{FP} formulations shown in \cite{ShenTSP2018} to satisfy the \ac{KKT} condition of the original \ac{SCCR} problem.
We therefore adopt here the \ac{FP} approach, which can be succinctly explained as follows.

Consider a generic maximization problem with an \ac{SCCR} objective, such as
\vspace{-1ex}
\begin{subequations}
\label{eqn:GeneralFP}
\begin{eqnarray}
\label{eqn:GeneralFPObj}
\mathop {\mathrm{maximize}} \limits_{\bm{u}}&& \sum^{M}_{m=1}\frac{f_m(\bm{u})}{g_m(\bm{u})}\\
\mathrm{subject\,to}&& \bm{u}\in \mathcal{U}, 
\end{eqnarray}
\end{subequations}
where $f_m(\bm{u})$ and $g_m(\bm{u})$ denote arbitrary \emph{nonnegative} and \emph{strictly positive} scalar functions, respectively, and $\bm{u}$ is a vector variable to be optimized subject to a feasible set $\mathcal{U}$.

The \ac{QT} \cite{ShenTSP2018} casts the latter problem into the equivalent
\begin{subequations}
\label{eqn:GeneralFP_QT}
\begin{eqnarray}
\label{eqn:GeneralFP_QTObj}
\mathop {\mathrm{maximize}} \limits_{\bm{u}}&& \sum^{M}_{m=1}2\beta_{m}\sqrt{f_m(\bm{u})} - \beta^2_{m}g_m(\bm{u})\\
\mathrm{subject\,to}&& \bm{u}\in \mathcal{U},\:\: \beta_{m}\in\mathbb{R},
\end{eqnarray}
\end{subequations}
where
\begin{equation}
\label{eq:beta}
\beta_{m} \triangleq \frac{\sqrt{f_m(\bm{u})}}{g_m(\bm{u})},
\end{equation}
is a scaling quantity iteratively updated for each point $\bm{u}$ and designed to ensure that, at that pivot point, the original objective function in equation \eqref{eqn:GeneralFPObj} is equivalent to the transformed function given in equation \eqref{eqn:GeneralFP_QTObj}.

In light of the above, the \ac{QT} applied to equation \eqref{eqn:L0Approx} yields \vspace{-2ex}
\vspace{-1ex}
\begin{subequations}
\label{eqn:L0Approx_FP}
\begin{align}
\|\mathbf{x}\|_{0} &\approx L -  \Big(\sum^{L}_{i=1}2\beta_i\sqrt{\alpha} - \beta^2_i({|x_{i}|^2 + \alpha})\Big)\\[-1ex]
& = \sum^{L}_{i=1}\beta^2_i |x_{i}|^2 + \underbrace{L - \Big(\sum^{L}_{i=1}2\beta_i\sqrt{\alpha} + \alpha\Big),}_{\text{independent of }\mathbf{x}}
\label{eqn:L0Approx_FP_B}
\end{align}
\end{subequations}
where we remark that the latter terms independent of $\mathbf{x}$ in equation \eqref{eqn:L0Approx_FP_B} can be discarded in the context of a minimization problem on the variable $\mathbf{x}$.

\begin{figure}[b!]
\centering
\includegraphics[width=0.45\columnwidth]{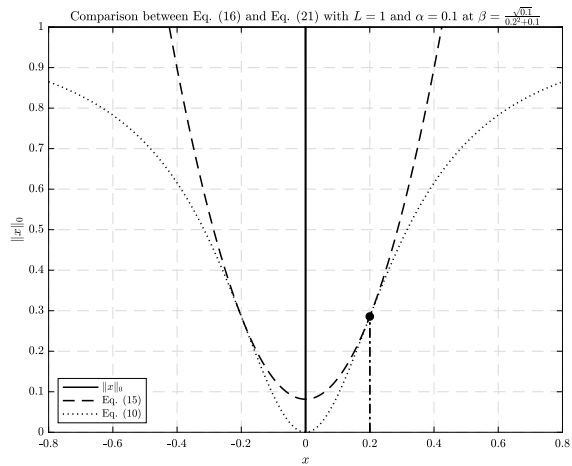}
\caption[]{Illustration of the relationship between equations \eqref{eqn:L0Approx} and \eqref{eqn:L0Approx_FP}, for the scalar case ($i.e.$, $N=1$), with $\alpha = 0.1$ and $\beta$ set to make both equations identical at $x = 0.2$.}
\label{fig:l0-normApproximation_comp}
\end{figure}

Figure \ref{fig:l0-normApproximation_comp} illustrates how equation \eqref{eqn:L0Approx_FP} in fact majorizes equation \eqref{eqn:L0Approx}, with equality at the pivot point ensured at each iteration by the optimal values of $\beta_{m}$ as per equation \eqref{eq:beta}.

Substituting equation \eqref{eqn:L0Approx_FP}, with the constant terms discarded, into equation \eqref{eqn:DAZF_form1} yields
\begin{equation}
\label{eqn:L0L2OP_ZF} 
{\mathop {\mathrm{minimize}} \limits _{\bm{s} \in \mathbb {C}^{N_t}}} \:\:{\lambda\sum^{2^{b}}_{i=1}\sum^{N_t}_{j=1}\beta^2_{i,j} |s_{j}-c_{i}|^2 + \|\bm{y} - \bm{H} \bm{s} \|_{2}^{2}},
\end{equation}
with
\begin{equation}
\label{eqn:beta_ZF}
\beta_{i,j} = \frac{\sqrt{\alpha}}{|s_j - c_i|^2 + \alpha},\:\:
\forall\;i\in\{1,\cdots,2^b\},j\in\{1,\cdots,N_t\}.
\end{equation}

Next, define the quantities
\vspace{-1ex}
\begin{subequations}
\begin{equation}
\label{eqn:bvecZF}
\bm{b} \triangleq \sum _{i=1}^{2^b}c_i\left[\beta^2_{i,1},\beta^2_{i,2},\ldots,\beta^2_{i,N_t}\right]^\mathrm{T},
\vspace{-1ex}
\end{equation}
\begin{equation}
\label{eqn:BmatZF}
\bm{B} \triangleq \sum _{i=1}^{2^b}\mathrm{diag}(\beta^2_{i,1},\beta^2_{i,2},\ldots,\beta^2_{i,N_t}) {\succ 0},
\vspace{-1ex}
\end{equation}
\end{subequations}
such that equation \eqref{eqn:L0L2OP_ZF} can be written as
\begin{equation}
\label{eqn:IDLS_VecComplex}
{\mathop {\mathrm{minimize}} \limits _{\bm{s} \in \mathbb {C}^{N_t}}}\:\: \lambda(\bm{s}^\mathrm{H}\bm{B}\bm{s} - 2\Re\big\{\bm{b}^\mathrm{H}\bm{s}\big\}) + \|\bm{y} - \bm{H} \bm{s} \|_{2}^{2},
\end{equation}
which again can be expressed more compactly by expanding the latter quadratic term, namely
\begin{equation}
\label{eqn:PenalizedZF}
{\mathop {\mathrm{minimize}} \limits _{\bm{s} \in \mathbb {C}^{N_t}}}\:\: \bm{s}^\mathrm{H}(\lambda\bm{B} + \bm{H}^\mathrm{H}\bm{H})\bm{s} - 2\Re\big\{(\lambda\bm{b}^\mathrm{H}+\bm{y}^\mathrm{H}\bm{H})\bm{s}\big\}.
\end{equation}

We remark that the formulation in equation \eqref{eqn:PenalizedZF} is an \emph{original} (never before proposed or derived), simple, convex, quadratic minimization variation of the \ac{ML}-derived mixed-norm regularized formulation of equation \eqref{eqn:L0L2OP_complex}, in which the $\ell_0$-norm term present in equation \eqref{eqn:L0L2OP_complex} is \emph{not} replaced by $\ell_1$-norm as done in various previous works, but instead approximated with arbitrary tightness by choosing $\alpha$ sufficiently small, and which therefore differs fundamentally from \ac{SotA} methods such as the \ac{SCSR} \cite{HayakawaAccess2018} and the \ac{SOAV} \cite{HayakawaTWC2017} on its ability to deliver performances close to that of the \ac{ML} receiver.
\vspace{-2ex}

Finally, the formulation in equation \eqref{eqn:PenalizedZF} establishes a general optimization framework based on which various \ac{ML}-like algorithms can be derived for the detection of symbols in large-scale systems, which thanks to their built-in ability to enforce constellation-compliant solutions, will prove very resilient to overloading, as well as other factors of practical relevance.

\vspace{-2ex}
\subsection[]{The Iterative Discrete Least Square Solution}
\label{sec:IDLSSolution}

For future convenience, let us define the function
\begin{equation}
\label{eqn:PenalizedZFObjective}
q(\bm{s}) \triangleq \bm{s}^\mathrm{H}(\lambda\bm{B} + \bm{H}^\mathrm{H}\bm{H})\bm{s} - 2\Re\big\{(\lambda\bm{b}^\mathrm{H}+\bm{y}^\mathrm{H}\bm{H})\bm{s}\big\},
\end{equation}
which is in fact the objective function in the optimization problem described by equation \eqref{eqn:PenalizedZF}.

Given that $q(\bm{s})$ is quadratic on $\bm{s}$, the latter problem can be readily solved in closed-form by setting its Wirtinger derivative \cite{AreSP2007} with respect to $\bm{s}$ equal to $0$, that is,
\begin{equation}
\label{eq:derivative_PenalizedZF}
\frac{\partial q(\bm{s})}{\partial\bm{s}^*} = (\lambda\bm{B} + \bm{H}^\mathrm{H}\bm{H})\bm{s} - \big(\lambda\bm{b} + \bm{H}^\mathrm{H}\bm{y}\big)=0,
\end{equation}
which readily yields the solution
\begin{subequations}
\label{eq:PenalizedZF_sopt}
\begin{equation}
\label{eq:PenalizedZF_soptComplex}
\bm{s} = \big(\lambda\bm{B} + \bm{H}^\mathrm{H}\bm{H}\big)^{-1}\big(\lambda\bm{b} + \bm{H}^\mathrm{H}\bm{y}\big).
\vspace{-0.5ex}
\end{equation}
  
Obviously, a real-domain equivalent of equation \eqref{eq:PenalizedZF_sopt} -- which will prove convenient later -- can also be obtained following the same steps as above, yielding 
\begin{equation}
\label{eq:PenalizedZF_soptReal}
\mathbf{s} = \big(\lambda\mathbf{B} + \mathbf{H}^\mathrm{T}\mathbf{H}\big)^{-1}\big(\lambda\mathbf{b} + \mathbf{H}^\mathrm{T}\mathbf{y}\big),
\vspace{-0.5ex}
\end{equation}
\end{subequations}
where we remind that $\mathbf{y}$, $\mathbf{H}$ and $\mathbf{s}$ are as previously defined in equation \eqref{eq:Complex2RealMapping}, and the majorization pivot quantities $\beta_{i,j}$, $\mathbf{b}$ and $\mathbf{B}$ are respectively redefined as
\begin{equation}
\label{eqn:beta_ZFReal}
\beta_{i,j} \triangleq \frac{\sqrt{\alpha}}{(s_j\! -\! p_i)^2\! +\! \alpha}\text{ with }
\begin{cases}
  i\in\{1,\cdots\!,2^{b/2}\}\\
  j\in\{1,\cdots\!,2N_t\}
\end{cases}\hspace{-2ex},
\end{equation}
and 
\begin{subequations}
\label{eqn:BMatZFReal}
\begin{equation}
\label{eqn:BMatZFReal_b}
\mathbf{b} \triangleq \sum _{i=1}^{2^{\frac{b}{2}}}p_i\left[\beta^2_{i,1},\beta^2_{i,2},\ldots,\beta^2_{i,2N_t}\right]^\mathrm{T},
\end{equation}
\begin{equation}
\label{eqn:BMatZFReal_B}
\mathbf{B} \triangleq \sum _{i=1}^{2^\frac{b}{2}}\mathrm{diag}(\beta^2_{i,1},\beta^2_{i,2},\ldots,\beta^2_{i,2N_t}) {\succ 0},
\end{equation}
\end{subequations}
with $p_i\in\mathcal{P} \triangleq \Re\{\mathcal{C}\}$.   

We emphasize the remarkably simple structure of the receiver described by equation \eqref{eq:PenalizedZF_sopt}, which is characterized by the mere iteration -- over which $\bm{b}$, $\bm{B}$, and consequently $\bm{s}$ are updated -- of an expression that is linear on the input $\bm{y}$, and which relies on the inversion\footnote{Since the matrix $\lambda\bm{B} + \bm{H}^\mathrm{H}\bm{H}$ is only updated by the term $\lambda\bm{B}$ at each iteration, its inverse can be accelerated employing tracking methods, $e.g.$ \cite{RosarioSPL2016}.} of a matrix guaranteed to be invertible due to the positive definiteness of the term $\lambda\bm{B}$.

In fact, one may immediately recognize that equation \eqref{eq:PenalizedZF_sopt} is akin to the conventional linear \ac{ZF} receiver, except for the penalization factor $\lambda$ and the dependence on the iteratively-computed regularization terms $\bm{b}$ and $\bm{B}$, which together enforce the constellation compliance of the solution.
In particular, with $\lambda = 0$, equation \eqref{eq:PenalizedZF_sopt} reduces to a conventional \ac{ZF} receiver, with $(\bm{H}^\mathrm{H}\bm{H})^{-1}\bm{H}^\mathrm{H}$ yielding the pseudo-inverse of the channel $\bm{H}$.

Unlike the conventional \ac{ZF} receiver, however, which is known to perform poorly if the channel matrix $\bm{H}$ is rank deficient -- as is the case of overloaded systems or spatially-correlated channels -- the iterative discrete least square linear detector here derived and summarized by equation \eqref{eq:PenalizedZF_sopt} is the solution of the optimization problem described by equation \eqref{eqn:PenalizedZF}, which in turn was obtained by a tight relaxation of the \ac{ML} problem of equation \eqref{eqn:L0L2OP_complex}.
It is thus expected and confirmed via computer simulations (see Subsection \ref{sec:performance1}) that the iterative linear detector here presented delivers \ac{ML}-derived performance, free of error floors and with improved robustness to overloading.
Finally, since $(\lambda\bm{B} + \bm{H}^\mathrm{H}\bm{H}) \succ 0$, the continuous and quadratic objective function $q(\bm{s})$ of equation \eqref{eqn:PenalizedZF} is differentiable with respect to $\bm{s}$, such that its solution given by equation \eqref{eq:PenalizedZF_sopt} has guaranteed convergence due to \cite{ShenTSP2018}.

All in all, it can be said that the receiver given by equation \eqref{eq:PenalizedZF_sopt} is a true generalization of the classical \ac{LS} receiver, which adheres to the discrete constellation in a continuous, asymptotically exact and absolutely converging manner.
Alluding to this fact, we refer to our scheme as the \ac{IDLS} framework.



\vspace{-1ex}
\subsection[]{Optimal Regularization Parameter of \ac{IDLS} Framework}
\label{sec:auto_parameterization}

Regularized optimization methods have grown in popularity in recent years \cite{WenAccess2018} due to their ability to solve complex problems at relatively low computational cost.
In these methods, penalized terms are added to an objective function in order to balance multiple desired features in the solution.
Well-known examples are \ac{CS} methods \cite{RauhutBook:2013} and the classic \ac{LASSO} estimator, in which sparse solutions are desired and ensured by  the $\ell_0$- or the $\ell_1$-norm, and matrix completion methods \cite{CandesMC09} in which low rank solutions are desired and ensured by the nuclear norm.

An important reason for the performance and robustness of such methods is that the choice of regularization parameter is not too sensitive, such that it can be dealt with offline  \cite{HassibiTIT2018} via dedicated techniques.
To cite a couple of examples, an approach to optimize regularization parameters based on deep unfolding was proposed in \cite{ItoTSP19}, and in \cite{MECKLENBRAUKER2017204} another method is proposed based on duality theory.

Despite the existence of these and several other techniques to optimize regularization parameters, for the sake of completeness we contribute also to this issue by proposing a new algorithm to optimize the penalization parameter $\lambda$ employed in the \ac{IDLS} detection framework presented above.
To this end, first  the real-valued equivalent of the fundamental \ac{ML}-derived optimization problem described in equation \eqref{eqn:IDLS_VecComplex}, from which our framework has been obtained, which is given by
\vspace{-0.5ex}
\begin{equation}
\label{eqn:IDLS_VecReal}
{\mathop {\mathrm{minimize}} \limits _{\mathbf{s} \in \mathbb {R}^{2N_t}}}\:\: \mathbf{s}^\mathrm{T}\mathbf{B}\mathbf{s} - 2\mathbf{b}^\mathrm{T}\mathbf{s} + \frac{1}{\lambda}\|\mathbf{y} - \mathbf{H} \mathbf{s} \|_{2}^{2},
\vspace{-0.5ex}
\end{equation}
where the penalization parameter has been moved to the $\ell_2$-norm term of the objective (only for future convenience and without prejudice to the formulation itself).

Now, notice that equation \eqref{eqn:IDLS_VecReal} is merely a regularized mixed norm variation of the original \ac{ML} problem described by equation \eqref{eqn:L0ML_real}, and observe that in order to satisfy the equality constraint of that formulation, the term $\sum^{2^{b/2}}_{i=1} \|\mathbf{s} - x_i\mathbf{1}\|_0$ in equation \eqref{eqn:L0ML_const_real} -- which corresponds to the term $\mathbf{s}^\mathrm{T}\mathbf{B}\mathbf{s} - 2\mathbf{b}^\mathrm{T}\mathbf{s}$ in equation \eqref{eqn:IDLS_VecReal} -- needs to be globally minimized.

It follows from the above that the regularized $\ell_2$-norm term in equation \eqref{eqn:IDLS_VecReal} can be placed as a constraint, with no penalty to the optimality of the formulation.
In other words, the \ac{ML}-derived formulation of equation \eqref{eqn:IDLS_VecReal} -- and by extension to the original \ac{ML} formulation of equation \eqref{eqn:L0ML_real} -- are equivalent to the following real-valued \ac{QCQP-1} formulation
\begin{subequations}
\label{eqn:P4_OP1}
\begin{eqnarray}
\label{eqn:P4_OP1_object}
\hspace{-5ex}&{\mathop {\mathrm{minimize}} \limits_{\mathbf{s} \in \mathbb {R}^{2N_t}}}&\!\!\mathbf{s}^\mathrm{T}\mathbf{B}\,\mathbf{s} - 2\,\mathbf{b}^\mathrm{T}\mathbf{s} \\
 \label{eqn:P4_OP1_const}
 \hspace{-5ex}&\mathrm{subject\,to}&\!\!   \mathbf{s}^\mathrm{T}\mathbf{H}^\mathrm{T}\mathbf{H}\,\mathbf{s} - 2\, \mathbf{y}^\mathrm{T} \mathbf{H}\,\mathbf{s}+ \mathbf{y}^\mathrm{T}\mathbf{y} - \delta \leq 0,
\end{eqnarray}
\end{subequations} 
where the constraint \eqref{eqn:P4_OP1_const} obeys Slater's condition.


We remark that the bounding parameter $\delta$ (a.k.a. ``search ball radius'') establishes the tightness within  which the squared distance $\|\mathbf{y} - \mathbf{H} \mathbf{s} \|_{2}^{2}$ is made to adhere, and is typically determined by the noise power \cite{LuzziTIT2013}, since noise variance is  standardly available in practical systems  \cite{5GZaidi}.

Setting therefore $\delta = \sigma^2_{\mathbf{n}}$ hereafter, the quadratic function in \eqref{eqn:P4_OP1_const} becomes
\begin{equation}
\label{eq:SlatersCondition}
k(\mathbf{s}) \triangleq \mathbf{s}^\mathrm{T}\mathbf{H}^\mathrm{T}\mathbf{H}\,\mathbf{s} - 2\, \mathbf{y}^\mathrm{T} \mathbf{H}\,\mathbf{s}+ \mathbf{y}^\mathrm{T}\mathbf{y} - \sigma^2_{\mathbf{n}}.
\vspace{-0.5ex}
\end{equation}

All that is left for us to do then is to obtain an efficient method to solve equation \eqref{eqn:P4_OP1}, which among many alternatives can be achieved by applying the result presented in \cite[Th.3.3]{MorePMS1993}.
Brought to the context hereby, that result states that if there exists a minimizer $\bar{\mathbf{s}}$ of equation \eqref{eqn:P4_OP1_object} satisfying the constraint \eqref{eqn:P4_OP1_const}, then $\bar{\mathbf{s}}$ is the global solution to equation \eqref{eqn:P4_OP1} if and only if (iff) there exists a parameter $\mu^\text{opt}\geq 0$ such that the following \ac{KKT} conditions are satisfied
\begin{subequations}
\label{eq:GEV_KKT}
\begin{eqnarray}
\label{eq:GEV_KKT1}
&(\mathbf{B} + \mu^\text{opt} \mathbf{H}^\mathrm{T}\mathbf{H})\,\bar{\mathbf{s}} = (\mathbf{b} + \mu^\text{opt} \mathbf{H}^\mathrm{T}\mathbf{y}),&\\
\label{eq:GEV_KKT2}
&k(\bar{\mathbf{s}}) \leq 0,&\\
\label{eq:GEV_KKT3}
&\mu^\text{opt}k(\bar{\mathbf{s}}) = 0.&
\end{eqnarray} 
\end{subequations}

In recognition to the outstanding work presented in \cite{MorePMS1993}, we refer to this formulation of the \ac{QCQP-1} problem of equation \eqref{eqn:P4_OP1} as Mor{\'e}'s Theorem, which in fact admits two distinct cases.
The first is when $\mu^\text{opt}=0$, in which case equation \eqref{eq:GEV_KKT1} reduces to equation \eqref{eqn:P4_OP1_object}, with unique global minimum at $\bar{\mathbf{s}} = \mathbf{B}^{-1}\mathbf{b}$, which is obviously a ``solution'' of no relevance since it is independent of the input.
The second and only relevant case is when $\mu^\text{opt} > 0$, in which case equations \eqref{eq:GEV_KKT2} and \eqref{eq:GEV_KKT3} both coincide and reduce to $k(\bar{\mathbf{s}})=0$, such that Mor{\'e}'s Theorem then yields
\begin{subequations}
\label{eq:GEV_KKT_v2}
\begin{eqnarray}
\label{eq:GEV_KKT_v2_1}
&\bar{\mathbf{s}} = (\mathbf{B} + \mu^\text{opt} \mathbf{H}^\mathrm{T}\mathbf{H})^{-1}(\mathbf{b} + \mu^\text{opt} \mathbf{H}^\mathrm{T}\mathbf{y}),&\\
\label{eq:GEV_KKT_v2_2}
&\bar{\mathbf{s}}^\mathrm{T}(\mathbf{H}^\mathrm{T}\mathbf{H}\,\bar{\mathbf{s}} - \mathbf{H}^\mathrm{T}\,\mathbf{y}) - \mathbf{y}^\mathrm{T} \mathbf{H}\,\bar{\mathbf{s}} + \mathbf{y}^\mathrm{T}\mathbf{y} - \sigma^2_{\mathbf{n}} = 0,&
\end{eqnarray} 
\end{subequations}
where we have in equation \eqref{eq:GEV_KKT_v2_1} inverted equation \eqref{eq:GEV_KKT1}, and in equation \eqref{eq:GEV_KKT_v2_2} expressed $k(\bar{\mathbf{s}})=0$ explicitly, with the term $2\, \mathbf{y}^\mathrm{T} \mathbf{H}\,\bar{\mathbf{s}}$ expanded and slightly rearranged.

A comparison of equations \eqref{eq:PenalizedZF_sopt} and \eqref{eq:GEV_KKT_v2_1} reveals that, except for the complex and real domains, respectively, both are identical if $\mu^\text{opt} = 1/\lambda$.
There is, however, a crucial  difference between both equations, namely, that in equation \eqref{eq:GEV_KKT_v2_1}, the parameter $\mu^\text{opt}$ results not from a regularization -- as indeed the \ac{QCQP-1} formulation of equation \eqref{eqn:P4_OP1} is not regularized -- but rather from the \ac{KKT} conditions required to solve equation \eqref{eqn:P4_OP1}, such $\mu^\text{opt}$ is an integral part of such solution.

In addition, notice that equation \eqref{eq:GEV_KKT_v2_1} is tied to the accompanying equation \eqref{eq:GEV_KKT_v2_2}, in the sense that \emph{both} must be simultaneously satisfied in order for the solution to hold.
In other words, the solution of the system of equations \eqref{eq:GEV_KKT_v2} yields within it the optimum KKT parameter $\mu^\text{opt}$, which in turn determines the optimum regularization parameter $\lambda^\text{opt} = 1/\mu^\text{opt}$ required in equation \eqref{eq:PenalizedZF_sopt}.

To accomplish this task, we first introduce the scaling quantity $\rho$ and the auxiliary vector $ \bar{\mathbf{s}} \triangleq \rho\,\mathbf{x}_1$, such that equations \eqref{eq:GEV_KKT_v2} can be rewritten as
\begin{subequations}
\begin{equation}
\label{eq:GEV_KKT_v3_1b}
\mathbf{x}_1^\mathrm{T} = \rho\,(\mathbf{b} + \mu^\text{opt} \mathbf{H}^\mathrm{T}\mathbf{y})^\mathrm{T}(\mathbf{B} + \mu^\text{opt} \mathbf{H}^\mathrm{T}\mathbf{H})^{-1},
\end{equation}
\begin{equation}
\label{eq:GEV_KKT_v3_2a}
(\mathbf{y}^\mathrm{T}\mathbf{y} - \sigma^2_{\mathbf{n}})\rho - \mathbf{y}^\mathrm{T} \mathbf{H}\,\mathbf{x}_1 + \frac{1}{\rho}\mathbf{x}_1^\mathrm{T}(\mathbf{H}^\mathrm{T}\mathbf{H}\,\mathbf{x}_1 - \rho\mathbf{H}^\mathrm{T}\mathbf{y})= 0,
\end{equation}
\end{subequations}
where, for future convenience, we exploited the symmetry in $(\mathbf{B}+ \mu^\text{opt} \mathbf{H}^\mathrm{T}\mathbf{H})$ to transpose equation \eqref{eq:GEV_KKT_v2_1} into equation \eqref{eq:GEV_KKT_v3_1b}, and rearranged equation \eqref{eq:GEV_KKT_v2_2} into equation \eqref{eq:GEV_KKT_v3_2a}.

Next, using equation \eqref{eq:GEV_KKT_v3_1b} in place of $\mathbf{x}_1^\mathrm{T}$ in the last term of equation \eqref{eq:GEV_KKT_v3_2a}, and rearranging the terms yields
\begin{equation}
\label{eq:GEV_KKT_v3_2b}
\underbrace{(\mathbf{y}^\mathrm{T}\mathbf{y}\! -\! \sigma_n^2)}_{\triangleq q_{11}}\,\rho - \underbrace{\,\mathbf{y}^\mathrm{T} \mathbf{H}\,}_{\triangleq \mathbf{q}_{12}}\,\mathbf{x}_1 + \underbrace{\mathbf{b}^\mathrm{T}}_{\triangleq \mathbf{q}_{13}}\mathbf{x}_2  = \mu^\text{opt} \underbrace{(-\mathbf{y}^\mathrm{T}\mathbf{H})}_{\triangleq \mathbf{p}_{13}}\mathbf{x}_2,
\end{equation}
where again for future convenience we have implicitly defined the quantities $q_{11}$, $\mathbf{q}_{12}$, $\mathbf{q}_{13}$, $\mathbf{p}_{13}$, and
\begin{equation}
\label{eq:GEV_KKT_v3_3a}
\mathbf{x}_2 \triangleq(\mathbf{B} + \mu^\text{opt} \mathbf{H}^\mathrm{T}\mathbf{H})^{-1}(\mathbf{H}^\mathrm{T}\mathbf{H}\,\mathbf{x}_1 - \rho\mathbf{H}^\mathrm{T}\mathbf{y}),
\end{equation}
which in turn can be rearranged as
\begin{equation}
\label{eq:GEV_KKT_v3_3b}
\underbrace{-\mathbf{H}^\mathrm{T}\mathbf{y}}_{=\mathbf{q}_{12} ^\mathrm{T}}\rho + \underbrace{\mathbf{H}^\mathrm{T}\mathbf{H}}_{\triangleq\mathbf{q}_{22}}\,\mathbf{x}_1 + \underbrace{(-\mathbf{B})}_{\triangleq\mathbf{q}_{23}}\mathbf{x}_2 = \mu^\text{opt} \underbrace{\mathbf{H}^\mathrm{T}\mathbf{H}}_{\triangleq\mathbf{p}_{23}}\, \mathbf{x}_2,
\end{equation}
where we have highlighted the reappearance of the previously defined quantity $\mathbf{q}_{12}$ and implicitly defined the further quantities $\mathbf{q}_{22}$, $\mathbf{q}_{13}$ and $\mathbf{p}_{23}$ for future convenience.

Finally, notice that equation \eqref{eq:GEV_KKT_v3_1b} can also be rewritten as
\begin{equation}
\label{eq:GEV_KKT_v3_1a}
\underbrace{\mathbf{b}}_{=\mathbf{q}_{13}^\mathrm{T}}\rho + \underbrace{(-\mathbf{B})}_{=\mathbf{q}_{23} ^\mathrm{T}}\mathbf{x}_1 = \mu^\text{opt}\underbrace{(-\mathbf{H}^\mathrm{T}\mathbf{y})}_{=\mathbf{p}_{13}^\mathrm{T}}\rho + \mu^\text{opt} \underbrace{\mathbf{H}^\mathrm{T}\mathbf{H}}_{=\mathbf{p}_{23}^\mathrm{T}}\mathbf{x}_1,
\end{equation}
where we once more highlighted the reappearance of the quantities 
$\mathbf{q}_{13}$, $\mathbf{q}_{23}$, $\mathbf{p}_{13}$ and $\mathbf{p}_{23}$.

Now define the vector $\mathbf{x} \triangleq [\rho, \mathbf{x}^\mathrm{T}_1, \mathbf{x}^\mathrm{T}_2]^\mathrm{T}$ and notice that the collection of equations \eqref{eq:GEV_KKT_v3_2b}, \eqref{eq:GEV_KKT_v3_3b} and \eqref{eq:GEV_KKT_v3_1a}, in that order, can be written compactly as  
\begin{equation}
\label{eq:GEP_Original}
{\left[\begin{array}{@{\,}c@{\,}c@{\,}c@{}}
q_{11} & \mathbf{q}_{12} & \mathbf{q}_{13}\\
\mathbf{q}_{12} ^\mathrm{T} & \mathbf{q}_{22} & \mathbf{q}_{23}\\
\mathbf{q}_{13} ^\mathrm{T} & \mathbf{q}_{23} ^\mathrm{T} & \mathbf{0}_{2N_t}\\
\end{array}\right]}\!\!\cdot\!\!
\left[\begin{array}{@{\,}c@{\,}}
\rho \\
\mathbf{x}_{1} \\
\mathbf{x}_{2} \\
\end{array}\right]
 = \mu^\text{opt}
{\left[\begin{array}{@{\,}c@{\,}c@{\,}c@{}}
0 &  \mathbf{0}_{1\times 2N_t} & \mathbf{p}_{13}\\
\mathbf{0}_{2N_t\times 1}&  \mathbf{0}_{2N_t}& \mathbf{p}_{23}\\
\mathbf{p}_{13}^\mathrm{T} & \mathbf{p}_{23}^\mathrm{T} & \mathbf{0}_{2N_t}\\
\end{array}\right]}\!\!\cdot\!\!
\left[\begin{array}{@{\,}c@{\,}}
\rho \\
\mathbf{x}_{1} \\
\mathbf{x}_{2} \\
\end{array}\right]_{_{\!}}\!,
\end{equation}
or simply
\begin{equation}
\label{eq:GEP_OriginalSimple}
\mathbf{Q}\,\mathbf{x} = \mu^\text{opt}\mathbf{P}\,\mathbf{x},
\end{equation}
with
\begin{subequations}
\label{eqn:MatricesQP}
\begin{equation}
\label{eqn:MatrixQ}
\mathbf{Q} \triangleq 
\left[\begin{array}{@{\,}c c c@{\,}}
\mathbf{y}^\mathrm{T}\mathbf{y} - \sigma^2_{\mathbf{n}} &  -  \mathbf{y}^\mathrm{T} \mathbf{H}& \mathbf{b}^\mathrm{T}\\
-\mathbf{H}^\mathrm{T}\mathbf{y}&  \mathbf{H}^\mathrm{T}\mathbf{H}& - \mathbf{B} \\
\mathbf{b} & -\mathbf{B} & \mathbf{0}_{2N_t}
\end{array}\right]\!,\\
\end{equation}
\begin{equation}
\label{eqn:MatrixP}
\mathbf{P} \triangleq
\left[\begin{array}{c c c}
0 &  \mathbf{0}_{1\times 2N_t}&  -\mathbf{y}^\mathrm{T}\mathbf{H}\\
\mathbf{0}_{2N_t\times 1}&  \mathbf{0}_{2N_t}&    \mathbf{H}^\mathrm{T}\mathbf{H}\\
-\mathbf{H}^\mathrm{T}\mathbf{y}& \mathbf{H}^\mathrm{T}\mathbf{H}& \mathbf{0}_{2N_t}
\end{array}\right]\!.
\end{equation}
\end{subequations}

One can readily recognize that equation \eqref{eq:GEP_OriginalSimple} defines a generalized eigenvalue problem \cite{MyListOfPapers:Golub1996} over the pencil defined by the pair of matrices $(\mathbf{Q},\mathbf{P})$.
In other words, the solution of the system of equations in \eqref{eq:GEV_KKT_v2}, and therefore of the \ac{QCQP-1} problem described by equation \eqref{eqn:P4_OP1}, is among the generalized eigenvalues of the pencil $(\mathbf{Q},\mathbf{P})$.

Problems described by a quadratic program with a single quadratic constraint, such that the one dealt with here, were studied thoroughly in \cite{AdachiMP2019}.
It was shown thereby, in particular in \cite[Lem.3 and Th.4]{AdachiMP2019}, that in fact the solution of the \ac{QCQP-1} extracted from equation \eqref{eq:GEP_Original} is given by its \emph{smallest} generalized eigenpair.

It was also shown thereby, however, that such a solution is also equivalent to the \emph{largest} finite real generalized eigenvalue of the M\"obius-transform equivalent of equation \eqref{eq:GEP_Original}, namely
\begin{equation}
\label{eqn:Transformed_GEP}
\mathbf{P}\,\mathbf{x} = \lambda^\text{opt}\mathbf{Q}\,\mathbf{x},
\end{equation}
where $\lambda^\text{opt} = 1/\mu^\text{opt}$.

We remark that the computation of the dominant generalized eigenvalue of a pencil of symmetric matrices can be done accurately, stably and at relatively low complexity, since many classic \cite{Golub1996} and modern \cite{MyListOfPapers:CaiEUSIPCO2020, MyListOfPapers:ZhiqiangUAI2020, MyListOfPapers:RommesMoC2007} algorithms exist to that end.
In fact, since the matrix $\mathbf{P}$ is constant over the iterations of the \ac{IDLS} detector, while only the last block column and row of the the matrix $\mathbf{Q}$ is updated in that process, the recursive adaptation of $\lambda^\text{opt}$ from a previous to a next iteration is also possible via Jacobian techniques such as $e.g.$ \cite{MyListOfPapers:RommesMoC2007}.
Finally, and better still, as shall be numerically shown in Section \ref{sec:performance2}, it is found in practice that
$\lambda^\text{opt}$ converges quickly and stably, varying only minimally over different channel realizations and signal transmissions, such that the values calculated in previous runs of the detector can be reused.

All in all, we have therefore arrived at the elegant conclusion that the optimum regularization parameter $\lambda^\text{opt}$ required to evaluate equation \eqref{eq:PenalizedZF_soptReal} is given by the \emph{largest} finite generalized eigenvalue of the pencil $(\mathbf{P},\mathbf{Q})$, that is
\begin{equation}
\label{eq:OptimizedParameter}
\lambda^\text{opt} = {\rm maxeig}(\mathbf{P},\mathbf{Q}).
\end{equation}

\begin{algorithm}[H]
\SetKwInOut{Input}{Input}
\nonl\quad\\[-1ex]
{\dosemic\nonl{\bf External Input:}\\
Received signal ${\mathbf{y}}$, channel matrix ${\mathbf{H}}$
and noise power $\sigma^2_{\mathbf{n}}$.\\
}
\dosemic\nonl{\bf Internal Parameters:}\\
Maximum number of iterations $k_\text{max}$;\\
\dosemic\nonl convergence threshold $\varepsilon\ll 1$ and shaping parameter $\alpha\ll 1$.\\
\dosemic\nonl{\bf Initialization:}\\
Set iteration counter $k = 0$;\\
\dosemic\nonl
Set initial solution to $\mathbf{s}^{(k)}=(\mathbf{H}^\mathrm{T}\mathbf{H})^{-1}\mathbf{H}^\mathrm{T}\mathbf{y}$\\
\Repeat{$k > k_\textup{max}$ \textup{or} $\varepsilon_k < \varepsilon$}{
Increase iteration counter $k = k + 1$\\
\noindent Update $\beta_{i,j}\forall\;i,j$, as in equation \eqref{eqn:beta_ZFReal}\\
Construct $\mathbf{b}$ and $\mathbf{B}$ from equations \eqref{eqn:BMatZFReal_b} and \eqref{eqn:BMatZFReal_B}\\
Construct $\mathbf{Q}$ and $\mathbf{P}$ from equations \eqref{eqn:MatrixQ} and \eqref{eqn:MatrixP}\\
Obtain $\lambda^{\text{opt}(k)}$ as in equation \eqref{eq:OptimizedParameter}\\
Update $\mathbf{s}^{(k)}$ as in equation \eqref{eq:PenalizedZF_soptReal}\\
Calculate $\varepsilon_k = \norm{\mathbf{s}^{(k)}-\mathbf{s}^{(k-1)}}_2$
}
\caption[]{Iterative Discrete Least Square Detector}
\label{alg:idls}
\end{algorithm}
\setlength{\textfloatsep}{4pt}
  
Finally, notice that although the result summarized by equation \eqref{eq:OptimizedParameter} is derived here for the particular case of the real-domain \ac{ML}-derived formulation problem in equation \eqref{eqn:IDLS_VecReal}, both the mathematical equivalence between the latter and equation \eqref{eqn:PenalizedZF} and the fact that the generalized eigenvalues of Hermitian pencils are real \cite{Golub1996, FarahMScThesisUniBirm2012}, implicate that the result holds also for the optimization of the penalization parameter for the complex-domain formulation of the problem given in equation \eqref{eqn:PenalizedZF}, and consequently also for its complex-valued \ac{IDLS} solution given by equation \eqref{eq:PenalizedZF_soptComplex}, sufficing it in this case to construct corresponding complex Hermitian variations of the matrices $\mathbf{Q}$ and $\mathbf{P}$, with the quantities $\mathbf{y}$, $\mathbf{H}$, $\mathbf{b}$ and $\mathbf{B}$ replaced by their complex-valued counterparts, and accordingly with the transpose operation $^\text{T}$ replaced by the Hermitian $^\text{H}$.

With that, for the convenience of the reader, we summarize the overall \ac{IDLS} scheme in the form of a pseudo-code in Algorithm \ref{alg:idls} adopting the real-valued variation for mathematical consistency, since the proofs of \cite[Th.3.3]{MorePMS1993} and \cite[Lem.3 and Th.4]{AdachiMP2019} were offered only for real matrices.

\subsection{Performance Assessment}
\label{sec:performance1}

In this subsection we offer a computer simulation-based performance evaluation of the \ac{IDLS} scheme described above.

In order to focus on the gains achieved by the new method over \ac{SotA}, we conduct the assessment here under ideal receiver conditions, namely, under the assumption that no distortions due to hardware impairments exists, and that perfect \ac{CSI} is available at the receiver, leaving those factors to be addressed later in the subsequent section, when the our proposed framework will be also revisited to yield \ac{IDLS} variations robust to such distortions.

Aside from these receiver-related limitations, however, system level conditions such as fading, channel correlation and overloading, which typically contribute to the performance deterioration of multiuser symbol detection, are considered.
In particular, the following two distinct flat-fading channel models will me employed.

\begin{itemize}
  \item \textit{Uncorrelated Rayleigh Fading Model}: This is the most commonly used channel model to evaluate ideal detection performance in the wireless literature. Each element of the channel matrix $\bm{H}$ is assumed to be an \ac{i.i.d.} circularly symmetric complex Gaussian random variable with zero mean and variance $1$, $i.e.$, $h_{m,n}\sim\mathcal{CN}(0,1)$ where $h_{m,n}$ denotes the element in the $m$-th row and $n$-th column of $\bm{H}$. This model captures, among others, the channel conditions of the uplink of a cell-free \ac{MIMO} system serving distributed single-antenna users with either \ac{OMA} or \ac{NOMA} protocol \cite{EmilTWC20}, in the fully loaded and overloaded cases, respectively.
  \item \textit{Exponentially-correlated Jakes Fading Model}: In this case, the spatial correlation between antenna elements is captured via the classical Jakes correlation model, in which the block-fading matrix $\bm{H}\in\mathbb{C}^{N_r\times N_t}$ is described via the following doubly-correlated channel matrix:
  \begin{equation}
  \label{eq:correlatedchannel}
  \bm{H} = \bm{\Phi}^{\frac{1}{2}}_r\bm{H}_\text{\ac{i.i.d.}}\bm{\Phi}^{\frac{1}{2}}_t,
  \vspace{-0.5ex}
  \end{equation}
  where $\bm{H}_\text{\ac{i.i.d.}}$ denotes an uncorrelated \ac{i.i.d.} zero mean, unit variance complex Gaussian matrix, and with the $(k,\ell)$-th element of the spatial correlation signatures $\bm{\Phi}_r$ and $\bm{\Phi}_t$ given by $J_0\big(\frac{2\pi f_c d |k-\ell|}{c}\big)$, where $f_c = 5$ [GHz] depicts the carrier frequency, $d$ is defined as the corresponding half antenna spacing, $c$ denotes the speed of light, and  $J_0(\cdot)$ is the zeroth order Bessel function of the first kind\cite{HyundongTWC03}.
\end{itemize}

\vspace{1ex}
\subsubsection{System Parameters}
\vspace{1ex}

As for the remaining system configuration, the following is considered.
The $\ell_0$-norm approximation tightness parameter is set to $\alpha = 0.1$.
Various loading conditions are simulated, ranging from fully loaded systems with $N_t = N_r = 100$, to moderately overloaded systems are subjected to either a $25\%$ of excess load, with $N_t/N_r = 1.25$, to severely overloaded systems with a $50\%$ of excess load, with $N_t/N_r = 1.5$.
Results are shown as a function of the \ac{$E_b/N_0$}, such that the corresponding \ac{AWGN} noise variance $\sigma^2_{{\bm{n}}}$ in a system with $N_t$ transmit antennas is given by $\sigma^2_{{\bm{n}}} = {N_t}/(b\cdot 10^{\frac{\rm \text{\ac{$E_b/N_0$}} [dB]}{10}})$.
Finally, it is assumed that the elements of the symbol vector $\bm{s}$ are sampled with uniform probability from a Gray-coded \ac{QPSK} modulation, such that $b=2$.

\vspace{1ex}
\subsubsection{Comparative Results}
\label{subsec:IDLSResults}
\vspace{1ex}

\begin{figure}[p]
  \centering
  \begin{subfigure}[b]{0.45\columnwidth}
  \centering
  \includegraphics[width=\columnwidth]{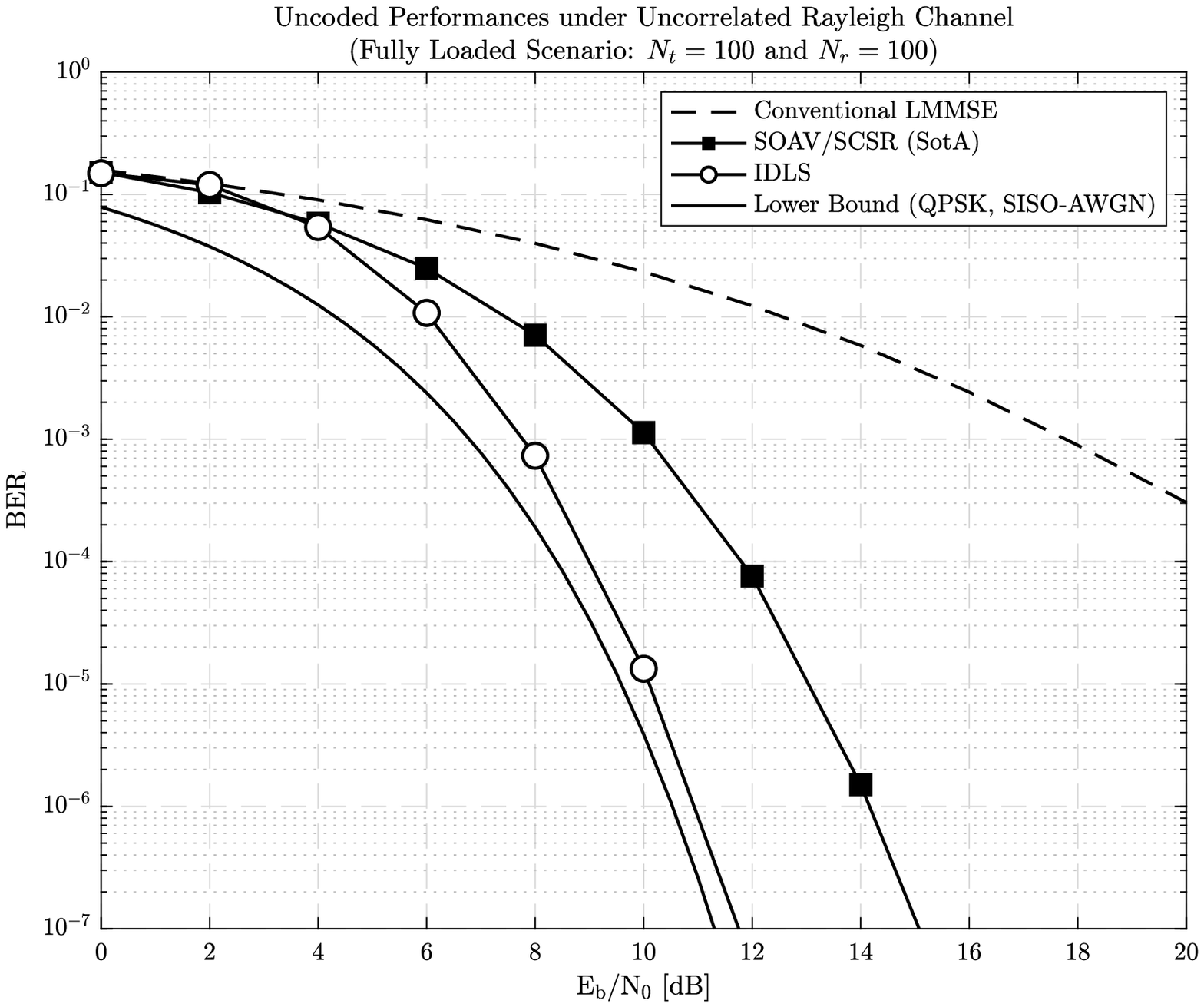}
  \caption{Fully Loaded Uncorrelated.}
  \label{fig:IDLSxSotA_FullyloadedUncorrelated}
  \vspace{1ex}
  \end{subfigure}
  \begin{subfigure}[b]{0.45\columnwidth}
  \centering
  \includegraphics[width=\columnwidth]{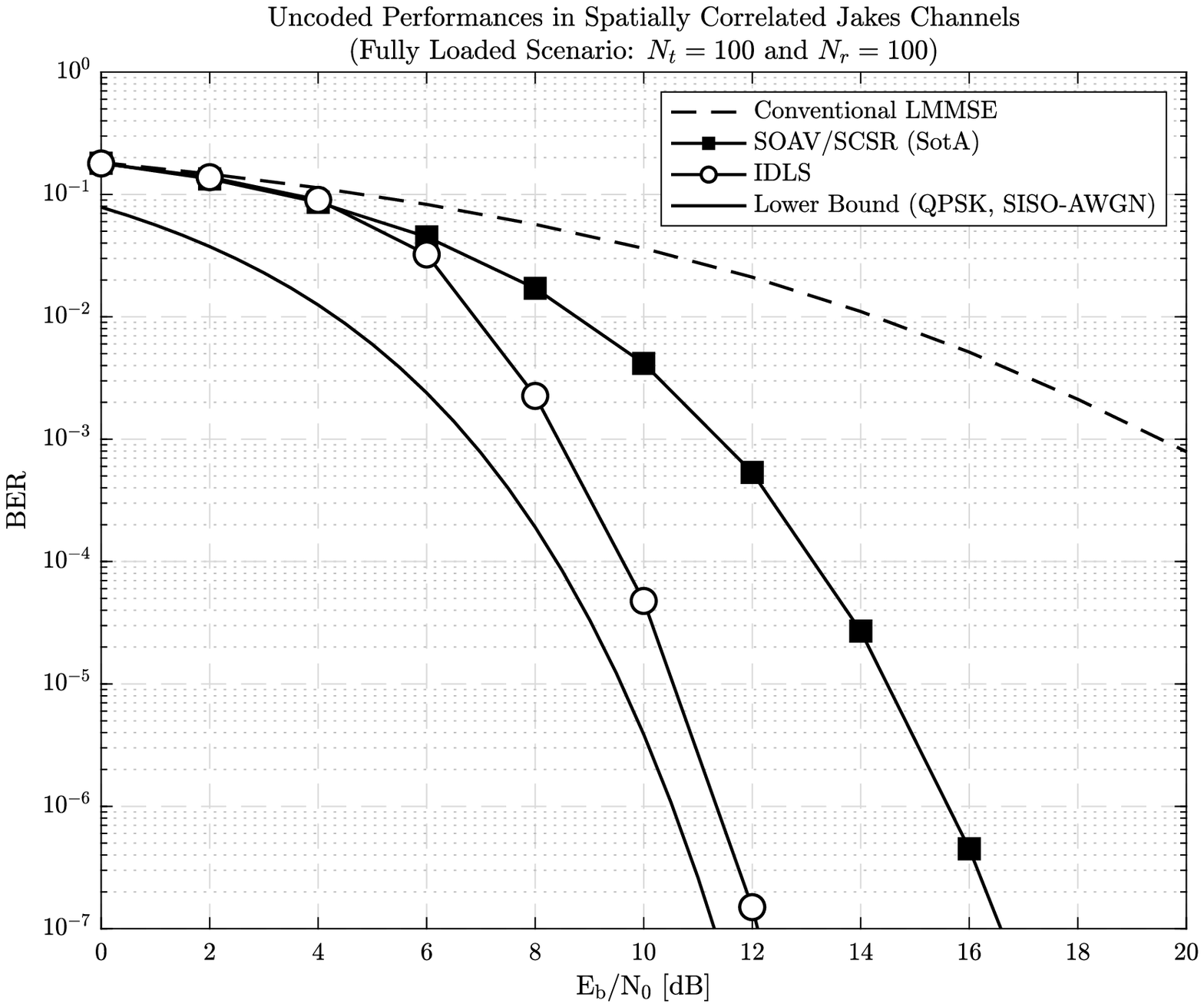}
  \caption{Fully Loaded Jakes Correlated.}
  \label{fig:IDLSxSotA_FullyloadedJakesCorrelated}
  \vspace{1ex}
  \end{subfigure}
  \begin{subfigure}[b]{0.45\columnwidth}
  \centering
  \includegraphics[width=\columnwidth]{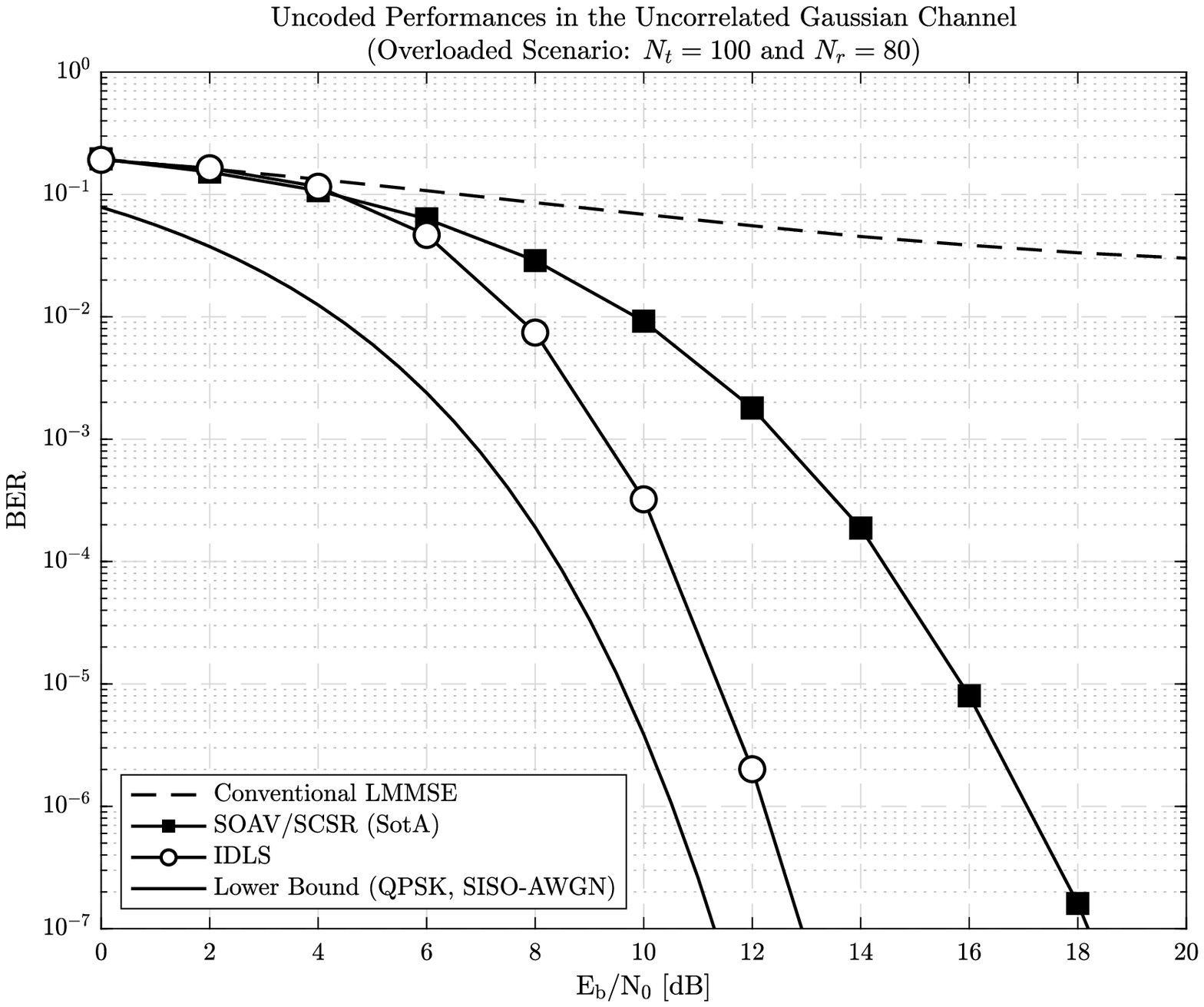}
  \caption{Moderately Overloaded Uncorrelated.}
  \label{fig:IDLSxSotA_ModeratelyOverloadedUncorrelated}
  \vspace{1ex}
  \end{subfigure}
  \begin{subfigure}[b]{0.45\columnwidth}
  \centering
  \includegraphics[width=\columnwidth]{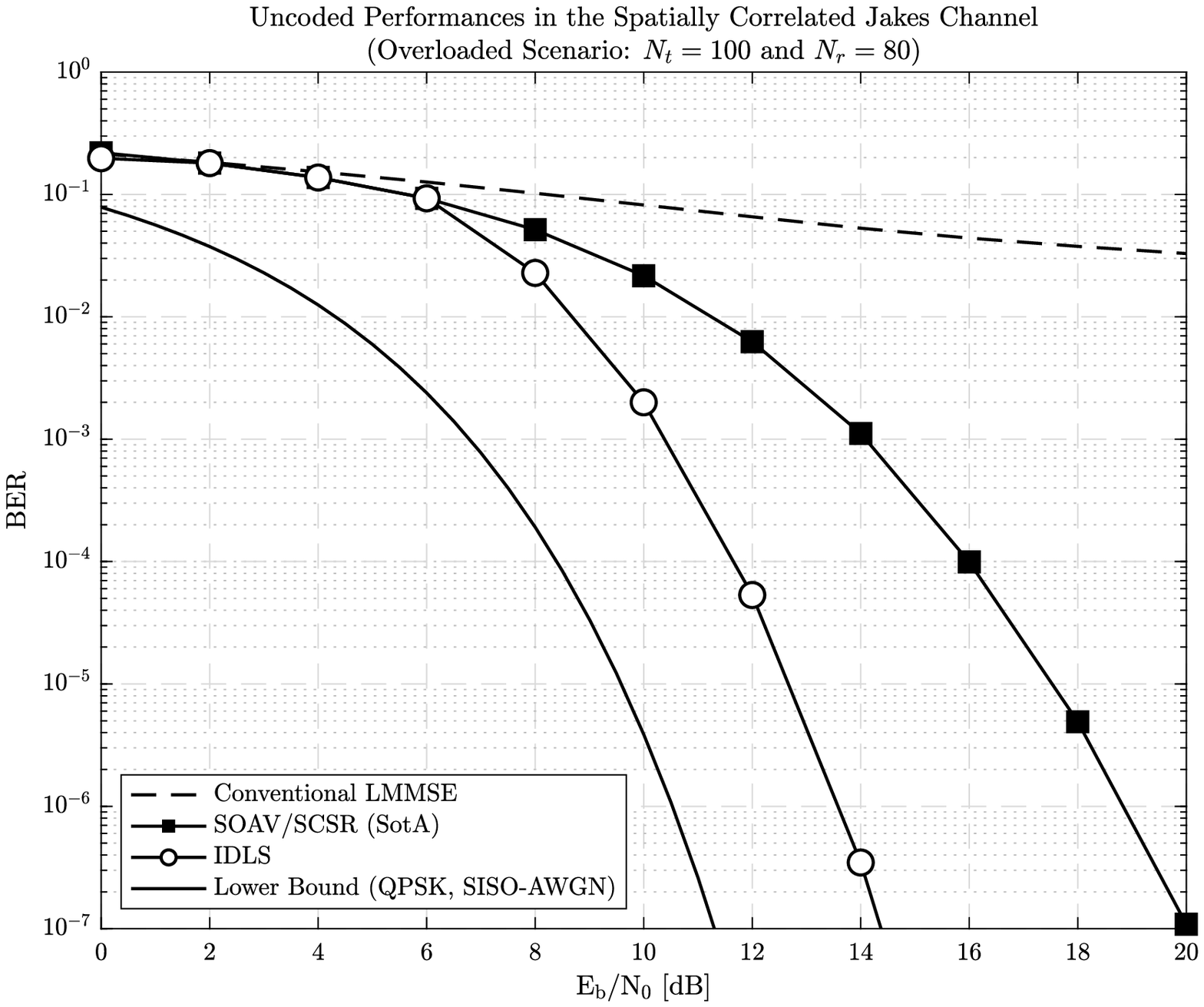}
  \caption{Moderately Overloaded Jakes Correlated.}
  \label{fig:IDLSxSotA_ModeratelyOverloadedJakesCorrelated}
  \vspace{1ex}
  \end{subfigure}
  \begin{subfigure}[b]{0.45\columnwidth}
  \centering
  \includegraphics[width=\columnwidth]{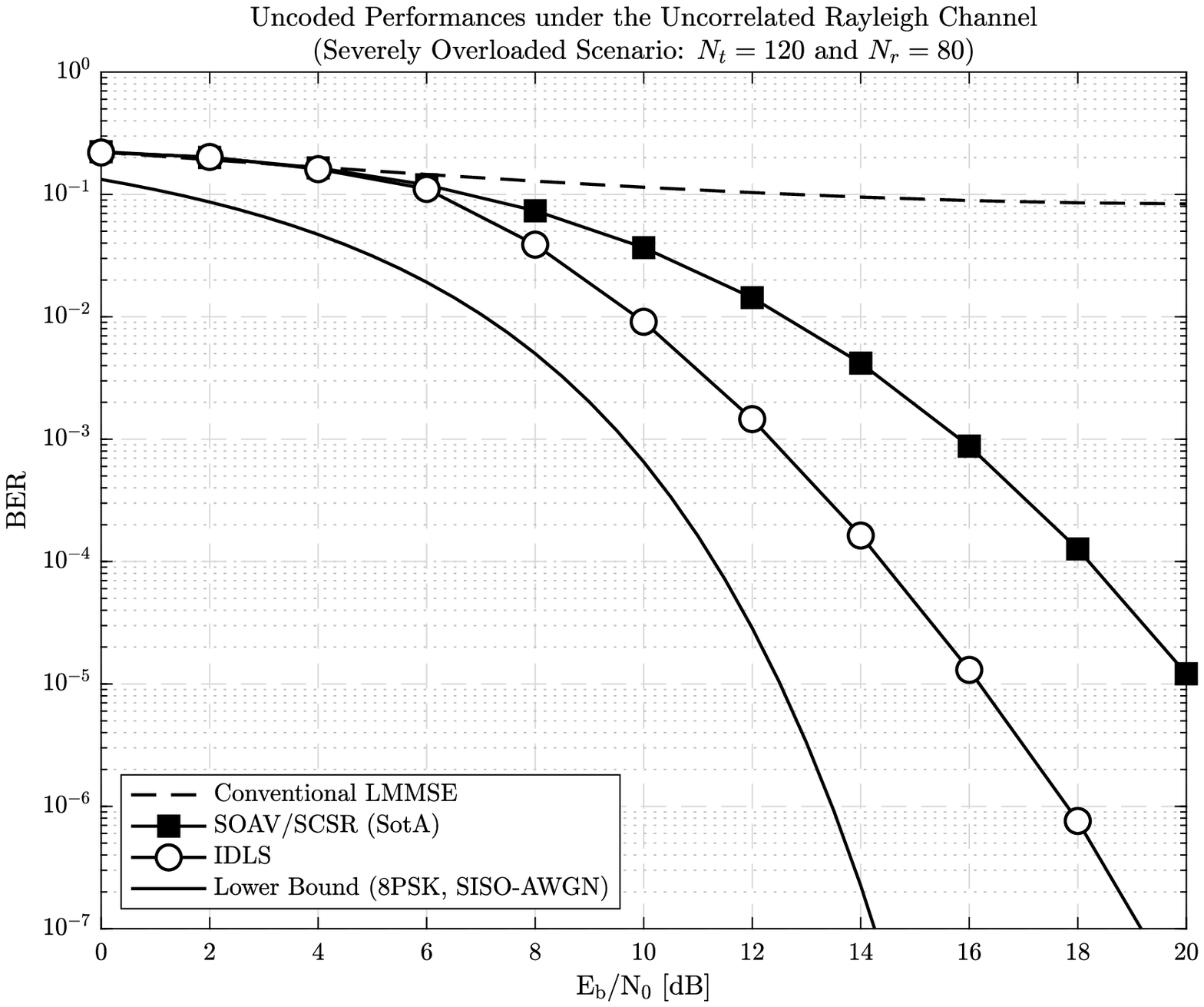}
  \caption{Severely Overloaded Uncorrelated.}
  \label{fig:IDLSxSotA_SeverelyOverloadedOneUnorrelated}
  \end{subfigure}
  \begin{subfigure}[b]{0.45\columnwidth}
  \centering
  \includegraphics[width=\columnwidth]{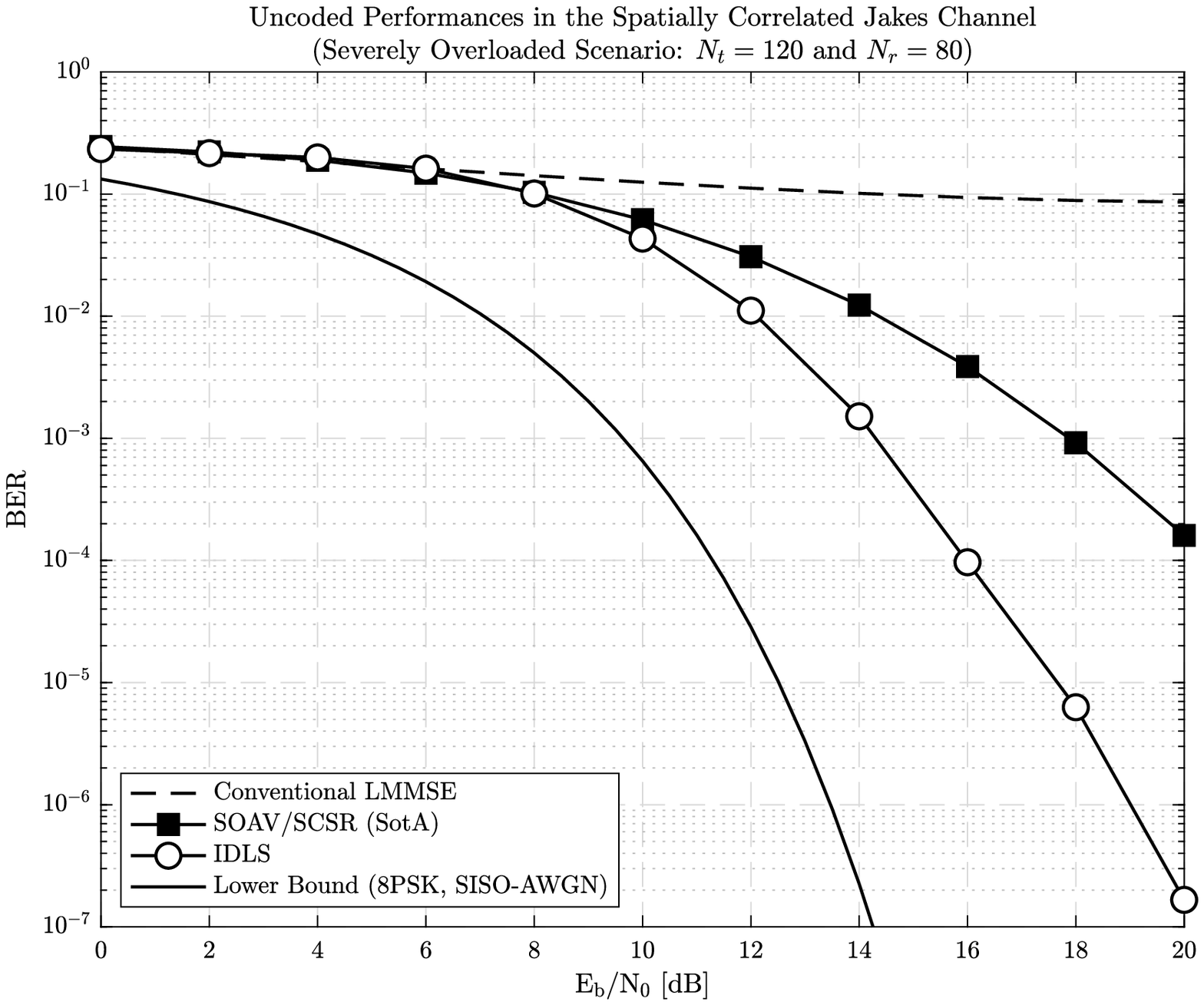}
  \caption{Severely Overloaded Jakes Correlated.}
  \label{fig:IDLSxSotA_SeverelyOverloadedJakesCorrelated}
  \end{subfigure}
  \caption[]{Uncoded \ac{BER} performance of the \acl{IDLS} described in  Algorithm \ref{alg:idls}, compared against conventional \ac{LMMSE} and \ac{SotA} multiuser detection schemes under various loading and channel conditions.}
  \label{fig:IDLSxSotA}\vspace{2ex}
\end{figure}

The set of simulated results are compactly displayed in Figure \ref{fig:IDLSxSotA}, which shows plots of the \ac{BER} performance achieved by the \ac{IDLS} multiuser symbol detection method described in the preceding subsections, compared against those of the conventional \ac{LMMSE}, low-complexity \ac{SOAV} and \ac{SCSR} \ac{SotA} receivers, as well as the \ac{SISO} AWGN lower bound, under various SNR, loading and channel correlation conditions.

The results altogether clearly demonstrate not only that the new \ac{IDLS} detector offers substantial gains over the \ac{SotA} and classic alternatives, but also that the proposed method exhibits a remarkable robustness to overloading and channel correlation conditions, which are known to be a fundamental cause of performance degradation in multiuser systems.
In particular, it is visible that in all cases the inclination of the \ac{BER} curves corresponding to the \ac{IDLS} method is the same as that achieved by in a \ac{SISO} with the same spectral efficiency.

We also call attention to the fact that all simulation results displayed are for systems of relatively large scales, which are therefore not tractable to other types of \ac{ML}-approaching receiver architectures based on techniques such as sphere detection \cite{ChenTWC2013}.
In other words, the results serve the additional purpose of effectively illustrating that the remarkable performances observed are achieved at a remarkably low complexity.

\begin{figure}[t]
  \centering
  \begin{subfigure}[b]{.45\columnwidth}
  \centering
  \includegraphics[width=0.95\columnwidth]{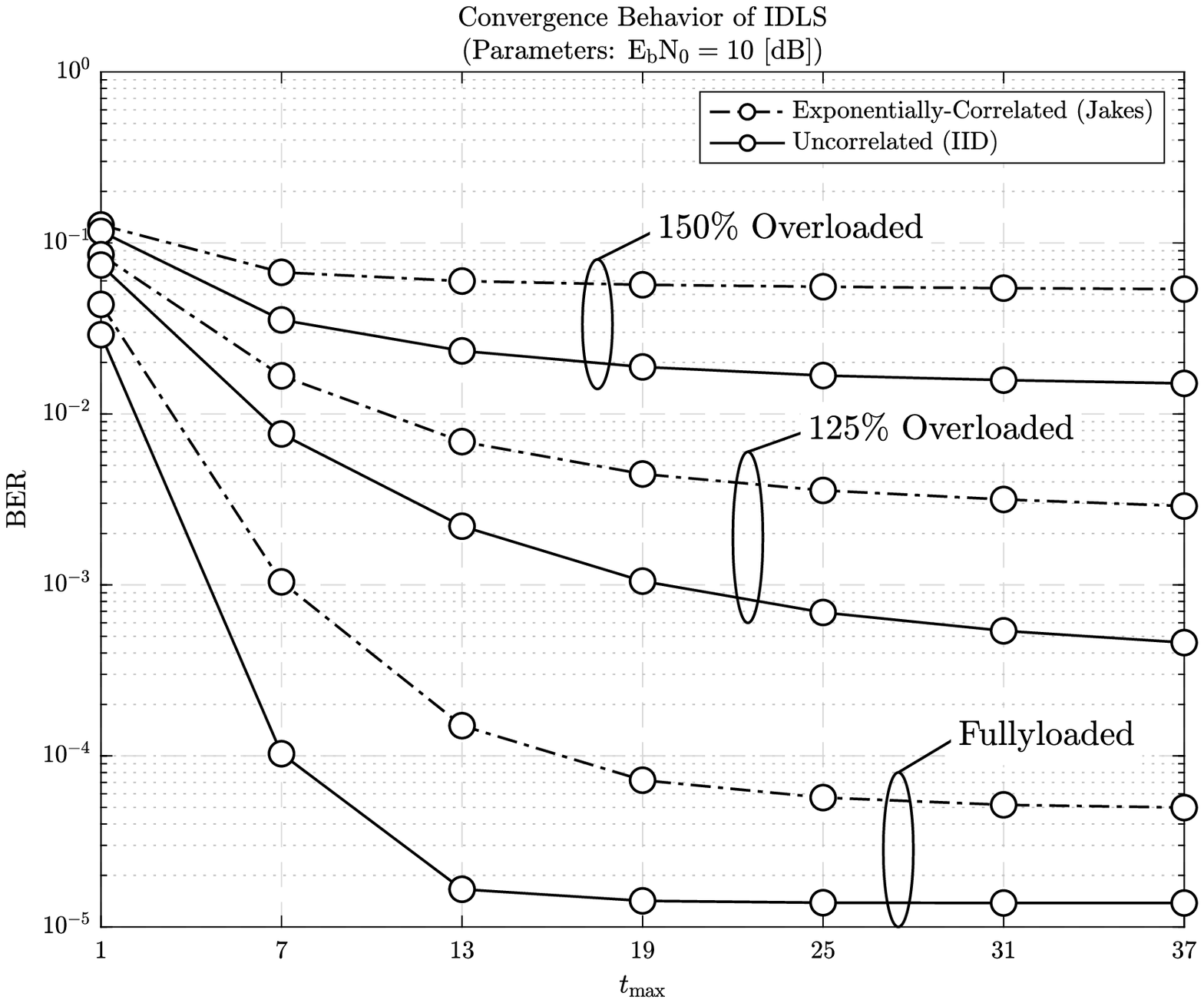}
  \caption[]{Convergence: \ac{BER} $\times$ No. of Iterations.}
  \label{fig:Convergence}
  \vspace{1ex}
  \end{subfigure}
  \begin{subfigure}[b]{.45\columnwidth}
  \centering
  \includegraphics[width=0.95\columnwidth]{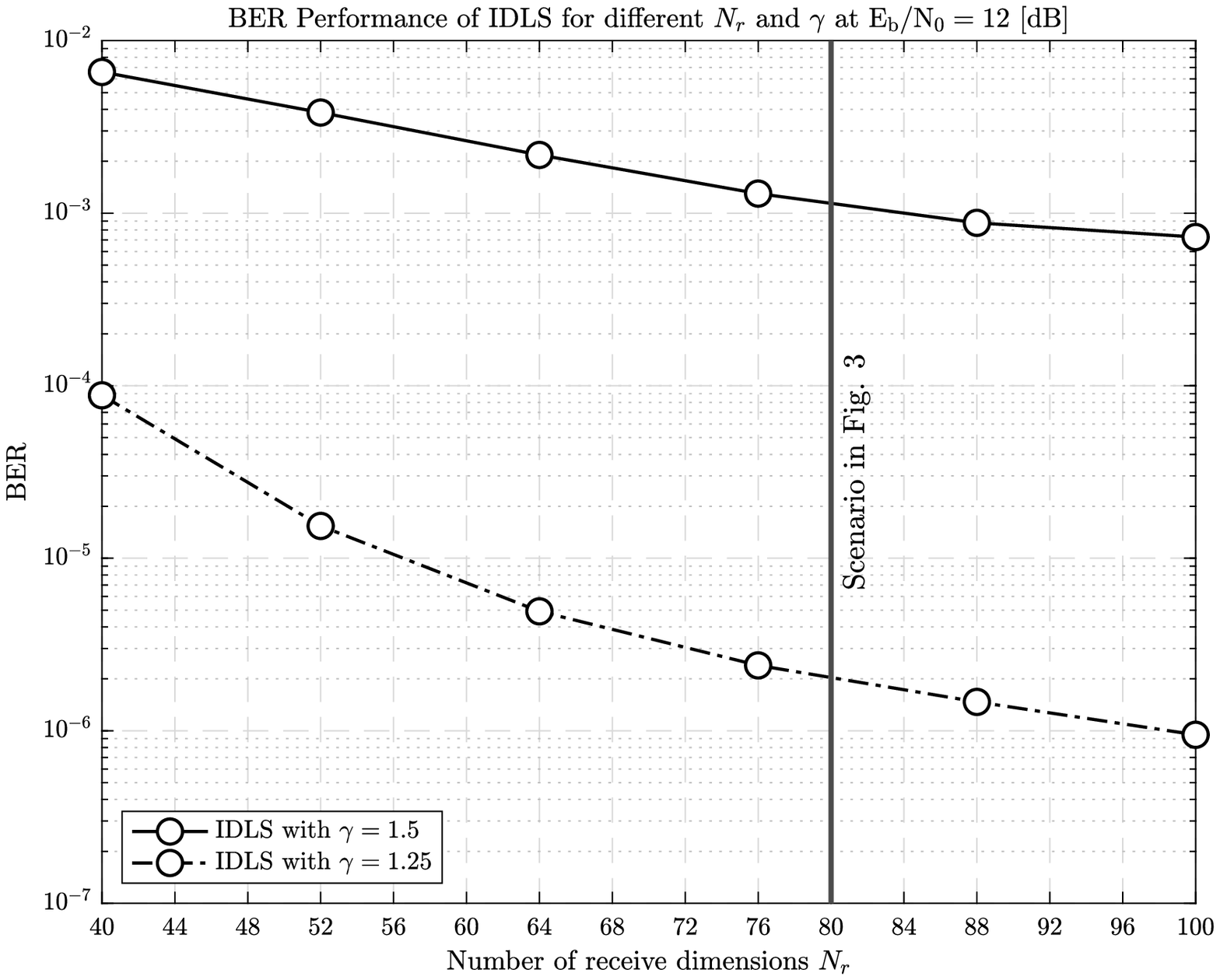}
  \caption{Asymptotic: \ac{BER} $\times$ System Size.}
  \label{fig:Asymptotics}
  \end{subfigure}
  \caption[]{Convergence and asymptotic behaviors of the \acl{IDLS} detector described in Algorithm \ref{alg:idls}.}
  \label{fig:Convergence_and_AsymtoticsUncorrelated}
  \vspace{1ex}
  \includegraphics[width=0.45\columnwidth]{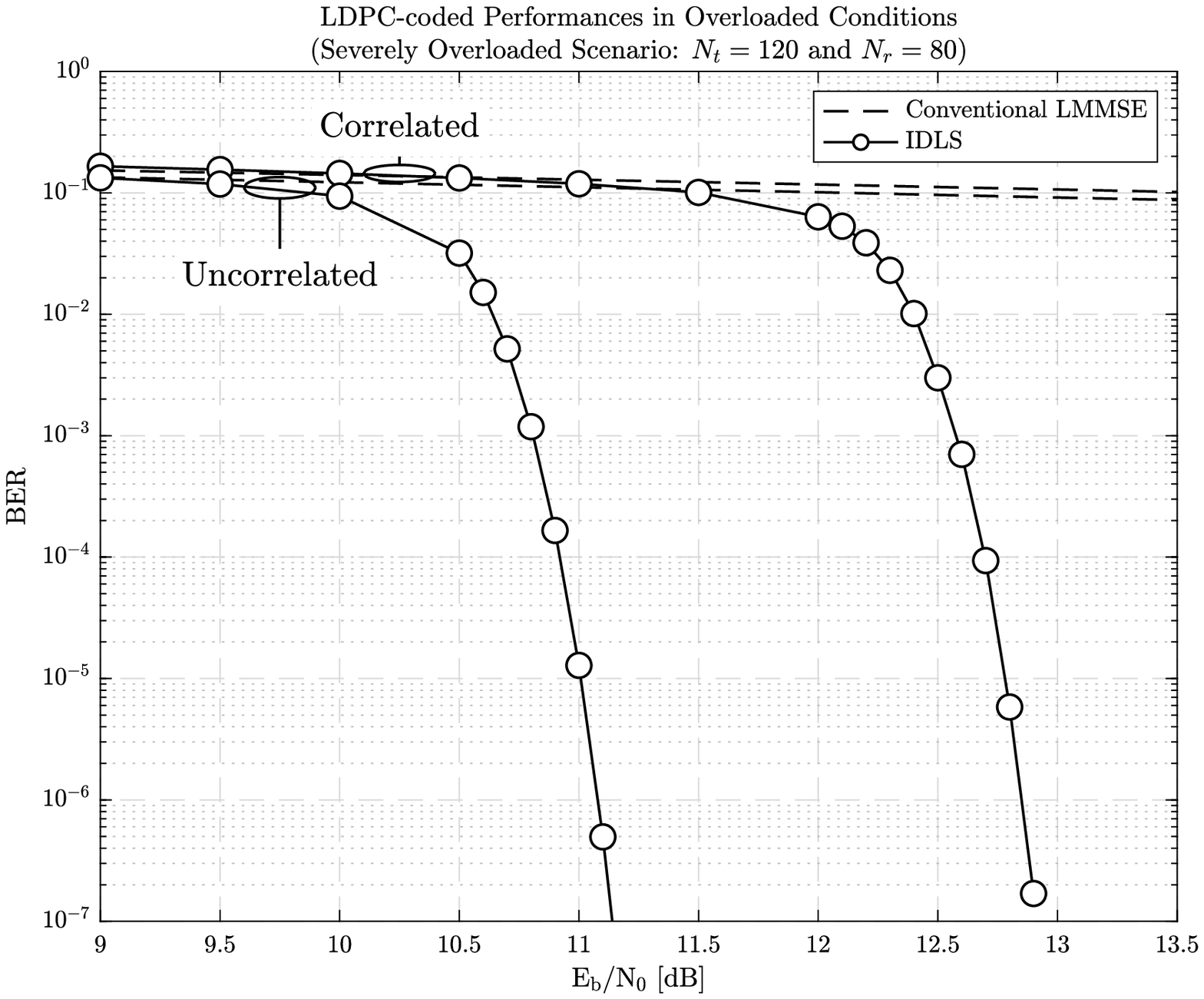}
  \caption[]{\ac{BER} performance of \ac{IDLS} and \ac{LMMSE} receivers combined with the rate-$1/2$ \ac{LDPC} of DVB-S2 standard.}
  \label{fig:CodedBER}
  \end{figure}

In order to further highlight this high-performance-at-low-complexity feature of the \ac{IDLS} framework, plots depicting its convergence and asymptotic behaviors, as a function of the number of iterations and system size, respectively, are shown in Figure 
\ref{fig:Convergence_and_AsymtoticsUncorrelated}.
The results of Figure \ref{fig:Convergence} indicate that less than $25$ iterations are sufficient for \ac{IDLS} to converge in almost all the conditions considered, with $37$ iterations sufficing for all the cases.

Interestingly (but non-surprisingly), it is also found that the number of iterations required until convergence is not strongly affected by the dimension of the system or loading conditions, as shown in the figure.

As for asymptotic behavior, it can be seen from Figure \ref{fig:Asymptotics} that indeed the \ac{BER} performance of the \ac{IDLS} detector enjoys the array or dimension gain as expected.
This is due to the fact that owing to the $\ell_0$-norm formulation embedded into \ac{IDLS}, a large system condition facilitates the detection performance of \ac{IDLS} as is the case with existing compressive sensing algorithms.

This remarkable resilience of the \ac{IDLS} detector against \ac{ISI} and \ac{MUI}, demonstrated when operating over uncoded symbols, which shows no signs of error floors even under severe overloading conditions and subjected to the further rank deficiencies resulting from channel correlation, motivates us to probe the performance of the method when aided by channel coding.
To that end, we compare in Figure \ref{fig:CodedBER} the \ac{BER} achieved by the \ac{IDLS} detector against that of the conventional \ac{LMMSE} receiver in a system with overloading ratio $\gamma=1.5$, this time equipping both receivers with the standardized DVB-S2 \ac{LDPC} of rate $1/2$, decoded with $50$ sum-product iterations.
The figure indicates that, despite the severe overloading conditions, \ac{LDPC}-coded \ac{IDLS} exhibits the waterfall \ac{BER} curves typical of receivers free of \ac{ISI} or \ac{MUI}.

In comparison, the conventional \ac{LMMSE} receiver is, as expected, simply not applicable in such scenarios. 
We attribute the notable performance gain of the \ac{IDLS} detector over the \ac{LMMSE} to its ability to close the ``gap'' that exists between the assumption of continuity required to enable invoking the Wirtinger derivative and the reality that the solution must belong to a discrete constellation.
As touched upon in Section \ref{sec:IDLSSolution}, in the derivation of the classical \ac{LMMSE}, this inconsistency is left unaddressed \cite{ShaoshiCST2015}, while in the proposed method, the solution is obtained over a space which is strictly speaking continuous on $\bm{s}$, but which at the same time has a strong ``bias'' towards solutions belonging to $\mathcal{C}$, imposed by the parameters $\alpha\ll 1$ and $\lambda > 0$.

It is this overall combination of high performance and low-complexity offered by the \ac{IDLS} scheme that motivates us, in the subsequent section, to further develop its robustness against various other factors of practical relevance such as noisy conditions, \ac{CSI} imperfection and hardware impairments, with great modularity, consequently extending the \ac{IDLS} method into a more general framework for the symbol detection of overloaded systems.

\section[]{Extensions for Robust \ac{IDLS} Detector}
\label{sec:practical}
\vspace{-0.5ex}

In the preceding section, we constructed a novel \ac{ML}-derived iterative discrete least squares detector which, taking advantage of a tight convexized approximation of the $\ell_0$-norm, enables the \ac{ISI}/\ac{MUI}-free detection of symbol vectors subjected to rank-deficient channels resulting from overloading or correlation.
The new scheme, which we refer to as the \ac{IDLS} detector, was found to yield \ac{BER} performances that are free of error floors and significantly superior to those of the classic \ac{LMMSE} receiver and of more recent \ac{SotA} methods.

The \ac{IDLS} detector was, however, derived and evaluated under the ideal assumption of perfect \ac{CSI} knowledge and impairment-free hardware, while in practice \ac{CSI} errors and hardware imperfection are possible additional causes of performance degradation.
And although \ac{CSI} and hardware imperfections can be mitigated by increasing the transmit power and quantity of pilot symbols, and the quality of \ac{RF} components, respectively, these measures come along with undesirable consequences such as the increase in energy consumption, communication delay and engineering costs.

We therefore proceed to show in this section that the \ac{IDLS} detector can be extended so as to incorporate robustness to the aforementioned factors.
In the process, it will become visible that, thanks to the flexibility of its formulation and the simplicity of its solution, the \ac{IDLS} scheme is indeed very convenient in enabling these and possibly further extensions, implicating therefore that the method can be seen as a generic framework for the design of robust, effective and low-complexity detectors for overloaded systems.

\vspace{-1ex}
\subsection{Mitigating Noisy Conditions}
\label{sec:DALMMSE}

In many cases, the effect of imperfections in \ac{CSI} or hardware are perceived in the form of an increase in the noise level experienced at the receiver, without the possibility for the receiver to estimate proportionality parameters that quantify the percentage of such noise increase relative to thermal and background \ac{RF} noise.

It makes sense, therefore, to start our effort of extending the \ac{IDLS} detector by considering a mechanism to combat a generic and uncategorized noise level, under the knowledge of only the aggregate noise power $\sigma_{\mathbf{n}}^2$.
To that end, recall the conventional linear \ac{MMSE} detector described by the equation
\begin{equation}
\label{eqn:MMSE}
\bm{s}^\text{MMSE} = (\bm{H}^\mathrm{H}\bm{H} + \sigma^2_{{\bm{n}}}\bm{I}_{N_r})^{-1}\bm{H}^\mathrm{H}\bm{y}.
\vspace{-0.5ex}
\end{equation}


Recall also the Woodbury inverse lemma \cite[Sec.2.7.3]{TeukolskyBook2007}, which applied to the context hereby establishes the equivalence
\begin{equation}
\label{eq:WoodburyLemma}
{\bm{H}}^\mathrm{H}({\bm{H}}{\bm{H}}^\mathrm{H} + \sigma^2_{{\bm{n}}}\bm{I}_{N_r})^{-1} = ({\bm{H}}^\mathrm{H}{\bm{H}} + \sigma^2_{{\bm{n}}}\bm{I}_{N_r})^{-1}{\bm{H}}^\mathrm{H}.
\vspace{-0.5ex}
\end{equation}

Next, notice that by force of the Woodbury inverse lemma, equation \eqref{eqn:MMSE} can in fact be recognized as the solution of the following regularized least square problem
\begin{equation}
\label{eqn:MMSE_LS}
{\mathop {\mathrm {argmin}} \limits_{\bm{s} \in {\mathbb{C}}^{N_t}}}\:\: \|{\bm{y}}  - {\bm{H}}\bm{s}\|^2_2 + \sigma^2_{{\bm{n}}}\|\bm{s}\|^2_2.
\vspace{-0.5ex}
\end{equation}

In light of the discussions of Subsection \ref{sect:ProblemSetup}, as well as the derivations and results of Section \ref{sect:sect3}, it can now readily understood that the well-known vulnerability of the classic \ac{LMMSE} to overloading and channel correlation stems from the lack, in equation \eqref{eqn:MMSE_LS}, of a constraint that enforces compliance of the solution to the prescribed symbol constellation $\mathcal{C}$, which can then be corrected by reformulating the latter as
\vspace{-0.5ex}
\begin{subequations}
\label{eqn:DAMMSE_LS}
\begin{eqnarray}
\label{eqn:DAMMSE_LSObj}
{\mathop {\mathrm{minimize}} \limits_{\bm{s} \in \mathbb {C}^{N_t}}} &&\!\!\!\!\|\bm{y} - \bm{H}\bm{s}\|^2_2 + \sigma^2_{{\bm{n}}}\|\bm{s}\|^2_2,\\[-2ex]
\label{eqn:DAMMSE_LSConstraint}
\mathrm{subject\,to} &&\!\!\!\! \sum^{2^{b}}_{i=1} \|\bm{s} - c_i\mathbf{1}\|_0 = N_t\cdot(2^{b}-1),
\end{eqnarray}
\end{subequations}
which in turn can be transformed into a regularized \ac{LS} variation, namely
\vspace{-1ex}
\begin{equation}
\label{eqn:DALMMSE_OP1}
{\mathop {\mathrm{min}} \limits _{\bm{s} \in \mathbb {C}^{N_t}}}\:\: \|{\bm{y}} - {\bm{H}}\bm{s}\|^2_2 + \sigma^2_{{\bm{n}}}\|\bm{s}\|^2_2 + \lambda\sum^{2^b}_{i=1}\|\bm{s} - c_i\mathbf{1}\|_0.
\vspace{-1ex}
\end{equation}

Succinctly, introducing the $\ell_0$-norm approximation into equation \eqref{eqn:DALMMSE_OP1} and subsequently applying the \ac{QT}, as described in Section \ref{sect:sect3}, we readily obtain
\vspace{-1ex}
\begin{equation}
\label{eqn:DALMMSE_OP3_weighted}
{\mathop {\mathrm{min}} \limits _{\bm{x} \in \mathbb {C}^{N_t}}}\:\: \|{\bm{y}} - {\bm{H}}\bm{s}\|^2_2 + \sigma^2_{{\bm{n}}}\|\bm{s}\|^2_2 + \lambda\sum^{2^b}_{i=1}
\sum^{N_t}_{j=1}\beta_{i,j}^2|s_j - c_i|^2,
\end{equation}
with $\beta_{i,j}$ given, as previously, by equation \eqref{eqn:beta_ZF}, and which can be rewritten as
\vspace{-0.5ex}
\begin{equation}
\label{eqn:DALMMSE_FINAL}
{\mathop {\mathrm {min}} \limits_{\bm{s} \in \mathbb {C}^{N_t}}}\:\:
\bm{s}^\mathrm{H}\Big({\bm{H}}^\mathrm{H}{\bm{H}} + \sigma^2_{{\bm{n}}}\bm{I}_{N_r}\!+ \lambda\bm{B}\Big)\bm{s} - 2\mathrm{Re}\Big\{\!\big({\bm{H}} ^\mathrm{H}{\bm{y}}\! +\! \lambda\bm{b}\big)^\mathrm{H}\bm{s}\Big\},
\end{equation}
with $\bm{b}$ and $\bm{B}$ as defined in equations \eqref{eqn:bvecZF} and \eqref{eqn:BmatZF}, respectively, and which in turn admits the iterative closed-form solution
\begin{subequations}
\label{eqn:opt_x_DALMMSE}
\begin{equation}
\label{eqn:opt_x_DALMMSEComplex}
\bm{s} =  \Big({\bm{H}}^\mathrm{H}{\bm{H}} + \sigma^2_{{\bm{n}}}\bm{I}_{N_r}+ \lambda\bm{B}\Big)^{-1}\big({\bm{H}} ^\mathrm{H}{\bm{y}}\! +\! \lambda\bm{b}\big),
\end{equation}
or equivalent in the real domain

\begin{equation}
\label{eqn:opt_x_DALMMSEReal}
\mathbf{s} =  \big({\mathbf{H}}^\mathrm{T}{\mathbf{H}} + \sigma^2_{{\mathbf{n}}}\mathbf{I}_{2N_r}+ \lambda\mathbf{B}\big)^{-1}\big({\mathbf{H}} ^\mathrm{T}{\mathbf{y}}\! +\! \lambda\mathbf{b}\big),
\end{equation}
in which case $\mathbf{s}$ and $\mathbf{H}$ are as in equation \eqref{eq:Complex2RealMapping}, while $\mathbf{b}$ and $\mathbf{B}$  are defined in equation \eqref{eqn:BMatZFReal}.
\end{subequations}

Note that if the noise variance is not available at the receiver ($i.e.,$ $\sigma^2_{{\bm{n}}}=0$), equation \eqref{eqn:opt_x_DALMMSE} reduces to the previously derived \ac{IDLS} detector described by equation \eqref{eq:PenalizedZF_sopt}, while setting $\lambda=0$ leads to the conventional linear \ac{MMSE} detector of equation \eqref{eqn:MMSE}. 
In other words, it can be concluded that equation \eqref{eqn:opt_x_DALMMSE} is generalization of both the \ac{IDLS} and the conventional linear \ac{MMSE} receivers in terms of noise- and constellation-awareness, respectively.

As for the regularization parameter $\lambda$, once again the same procedure described in Subsection \ref{sec:auto_parameterization} can be reproduced.
In particular, the real-valued equivalent of equation \eqref{eqn:DALMMSE_FINAL} can once again be cast into the \ac{QCQP-1}
\begin{subequations}
\label{eqn:QCQP_MMSE}
\begin{eqnarray}
\label{eqn:QCQP_MMSEObject}
\hspace{-5ex}&\hspace{-2ex}{\mathop {\mathrm{minimize}} \limits_{\mathbf{s} \in \mathbb {R}^{2N_t}}}&\!\!\mathbf{s}^\mathrm{T}\mathbf{B}\,\mathbf{s} - 2\,\mathbf{b}^\mathrm{T}\mathbf{s} \\
\hspace{-5ex}&\hspace{-2ex}\mathrm{subject\,to}&\!\!{\mathbf{s}^\mathrm{T}\!(\mathbf{H}^\mathrm{T}\mathbf{H}\! +\! \sigma^2_{\mathbf{n}}\mathbf{I}_{2N_r})\,\mathbf{s} \!-\! 2\, \mathbf{y}^\mathrm{T} \mathbf{H}\,\mathbf{s}\!+\! \mathbf{y}^\mathrm{T}\mathbf{y}\! - \!\delta}\! \leq\! 0,\nonumber\\
&&\label{eqn:QCQP_MMSEConst}\\[-4ex]
\nonumber
\end{eqnarray}
\end{subequations} 
where again $\delta=\sigma^2_{\mathbf{n}}$, and from which the combination of \ac{KKT} conditions with Mor{\'e}'s Theorem \cite{MorePMS1993} yields
\begin{subequations}
\label{eq:KKT_MMSE}
\begin{equation}
\label{eq:KKT_MMSE1}
\bar{\mathbf{s}} = (\mathbf{B} + \mu^\text{opt} (\mathbf{H}^\mathrm{T}\mathbf{H}\! +\! \sigma^2_{\mathbf{n}}\mathbf{I}_{2N_r}))^{-1}(\mathbf{b} + \mu^\text{opt} \mathbf{H}^\mathrm{T}\mathbf{y}),
\end{equation}
\begin{equation*}
\bar{\mathbf{s}}^\mathrm{T}(\underbrace{(\mathbf{H}^\mathrm{T}\mathbf{H}\! +\! \sigma^2_{\mathbf{n}}\mathbf{I}_{2N_r})}_{=\mathbf{q}_{22}}\,\bar{\mathbf{s}} - \mathbf{H}^\mathrm{T}\,\mathbf{y}) - \mathbf{y}^\mathrm{T} \mathbf{H}\,\bar{\mathbf{s}} + \mathbf{y}^\mathrm{T}\mathbf{y} - \sigma^2_{\mathbf{n}} = 0.
\vspace{-3.5ex}
\end{equation*}
\begin{equation}
\label{eq:KKT_MMSE2}
\end{equation}
\end{subequations}

For brevity, we omit the derivation details which are similar to those of Subsection \ref{sec:auto_parameterization}, but solving the system of equations \eqref{eq:KKT_MMSE} in the same manner described in Subsection \ref{sec:auto_parameterization} and applying the results of \cite[Lem.3 and Th.4]{AdachiMP2019} and the M\"obius-transform, one finally arrives once again at the optimum regularization parameter given in equation \eqref{eq:OptimizedParameter}, only with the entry identified in equation \eqref{eq:KKT_MMSE2} updated in the matrices $\mathbf{Q}$ and $\mathbf{P}$, which yields
\begin{subequations}
\label{eqn:MatricesQPMMSE}
\begin{equation}
\label{eqn:MatrixQMMSE}
\mathbf{Q} \triangleq 
\left[\begin{array}{@{\,}c c c@{\,}}
\mathbf{y}^\mathrm{T}\mathbf{y} - \sigma^2_{\mathbf{n}} &  -  \mathbf{y}^\mathrm{T} \mathbf{H}& \mathbf{b}^\mathrm{T}\\
-\mathbf{H}^\mathrm{T}\mathbf{y}&  \mathbf{H}^\mathrm{T}\mathbf{H} + \sigma^2_{\mathbf{n}}\mathbf{I}_{2N_r}& - \mathbf{B} \\
\mathbf{b} & -\mathbf{B} & \mathbf{0}_{2N_t}
\end{array}\right]\!,\\
\end{equation}
\begin{equation}
\label{eqn:MatrixPMMSE}
\mathbf{P} \triangleq
\left[\begin{array}{@{\,}c c@{\,} c@{\!}}
0 &  \mathbf{0}_{1\times 2N_t}&  -\mathbf{y}^\mathrm{T}\mathbf{H}\\
\mathbf{0}_{2N_t\times 1}&  \mathbf{0}_{2N_t}&    \mathbf{H}^\mathrm{T}\mathbf{H}+\sigma^2_{\mathbf{n}}\mathbf{I}_{2N_r}\\
-\mathbf{H}^\mathrm{T}\mathbf{y}& \mathbf{H}^\mathrm{T}\mathbf{H}+\sigma^2_{\mathbf{n}}\mathbf{I}_{2N_r}& \mathbf{0}_{2N_t}
\end{array}\right]_{_{\!}}\!.
\end{equation}
\end{subequations}

From the above, it is evident that replacing \eqref{eq:PenalizedZF_soptReal} by equation \eqref{eqn:opt_x_DALMMSEReal}, as well as equations \eqref{eqn:MatrixQ} and \eqref{eqn:MatrixP} by equations \eqref{eqn:MatrixQMMSE} and \eqref{eqn:MatrixPMMSE} yields a noise-robust extension of the \ac{IDLS} detector, which maintains the same overall structure and thus the same complexity, while offering improved \ac{BER} performance at low \acp{SNR}.
We also remark, however, that at high \acp{SNR} or if the noise power is unknown and thus set to zero, the term $\sigma^2_{{\mathbf{n}}}\mathbf{I}_{2N_r}$ in equations \eqref{eqn:opt_x_DALMMSEReal} and \eqref{eqn:MatricesQPMMSE} disappear, reducing them to equations \eqref{eq:PenalizedZF_soptReal} and \eqref{eqn:MatricesQP}, respectively, and consequently the detector itself back to the original \ac{IDLS} described in Section \ref{sect:sect3}.

Although $\sigma^2_{\mathbf{n}}$ appears also in the upper-left minor of the matrix $\mathbf{Q}$ given in equations \eqref{eqn:MatricesQP} and \eqref{eqn:MatricesQPMMSE}, the latter is a consequence of a particular choice of search ball radius $\delta$ made in equations \eqref{eqn:P4_OP1_const} and \eqref{eqn:QCQP_MMSEConst}, respectively. In case $\sigma^2_{\mathbf{n}}$ is unknown or negligibly small, another value of $\delta$ must be chosen in $\mathbf{Q}$, for instance, based on the minimum Euclidean distance between points in the constellation $\mathcal{P}$. 
In other words, taking advantage of lattice reduction approaches \textit{e.g.,} \cite{LuzziTIT2013}, a robust choice of $\delta$   can be made as a function of the dominant eigenvalue of the channel matrix if $\sigma^2_{\mathbf{n}}$ is unknown.

With the modularity of the \ac{IDLS} framework well identified, we proceed to further extend it to mitigate imperfect \ac{CSI} and hardware impairments as previously announced.

\vspace{-1ex}
\subsection[]{Mitigating Imperfect \ac{CSI} and Hardware Impairments}

For the sake of brevity, in this Subsection we omit repetitive details and succinctly describe only the necessary modifications in the channel and system models, as well as in the fundamental equations of the \ac{IDLS} framework, required to identify how the latter can be extended to also mitigate imperfect \ac{CSI} and hardware impairments.

First, consider the well-known Gauss-Markov uncertainty model \cite{BehrangTSP2011} commonly used to incorporate the impact of \ac{CSI} errors in channel estimates, which is described by
\vspace{-0.5ex}
\begin{equation}
\label{eqn:GaussMarkov}
\bm{H} = \sqrt{1 - \tau^2}\hat{\bm{H}} + \tau\bm{E},
\vspace{-0.5ex}
\end{equation}
where $\hat{\bm{H}}$ denotes the estimate of the true channel matrix $\bm{H}$ ($i.e.,$ imperfect observation of $\bm{H}$), $\bm{E}$ corresponds to the error matrix whose distribution is associated with that of $\bm{H}$, and $\tau\in[0,1]$ denotes the Gauss-Markov uncertainty parameter that characterizes the \ac{CSI} estimation inaccuracy.

For the sake of completeness, we combine the \ac{CSI} error model of \eqref{eqn:GaussMarkov} with the channel correlation model described in equation \eqref{eq:correlatedchannel} to obtain \cite{BehrangTSP2011}
\vspace{-0.5ex}
\begin{align}
\bm{H} &= \bm{\Phi}^{\frac{1}{2}}_r\left(\sqrt{1 - \tau^2}\bm{H}_\text{\ac{i.i.d.}} + \tau\bm{E}_\text{\ac{i.i.d.}}\right)\bm{\Phi}^{\frac{1}{2}}_t\nonumber\\
& = \sqrt{1 - \tau^2}\underbrace{\bm{\Phi}^{\frac{1}{2}}_r\bm{H}_\text{\ac{i.i.d.}}\bm{\Phi}^{\frac{1}{2}}_t}_{\triangleq \hat{\bm{H}} \:\text{(known)}} + \tau\underbrace{\bm{\Phi}^{\frac{1}{2}}_r\bm{E}_\text{\ac{i.i.d.}}\bm{\Phi}^{\frac{1}{2}}_t}_{\triangleq\bm{E} \:\text{(unknown)}},
\label{eqn:GaussMarkov_corr}
\end{align}      
where $\bm{E}_\text{\ac{i.i.d.}}$ follows the circular symmetric complex Gaussian distribution with zero mean and unit variance while assuming that perfect (or considerably accurate) knowledge of the spatial correlation matrices $\bm{\Phi}_r$ and $\bm{\Phi}_t$ is available at the receiver, since such correlations are vary much slower than the instantaneous channel, and therefore can be estimated accurately even though the channel estimates $\hat{\bm{H}}$ themselves are imperfect \cite{BehrangTSP2011}.

Notice that equations \eqref{eqn:GaussMarkov} and \eqref{eqn:GaussMarkov_corr} indeed extend the perfect \ac{CSI} model employed in preceding sections of the article, such that setting $\tau = 0$, one returns simply to $\bm{H} = \hat{\bm{H}}$.
Next, we turn our attention to hardware impairments, which typically refer to  practical imperfections in \acp{PA}, \acp{DAC}, \acp{ADC}, I/Q mixers or other \ac{RF} chain components, and whose effect is to cause distortions in transmit signals that were shown in $e.g.$ \cite{SuzukiTWC08} to be well-modeled by i.i.d additive zero-mean Gaussian random variables, with variance proportional to the power of the undistorted signal.
To elaborate, in the presence of hardware impairment, an intended transmit symbol vector $\bm{s}\in\mathcal{C}^{N_t\times 1}$ is distorted into the transmit signal $\bm{x}\in\mathbb{C}^{N_t\times1}$ described by \cite{OmidTWC2018, IimoriTWC19}
\vspace{-0.5ex}
\begin{equation}
\label{eqn:hardwareimp}
\bm{x} = \bm{s} + \bm{w},
\vspace{-0.5ex}
\end{equation}
where $\bm{w}$ denotes an additive hardware distortion vector modeled as $\bm{w}\sim\mathcal{CN}\left(\bm{0},\eta\cdot{\rm diag}\left(\mathbf{C}_{\bm{s}}\right)\right)$ with $\eta$ denoting the RF distortion level parameter characterized by the quality of the \ac{RF} chain components and $\mathbf{C}_{\bm{s}}\triangleq\E{\bm{s}^{\rm H}\bm{s}}$.

Assuming that the elements of the intended symbol vector ${\bm{s}}$ are independent from each other and have unit power (\textit{i.e.,} $\mathbf{C}_{\bm{s}}=\mathbf{I}$), the received signal corresponding to the transmit signal in equation \eqref{eqn:hardwareimp} is given by
\vspace{-0.5ex}
\begin{align*}
{\bm{y}} =& {\bm{H}} {\bm{x}} + {\bm{n}}\nonumber\\[-3ex]
= & \underbrace{\sqrt{1\!-\!\tau^2}\hat{\bm{H}}\bm{s}}_\text{Intended} +
\overbrace{\tau \bm{E}\bm{s}\! +\! \tau\bm{E}\bm{w}}^\text{CSI imperfection} +\hspace{-8ex}\underbrace{\hspace{7.5ex}\sqrt{1\!-\!\tau^2}\hat{\bm{H}} \bm{w}}_\text{Hardware impairment} + \bm{n} \in\mathbb{C}^{N_r\times 1}\!\!.
\end{align*}
\vspace{-5ex}
\begin{equation}
\label{eqn:receivedsignalCSIE}
\end{equation}

It will prove convenient to normalize the received signal by the scalar $\sqrt{1-\tau^2}$, and rearrange the terms so as to yield
\begin{equation}
\label{eqn:normalizedY}
\bar{\bm{y}} = \hat{\bm{H}}\bm{s} + \underbrace{\hat{\bm{H}} \bm{w}+ \frac{\tau\bm{E}\bm{s}+\tau\bm{E}\bm{w} + \bm{n}}{\sqrt{1-\tau^2}}}_{\triangleq \tilde{\bm{n}}},
\end{equation}
where we have implicitly defined the total effective noise $\tilde{\bm{n}}$.

At this point it is worth mentioning that in the absence of knowledge of the Gauss-Markov uncertainty parameter $\tau$ and the \ac{RF} distortion level
$\eta$, a receiver would simply perceive the total effective noise $\tilde{\bm{n}}$ as a higher noise level with power $\sigma^2_{\tilde{\bm{n}}}$, such that the robust \ac{IDLS} scheme of Subsection \ref{sec:DALMMSE} could be in fact employed directly, as long as $\sigma^2_{\tilde{\bm{n}}}$ could be estimated, under the assumption of whiteness of $\tilde{\bm{n}}$.

On the other hand, if $\tau$ and $\eta$ can themselves be estimated, and under the knowledge of the spatial correlation matrices $\bm{\Phi}_r$ and $\bm{\Phi}_t$, a more effective mitigation of the \ac{CSI} and hardware imperfection can be achieved under the \ac{IDLS} framework, by incorporating the resulting knowledge.

First, as shown in Appendix \ref{app:CovDerivation}, the covariance matrix of 
the total effective noise $\tilde{\bm{n}}$ is given by
\begin{equation}
\label{eq:CovNoiseTilde}
\bm{\mathit \Sigma}_{\tilde{\bm{n}}} = \underbrace{\eta\hat{\bm{H}}\hat{\bm{H}}^{\mathrm{H}}\! +\! \frac{\tau^2}{1\!-\!\tau^2}(1\!+\!\eta)\Tr{\bm{\Phi}_t}\bm{\Phi}_r}_{\triangleq \bm{\mathit \Sigma}^\text{C}_{\tilde{\bm{n}}}} + \underbrace{\frac{\sigma^2_{{\bm{n}}}}{1\!-\!\tau^2}\mathbf{I}_{N_r}}_{\triangleq \bm{\mathit \Sigma}^\text{U}_{\tilde{\bm{n}}}},\!\!\!\!
\end{equation}
where, for future convenience, we have decomposed $\bm{\mathit \Sigma}_{\tilde{\bm{n}}}$ into a sum of the matrices $\bm{\mathit \Sigma}^\text{C}_{\tilde{\bm{n}}}$ and $\bm{\mathit \Sigma}^\text{U}_{\tilde{\bm{n}}}$, the first corresponding to quantities that are subjected to correlation, and the second corresponding to terms that are not.

Next, we observe that the presence of the correlated covariance component $\bm{\mathit \Sigma}^\text{C}_{\tilde{\bm{n}}}$ in $\bm{\mathit \Sigma}_{\tilde{\bm{n}}}$ implicates that, in general,
\begin{equation}
\label{eq:WoodburyLemmaNotSatisfied}
\hat{\bm{H}}^\mathrm{H}(\hat{\bm{H}}\hat{\bm{H}}^\mathrm{H} + \bm{\mathit \Sigma}_{\tilde{\bm{n}}})^{-1} \neq (\hat{\bm{H}}^\mathrm{H}\hat{\bm{H}} +\bm{\mathit \Sigma}_{\tilde{\bm{n}}})^{-1}\hat{\bm{H}}^\mathrm{H},
\end{equation}

\begin{figure*}
\setcounter{equation}{63}
\begin{subequations}
\label{eqn:MatricesQPRobust}
\begin{equation}
\label{eqn:MatrixQRobust}
\mathbf{Q}\! \triangleq \!\!
\left[\begin{array}{@{\,}c@{\qquad} c@{\qquad} c@{\,}}
\bar{\mathbf{y}}^\mathrm{T}(\mathbf{\Sigma}^\text{C}_{\tilde{\mathbf{n}}}+\mathbf{I}_{2N_r})^{-1}\bar{\mathbf{y}} - \sigma^2_{\mathbf{n}} &
 -\bar{\mathbf{y}}^\mathrm{T} (\mathbf{\Sigma}^\text{C}_{\tilde{\mathbf{n}}}\!+\!\mathbf{I}_{2N_r})^{-1}\hat{\mathbf{H}}& \mathbf{b}^\mathrm{T}\\
-\hat{\mathbf{H}}^\mathrm{T}\!(\mathbf{\Sigma}^\text{C}_{\tilde{\mathbf{n}}}\!+\!\mathbf{I}_{2N_r})^{-1}\bar{\mathbf{y}}&
\hat{\mathbf{H}}^\mathrm{T}\!(\mathbf{\Sigma}^\text{C}_{\tilde{\mathbf{n}}}\!+\!\mathbf{I}_{2N_r})^{-1}\hat{\mathbf{H}}\! +\! \tfrac{\sigma^2_{{\mathbf{n}}}}{1-\tau^2}\mathbf{I}_{2N_t}& - \mathbf{B} \\
\mathbf{b} & -\mathbf{B} & \mathbf{0}_{2N_t}
\end{array}\right]\!,\\
\end{equation}
\begin{equation}
\label{eqn:MatrixPRobust}
\mathbf{P} \triangleq
\left[\begin{array}{c@{\qquad} c@{\qquad} c}
0 &  \mathbf{0}_{1\times 2N_t}&  -\bar{\mathbf{y}}^\mathrm{T} (\mathbf{\Sigma}^\text{C}_{\tilde{\mathbf{n}}}\!+\!\mathbf{I}_{2N_r})^{-1}\hat{\mathbf{H}}\\
\mathbf{0}_{2N_t\times 1}&  \mathbf{0}_{2N_t}&    \hat{\mathbf{H}}^\mathrm{T}\!(\mathbf{\Sigma}^\text{C}_{\tilde{\mathbf{n}}}\!+\!\mathbf{I}_{2N_r})^{-1}\hat{\mathbf{H}}\! +\! \tfrac{\sigma^2_{{\mathbf{n}}}}{1-\tau^2}\mathbf{I}_{2N_t}\\
-\hat{\mathbf{H}}^\mathrm{T}\!(\mathbf{\Sigma}^\text{C}_{\tilde{\mathbf{n}}}\!+\!\mathbf{I}_{2N_r})^{-1}\bar{\mathbf{y}}& \hat{\mathbf{H}}^\mathrm{T}\!(\mathbf{\Sigma}^\text{C}_{\tilde{\mathbf{n}}}\!+\!\mathbf{I}_{2N_r})^{-1}\hat{\mathbf{H}}\! +\! \tfrac{\sigma^2_{{\mathbf{n}}}}{1-\tau^2}\mathbf{I}_{2N_t}& \mathbf{0}_{2N_t}
\end{array}\right]_{_{\!}}.
\end{equation}
\end{subequations}
\hrule
\end{figure*}

\setcounter{equation}{59}

\noindent such that the total effective noise $\tilde{\bm{n}}$ due to \ac{CSI} imperfection and hardware impairment does not enjoy the homoscedasticity assumed in the previous formulations of \ac{IDLS} described in preceding sections.

To circumvent this new challenge, we consider an extension of the \ac{IDLS} framework under the prism of the generalized total least square regression problem \cite{Huffel1991}, which we here modify to include our constellation-aware $\ell_0$-norm regularizer, yielding
\begin{align}
\label{eqn:GLS_L0_V2} 
{\mathop {\mathrm{min}} \limits _{\bm{s} \in \mathbb {C}^{N_t}}}\:\:&
(\bar{\bm{y}}  - \hat{\bm{H}} \bm{s})^\mathrm{H}(\bm{\mathit \Sigma}^\text{C}_{\tilde{\bm{n}}} + \bm{I}_{N_r})^{-1}(\bar{\bm{y}}  - \hat{\bm{H}} \bm{s})\\[-2ex]
&\hspace{15ex}+\frac{\sigma^2_{{\bm{n}}}}{1-\tau^2}\|\bm{s}\|^2_2 +  \lambda\sum _{i=1}^{2^b} \| \bm{s}-c_{i}\mathbf{1}\|_{0},\nonumber
\end{align}
which seeks to minimize the Mahalanobis distance between the output and input vectors while enforcing the constellation-compliance of the solution.

Proceeding succinctly hereafter, introducing the $\ell_0$-norm approximation of equation \eqref{eqn:L0Approx_FP} and applying the \ac{QT} into equation \eqref{eqn:GLS_L0_V2}, we obtain
\begin{subequations}
\begin{align}
\label{eqn:GLS_L0_V3} 
{\mathop {\mathrm{min}} \limits _{\bm{s} \in \mathbb {C}^{N_t}}} \:\:& (\bar{\bm{y}}  - \hat{\bm{H}} \bm{s})^\mathrm{H}(\bm{\mathit \Sigma}^\text{C}_{\tilde{\bm{n}}}+\bm{I}_{N_r})^{-1}(\bar{\bm{y}}  - \hat{\bm{H}} \bm{s})\\[-1ex]
&\hspace{12ex} +\frac{\sigma^2_{{\bm{n}}}}{1-\tau^2}\|\bm{s}\|^2_2 +  \lambda(\bm{s}^\mathrm{H}\bm{B}\bm{s} - 2\Re\{\bm{s}^\mathrm{H}\bm{b}\}),\nonumber
\end{align}

\begin{algorithm}[t!]
\SetKwInOut{Input}{Input}
\nonl\quad\\[-1ex]
{\dosemic\nonl{\bf External Input:}\\
Received signal ${\mathbf{y}}$, estimated channel matrix $\hat{\mathbf{H}}$,
noise power $\sigma^2_{\mathbf{n}}$, Gauss-Markov uncertainty parameter $\tau$ and hardware impairment parameter $\eta$.
}

\dosemic\nonl{\bf Internal Parameters:}\\
Maximum number of iterations $k_\text{max}$;\\
\dosemic\nonl convergence threshold $\varepsilon\ll 1$ and shaping parameter $\alpha\ll 1$.\\
\dosemic\nonl{\bf Initialization:}\\
Set iteration counter $k = 0$;\\
\dosemic\nonl
Normalize received signal as $\bar{\mathbf{y}} \triangleq \frac{\mathbf{y}}{\sqrt{1-\tau^2}}$;\\
\dosemic\nonl
Set initial solution to\\ $\mathbf{s}^{(k)}\!\!=\!\!\Big(\!\hat{\mathbf{H}}^\mathrm{T}(\mathbf{\Sigma}^\text{C}_{\tilde{\mathbf{n}}}+\mathbf{I}_{2N_r})^{-1}\hat{\mathbf{H}}\! +\! \tfrac{\sigma^2_{{\mathbf{n}}}}{1-\tau^2}\mathbf{I}_{2N_t}\!\Big)^{^{\!\!\!-\!1}}\!\!\!\!\big(\hat{\mathbf{H}}^\mathrm{T}(\mathbf{\Sigma}^\text{C}_{\tilde{\mathbf{n}}}+\mathbf{I}_{2N_r})^{-1}\bar{\mathbf{y}}\big)$\!.\\
\Repeat{$k > k_\textup{max}$ \textup{or} $\varepsilon_k < \varepsilon$}{
Increase iteration counter $k = k + 1$\\
\noindent Update $\beta_{i,j}\forall\;i,j$, as in equation \eqref{eqn:beta_ZFReal}\\
Construct $\mathbf{b}$ and $\mathbf{B}$ from equations \eqref{eqn:BMatZFReal_b} and \eqref{eqn:BMatZFReal_B}\\
Construct $\mathbf{Q}$ and $\mathbf{P}$ from equations \eqref{eqn:MatrixQRobust} and \eqref{eqn:MatrixPRobust}\\
Obtain $\lambda^{\text{opt}(k)}$ as in equation \eqref{eq:OptimizedParameter}\\
Update $\mathbf{s}^{(k)}$ as in equation \eqref{eqn:IDAGLSReal}\\
Calculate $\varepsilon_k = \norm{\mathbf{s}^{(k)}-\mathbf{s}^{(k-1)}}_2$
}
\caption[]{Robust IDLS Detector}
\label{alg:ridls}
\end{algorithm}
\setlength{\textfloatsep}{5pt}

\noindent or alternatively, in the real domain,
\begin{align}
\label{eqn:GLS_L0_V3Real} 
{\mathop {\mathrm{min}} \limits _{\mathbf{s} \in \mathbb {C}^{N_t}}} \:\:& (\bar{\mathbf{y}}  - \hat{\mathbf{H}} \mathbf{s})^\mathrm{T}(\mathbf{\Sigma}^\text{C}_{\tilde{\mathbf{n}}}+\mathbf{I}_{2N_r})^{-1}(\bar{\mathbf{y}}  - \hat{\mathbf{H}} \mathbf{s})\\[-1ex]
&\hspace{15ex} +\frac{\sigma^2_{{\mathbf{n}}}}{1-\tau^2}\|\mathbf{s}\|^2_2 +  \lambda(\mathbf{s}^\mathrm{T}\mathbf{B}\mathbf{s} - 2\mathbf{s}^\mathrm{T}\mathbf{b}),\nonumber
\end{align}
\end{subequations}
which in turn admit the respective solutions
\begin{subequations}
\label{eqn:IDAGLS}
\begin{align}
\label{eqn:IDAGLSComplex}
\bm{s} = \Big(\!\hat{\bm{H}}^\mathrm{H}(\bm{\mathit \Sigma}^\text{C}_{\tilde{\bm{n}}}+\bm{I}_{N_r})^{-1}\!\hat{\bm{H}}\! +\! \tfrac{\sigma^2_{{\bm{n}}}}{1-\tau^2}\bm{I}_{N_t}\! +\! \lambda\bm{B}\Big)^{\!\!-1}&\\[-0.5ex]
&\hspace{-17ex}\times\Big(\!\hat{\bm{H}}^\mathrm{H}(\bm{\mathit \Sigma}^\text{C}_{\tilde{\bm{n}}}+\bm{I}_{N_r})^{-1}\bar{\bm{y}} +\! \lambda\bm{b}\Big),\nonumber
\end{align}
and
\begin{align}
\label{eqn:IDAGLSReal}
\mathbf{s} = \!\Big(\hat{\mathbf{H}}^\mathrm{T}(\mathbf{\Sigma}^\text{C}_{\tilde{\mathbf{n}}}+\mathbf{I}_{2N_r})^{-1}\hat{\mathbf{H}} + \tfrac{\sigma^2_{{\mathbf{n}}}}{1-\tau^2}\mathbf{I}_{2N_t} + \lambda\mathbf{B}\Big)^{\!-1}&\\[-0.5ex]
&\hspace{-20ex}\times\big(\hat{\mathbf{H}}^\mathrm{T}(\mathbf{\Sigma}^\text{C}_{\tilde{\mathbf{n}}}+\mathbf{I}_{2N_r})^{-1}\bar{\mathbf{y}} +\! \lambda\mathbf{b}\big),\nonumber
\end{align}
\end{subequations}
where, as previously, all the straight up bold letters are the real-domain equivalent of their bold italic counterparts in the complex domain.

Finally, in order to optimize the regularization parameter, we once again transform the formulation given in equation \eqref{eqn:GLS_L0_V3Real} into a \ac{QCQP-1}, build the corresponding \ac{KKT} conditions, and apply Mor{\'e}'s Theorem \cite{MorePMS1993} to obtain
\vspace{-1ex}
\begin{subequations}
\label{eq:KKT_Robust}
\begin{eqnarray}
\label{eq:KKT_Robust}
\bar{\mathbf{s}} = \Big(\mathbf{B} + \mu^\text{opt} \big(\hat{\mathbf{H}}^\mathrm{T}(\mathbf{\Sigma}^\text{C}_{\tilde{\mathbf{n}}}+\mathbf{I}_{2N_r})^{-1}\hat{\mathbf{H}} + \tfrac{\sigma^2_{{\mathbf{n}}}}{1-\tau^2}\mathbf{I}_{2N_t}\big)\Big)^{-1}&&\\[-0.5ex]
&&\hspace{-30ex}\times\Big(\mathbf{b} + \mu^\text{opt} \hat{\mathbf{H}}^\mathrm{T}(\mathbf{\Sigma}^\text{C}_{\tilde{\mathbf{n}}}+\mathbf{I}_{2N_r})^{-1}\bar{\mathbf{y}}\Big),
\nonumber 
\end{eqnarray}
\vspace{-4ex}
\begin{align}
\bar{\mathbf{s}}^\mathrm{T}\Big(\overbrace{\!\!(\hat{\mathbf{H}}^\mathrm{T}\!(\mathbf{\Sigma}^\text{C}_{\tilde{\mathbf{n}}}\!+\!\mathbf{I}_{2N_r})^{-1}\hat{\mathbf{H}}\! +\! \tfrac{\sigma^2_{{\mathbf{n}}}}{1-\tau^2}\mathbf{I}_{2N_t})\!}^{=\mathbf{q}_{22}}\bar{\mathbf{s}}\! -\! \hat{\mathbf{H}}^\mathrm{T}\!(\mathbf{\Sigma}^\text{C}_{\tilde{\mathbf{n}}}\!+\!\mathbf{I}_{2N_r})^{-1}\bar{\mathbf{y}}\Big)&\nonumber\\
&\hspace{-53ex} \underbrace{- \bar{\mathbf{y}}^\mathrm{T} (\mathbf{\Sigma}^\text{C}_{\tilde{\mathbf{n}}}\!+\!\mathbf{I}_{2N_r})^{-1}\hat{\mathbf{H}}}_{=\mathbf{q}_{12}}\,\bar{\mathbf{s}} + \underbrace{\bar{\mathbf{y}}^\mathrm{T}(\mathbf{\Sigma}^\text{C}_{\tilde{\mathbf{n}}}+\mathbf{I}_{2N_r})^{-1}\bar{\mathbf{y}} - \sigma^2_{\bm{n}}}_{=q_{11}} = 0,
\nonumber\\[-3ex]
&\label{eq:KKT_Robust}
\end{align}
\end{subequations}
such that the matrices $\mathbf{Q}$ and $\mathbf{P}$ to be used in equation \eqref{eq:OptimizedParameter} in order to calculate $\lambda^\text{opt}$ are updated as in equation \eqref{eqn:MatricesQPRobust}, shown on top of the page.

Once again, we remark that in case perfect \ac{CSI} is available at the receiver ($i.e.$, when $\tau = 0$), and the transmitter is free of distortions due to hardware impairment ($i.e.$, when $\eta = 0$), equations \eqref{eqn:GLS_L0_V2} through \eqref{eqn:MatricesQPRobust} reduce to the corresponding equations in Subsection \ref{sec:DALMMSE}.

Likewise, setting the noise variance to zero further reverts the detector described in this subsection back to the fundamental \ac{IDLS} design introduced in Section \ref{sect:sect3}.
In other words, it is evident that the \ac{IDLS} variation introduced here is in fact an extension of the original method presented earlier, with embedded robustness to noise, \ac{CSI} imperfection and hardware impairment, for which it shall be referred to as the \emph{Robust} \ac{IDLS} detector.

For the convenience of the reader, a pseudo-code summarizing the Robust \ac{IDLS} detector is offered above in Algorithm \ref{alg:ridls}. 
It can be readily recognized that indeed Algorithm \ref{alg:ridls} is generalization of Algorithm \ref{alg:idls}.

\section[]{Performance Assessment: Robust \ac{IDLS}}
\label{sec:performance2}

In this section, we offer performance evaluation of the Robust \ac{IDLS} detector for different \ac{CSI} error and hardware impairment setups in order to illustrate the effectiveness and flexibility of the proposed framework.

Before we proceed to the \ac{BER} performance assessment of the Robust \ac{IDLS} detector compared to the state of the art, let us offer a few comments on the complexity of the framework in the following subsection.

\vspace{-1ex}
\subsection{Remarks on Complexity}

\begin{figure}[t!]
  \centering
  \includegraphics[width=0.45\columnwidth]{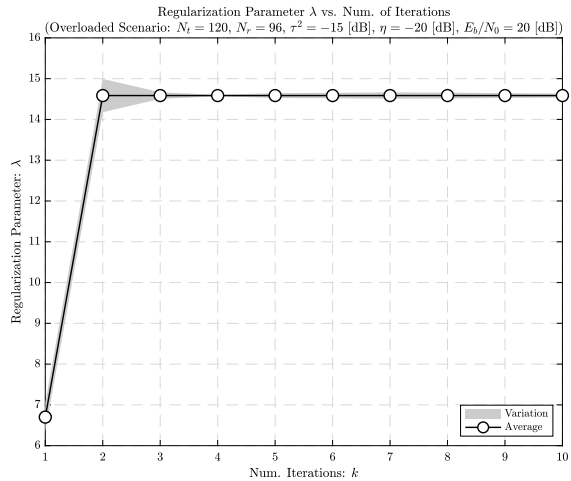}
  \caption{Optimized regularization parameter $\lambda^{\text{opt}(k)}$ averaged over channel realizations and signal transmissions with $N_t=120$, $N_r=96$, $\tau^2=-15$ [dB] and $\eta=-20$ [dB] as a function of algorithmic iteration $k$.}
  \label{fig:LambdaOpt}
\end{figure}

For starters, we emphasize that although the computation of the regularization parameter $\lambda^{\text{opt}(k)}$ obtained from the robust \ac{IDLS} was included among the repetitive steps executed in Algorithms \ref{alg:idls} and \ref{alg:ridls}, for the sake of completeness, in practice such step does not need to be repeated at every iteration of the \ac{IDLS} detector.
Indeed, notice that only the first row and column of the matrix $\mathbf{Q}$ is signal-dependent while the matrix $\mathbf{P}$ is constant for a given channel realization.
Consequently, it is both mathematically justified and empirically observed that the sequences of regularization parameters $\{\lambda^{\text{opt}(0)},\lambda^{\text{opt}(1)},\cdots,\lambda^{\text{opt}(k_{\max})}\}$ obtained from Algorithm \ref{alg:idls} and \ref{alg:ridls} do not vary significantly with channel realizations and signal transmissions, for fixed system size and \ac{SNR}.
This can be confirmed in Figure \ref{fig:LambdaOpt}, in which the values $\lambda^{\text{opt}(k)}$ is plotted against the iteration number. Both the average (line) and individual outcomes (gray shadow) are shown, for multiple signal and channel realizations.
It is found that the evolution of $\lambda^{\text{opt}(k)}$ is fast and stable, such that the results obtained over a given (or a few) different channel realizations and signal transmissions can be stored and used for subsequent uses.

That, allied with the fact that in \ac{SotA} methods the optimization of $\lambda$ is typically performed via exhaustive search, justifies us not to consider the optimization of $\lambda$ as a core contributor to the complexity of the \ac{IDLS} framework.
Consequently, the most expensive operation in the \ac{IDLS} framework is  the matrix inversion required to evaluate the closed-form iterative expression in equation \eqref{eq:KKT_Robust}, which requires $2N_t^3/3 + \mathcal{O}(N_t^2)$ \ac{flops} to complete, if carried out via a conventional matrix inversion algorithm such as the naive Gauss-Jordan elimination method \cite{Golub1996}.
However, since the quadratic coefficient matrices in equations \eqref{eq:PenalizedZF_sopt}, \eqref{eqn:opt_x_DALMMSE} and \eqref{eqn:IDAGLS} are Hermitian positive definite, the latter complexity can be halved by taking advantage of the Cholesky factorization, which transforms the latter equations into a linear system involving triangular matrices, resulting in an order of $N_t^3/3 + \mathcal{O}(N_t^2)$ \ac{flops} \cite{LloydSIAM1997}.

We have numerically found via Monte-Carlo simulations, for instance, that the Cholesky factorization approach suffices to efficiently solve the inversions in equations \eqref{eq:PenalizedZF_sopt}, \eqref{eqn:opt_x_DALMMSE} and \eqref{eqn:IDAGLS} under moderate problem setups ($i.e.$, $N_t = N_r \approx 100$) within an average time of hundreds of microseconds when using 64-bit MATLAB 2019a in a computer with an Intel Core i9 processor with a clock speed of 3.6GHz and 32GB of \ac{RAM}.
Thus, the operation in practice can be computed in the order of nanoseconds on \acp{FPGA}, which complies with latency requirements posed by various global standards including \ac{5G} \ac{NR}.

For systems of larger size ($e.g.$, $N_t\gg500$), numerous iterative algorithms including variates of the conjugate gradient methods, accelerated first/second-order gradient methods, and minimum residual methods exist \cite{MyListOfPapers:Golub1996}, which can also be utilized to solve the linear equations \eqref{eq:derivative_PenalizedZF}, \eqref{eqn:opt_x_DALMMSE}, and \eqref{eqn:IDAGLS}.
Note that efficient factorization-based inverse solvers are publicly available as native functions in numerical computing languages, $e.g.$, \texttt{mldivide} in MATLAB and \texttt{cho\_solve} in Python.
Finally, we remark that the proposed algorithm has a comparative worst-case complexity to the \ac{MMSE} detector which is currently employed in \ac{5G} systems \cite{5GZaidi}.

\vspace{-1.5ex}
\subsection[]{\ac{BER} Performance: \ac{IDLS} versus \ac{SotA} Alternatives}
\label{sec:performance2}

Next, we turn our attention to the evaluation of the \ac{BER} performance of Algorithm \ref{alg:ridls} in comparison to the conventional \ac{LMMSE} and recent \ac{SotA} alternatives, in mitigating the effect of imperfections such as \ac{CSI} errors and hardware impairments, inevitable in real-life systems.
For the sake of a more effective comparison, in particular in terms of capturing the gains achieved as a result of the \ac{IDLS} approach itself rather then the system model alone, and given that the derivation of the total effective noise covariance matrix described in Appendix \ref{app:CovDerivation} and summarized in equation \eqref{eq:CovNoiseTilde} is adjacent to the \ac{IDLS} detector, which therefore can also be utilized outside of the context of the \ac{IDLS} framework, we have fed the conventional \ac{LMMSE} estimator with $\bm{\mathit \Sigma}_{\tilde{\bm{n}}}$ as described by equation \eqref{eq:CovNoiseTilde}.

In other words, in the figures to follow, curves attributed to the ``conventional'' \ac{LMMSE} corresponds to the results obtained with the receiver
\begin{equation}
\label{eqn:MMSECorr}
\bm{s}^\text{MMSE} = \hat{\bm{H}}^\mathrm{H}(\hat{\bm{H}}\hat{\bm{H}}^\mathrm{H} + \bm{\mathit \Sigma}_{\tilde{\bm{n}}})^{-1}\bar{\bm{y}}.
\vspace{-0.5ex}
\end{equation}

Following \cite{OmidTWC2018} and \cite{IimoriTWC19}, it is assumed that the \ac{CSI} imperfection level $\tau^2$ varies from $-15$ [dB] to $-10$ [dB], while the hardware imperfection parameter $\eta$ takes from the interval $[-20, -10]$ [dB].
\vspace{-2ex}
\begin{figure}[t]
\centering
\begin{subfigure}[b]{0.45\columnwidth}
\centering
\includegraphics[width=\columnwidth]{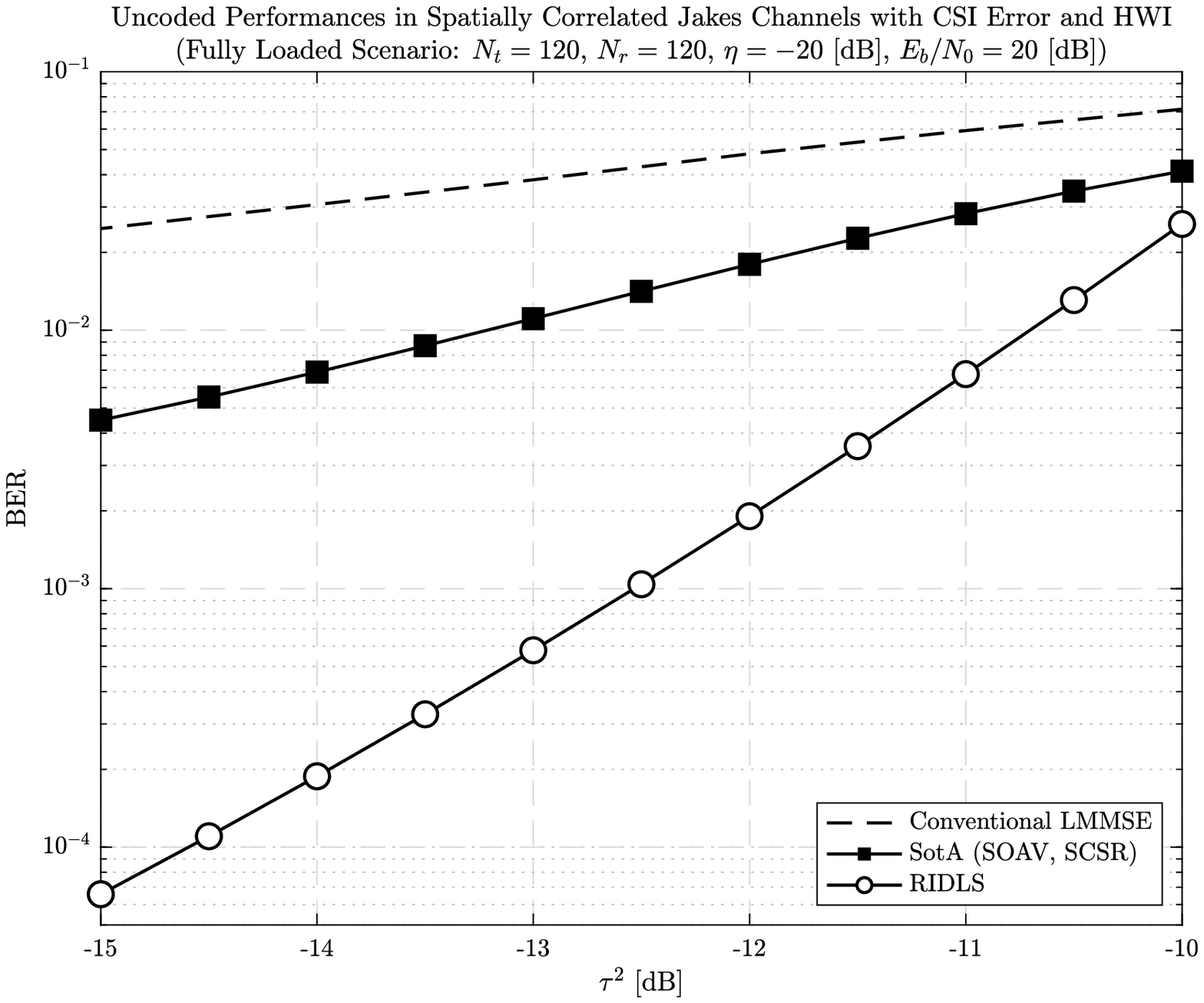}
\caption{Impact of CSI Error with Fixed Hardware Impairment Level.}
\label{fig:BERvsCSI_Full}
\end{subfigure}
\begin{subfigure}[b]{0.45\columnwidth}
\centering
\includegraphics[width=\columnwidth]{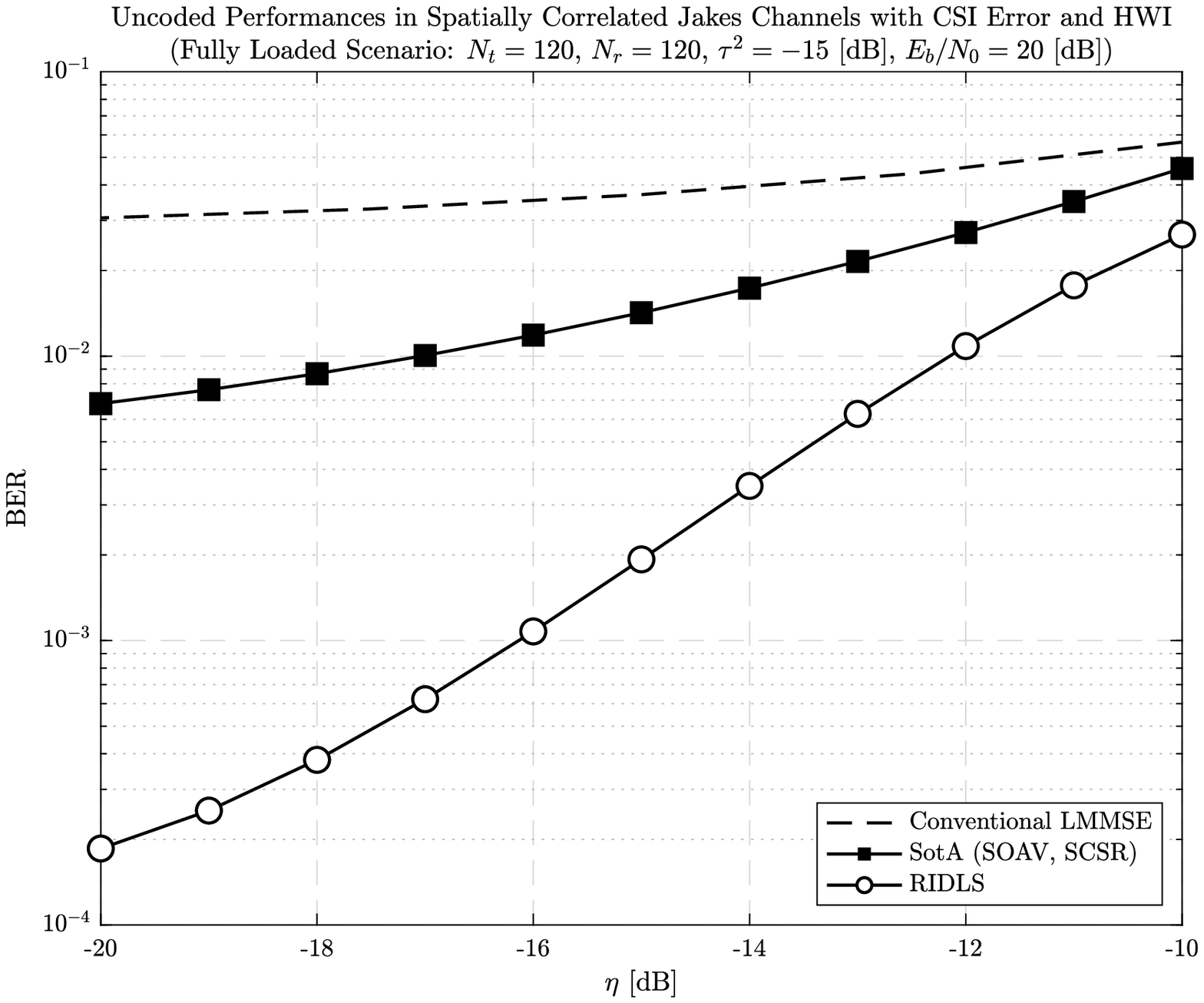}
\caption{Impact of Hardware Impairment with Fixed CSI Error Level.}
\label{fig:BERvsHW_Full}
\end{subfigure}
\caption[]{Uncoded \ac{BER} performance of Robust \ac{IDLS} detector compared to \ac{SotA} alternatives, under fully-loaded ($\gamma=1$) conditions in spatially correlated channels and subjected to different CSI error and hardware impairment levels.}
\label{fig:BER_Full}
\vspace{-2ex}
\end{figure}

Fully loaded scenarios will be set with $N_t = N_r = 120$, whereas overloaded scenarios will be set with $N_t=120$ and $N_r=96$, yielding an overloading ratio of $\gamma=1.25$ so that the results of Subsection \ref{subsec:IDLSResults} may serve as a reference.

Our comparisons start with Figure \ref{fig:BER_Full}, which shows the gains in \ac{BER} performance achieved by the robust \ac{IDLS} detector  over \ac{SotA} alternatives in a fully-loaded scenario (\textit{i.e.,} $N_t=N_r=120$) with different \ac{CSI} error and hardware imperfection condition, respectively.
It is found that the conventional \ac{LMMSE} estimator is significantly outperformed not only by the newly-proposed robust \ac{IDLS} detector, but also by the \ac{SOAV} and \ac{SCSR} schemes, which follows from the constellation-awareness that these techniques have in common, and which demonstrates its robustness of the approach against non-ideal conditions such as \ac{CSI} and hardware imperfections.
\vspace{-3.5ex}
\begin{figure}[t]
\centering
\begin{subfigure}[b]{0.45\columnwidth}
\centering
\includegraphics[width=\columnwidth]{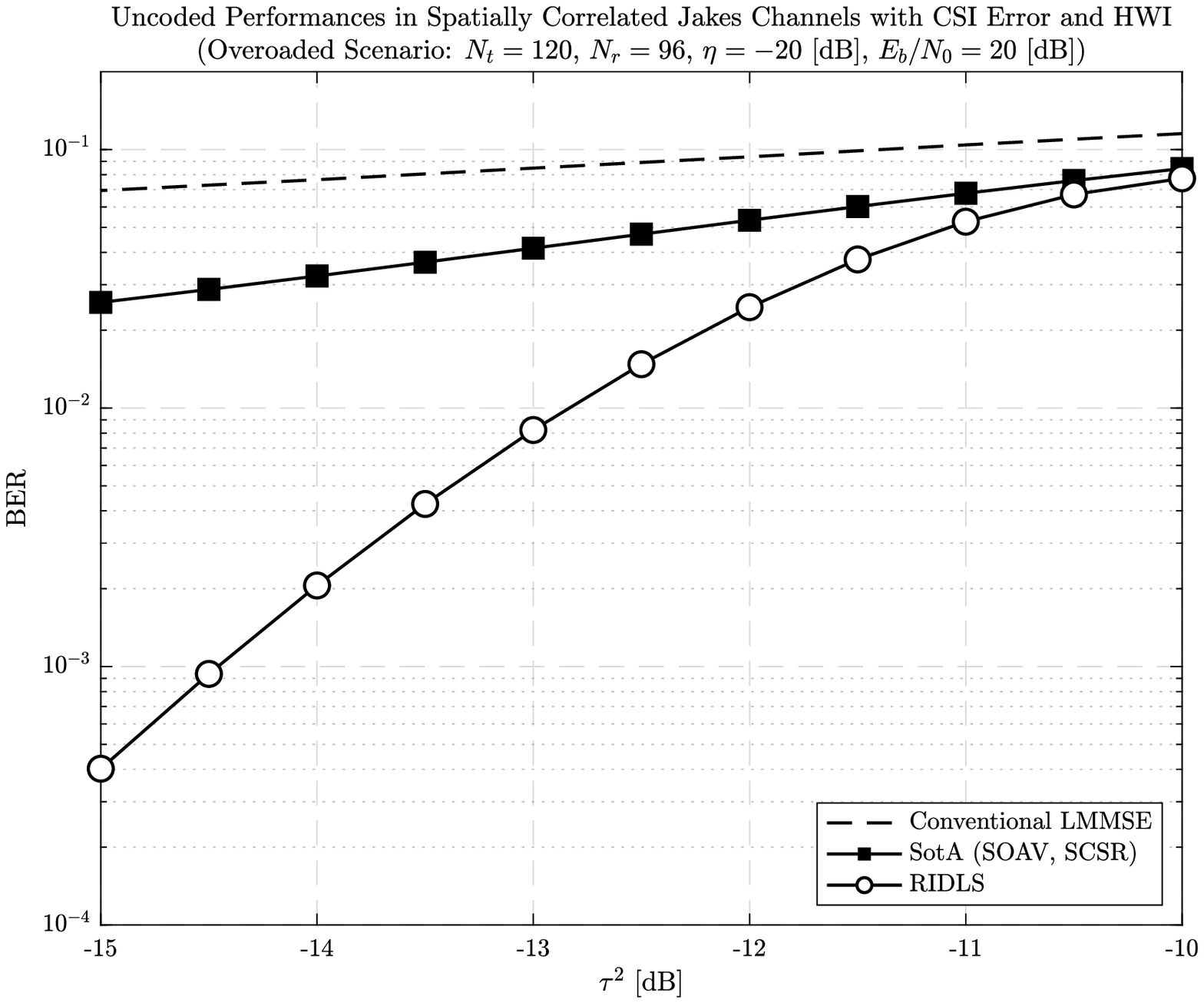}
\caption{Impact of CSI Error with Fixed Hardware Impairment Level.}
\label{fig:BERvsCSI_Over}
\end{subfigure}
\begin{subfigure}[b]{0.45\columnwidth}
\centering
\includegraphics[width=\columnwidth]{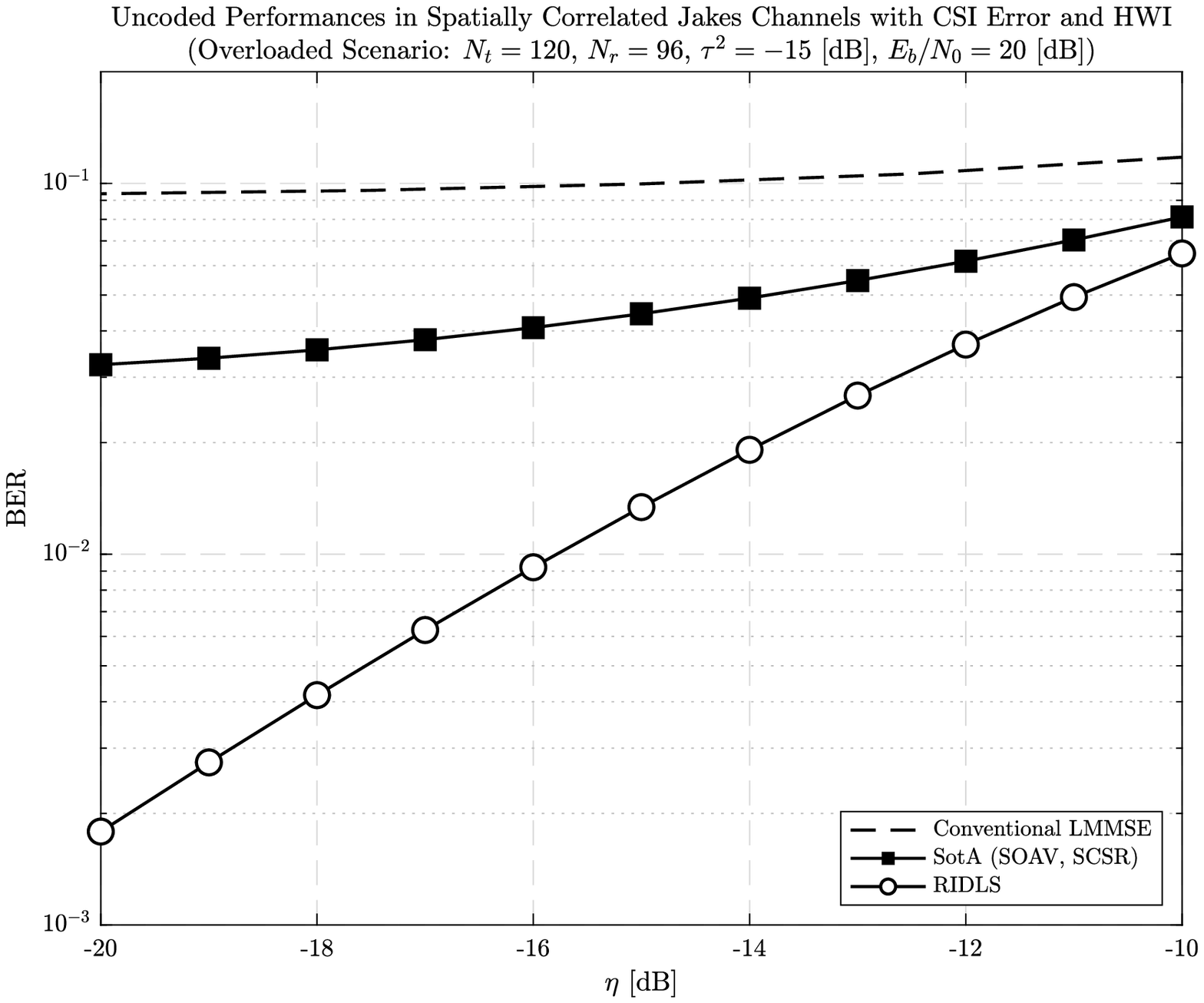}\vspace{-0.5ex}
\caption{Impact of Hardware Impairment with Fixed CSI Error Level.}
\label{fig:BERvsHW_Over}
\end{subfigure}
\caption[]{Uncoded \ac{BER} performance of Robust \ac{IDLS} detector compared to \ac{SotA} alternatives, under overloaded ($\gamma=1.25$) conditions in spatially correlated channels and subjected to different CSI error and hardware impairment levels.}
\label{fig:BER_Over}
\end{figure}

It can also be seen, however, that the robust \ac{IDLS} consistently outperforms all alternatives including \ac{SOAV} and \ac{SCSR} across the entire range of \ac{CSI} error and hardware imperfection levels, with an especially large gain over the latter over a wide range of values of $\tau$ and $\eta$.

Next, we examine the impact of overloading on the aforementioned gains in order to reveal the performance in a \ac{NOMA} setup.
To this end, Figure \ref{fig:BER_Over} compares the uncoded \acp{BER} achieved by a system with $N_t=120$ and $N_r=96$, yielding an overloading ratio of $\gamma = 1.25$.
It can be seen that while the harsher overloading has an overall impact on the performances of all schemes compared, as a consequence of the higher levels of \ac{ISI}, also exacerbated by channel and noise correlation resulting from the \ac{CSI} errors and hardware distortions, the rebust \ac{IDLS} detector continues to provide superior performance with substantial gains over all  alternatives.

These gains are a consequence of the incorporation of imperfection-aware extensions via the generalized \ac{LS} method into the \ac{IDLS} framework.
As confirmed by the figures, it is shown that the proposed framework is generic and compatible with various system setups in a plug-and-play fashion.

\vspace{-1ex}
\section{Conclusion}
\label{sec:conclusion}

We proposed a flexible framework for the symbol detection problem in overloaded \ac{NOMA} communication systems without assuming particular channel statistics.
The key concept of the proposed framework is the compliance of the solution to the prescribed symbol constellation $\mathcal{C}$, which is technically realized by the combination of a novel tight approximation of the $\ell_0$-norm, with a fractional programming technique that leads to a closed-form-based iterative solution dubbed the \acf{IDLS}. 

Due to its structure, it can be said that the proposed \ac{IDLS} detector is a generalization of the classic \ac{LS} counterpart for discrete signal detection.
Despite its low computational complexity, the \ac{IDLS} detector was shown to achieve superior performance compared to \ac{SotA} schemes in terms of \ac{BER}, regardless of the presence of channel correlation and severe overloading, approaching the absolute performance lower bound in fully loaded conditions without correlation.

The \ac{IDLS} detector was then extended to incorporate mitigation capabilities agains non-ideal factors, such as \ac{CSI} imperfection and hardware impairments, yielding the Robust \ac{IDLS} detector.
Due to the modular nature of its structure, which enables the construction of several variations depending on the desired mitigating features, the \ac{IDLS} can be considered as a novel framework for the design of detectors for \ac{NOMA} systems.


\vspace{-1ex}
\appendix[Derivation of CSI/HWI Effective Noise Covariance]
\label{app:CovDerivation}

Recall from equation \eqref{eqn:normalizedY} that the total effective noise, including the contributions due to \ac{CSI} imperfection and hardware impairment is given by
\begin{equation}
\label{eqn:CSIHWINoise}
\tilde{\bm{n}} \triangleq \frac{\tau\bm{E}\bm{s}+\tau\bm{E}\bm{w} + \bm{n}}{\sqrt{1-\tau^2}}.
\end{equation}

By definition we then have
\begin{eqnarray}
\label{eqn:intnoise_covariance}
&&\hspace{-4ex}\bm{\mathit \Sigma}_{\tilde{\bm{n}}} \triangleq \E{\tilde{\bm{n}}\tilde{\bm{n}}^\mathrm{H}}\\
&&\hspace{-2.5ex}= \mathbb{E}\!\left[\!\Big(\!\hat{\bm{H}} \bm{w} \!+\! \frac{\tau\bm{E}\bm{s}\!+\!\tau\bm{E}\bm{w}\! +\! \bm{n}}{\sqrt{1-\tau^2}}\!\Big)\Big(\!\hat{\bm{H}} \bm{w} \!+\! \frac{\tau\bm{E}\bm{s}\!+\!\tau\bm{E}\bm{w}\! +\! \bm{n}}{\sqrt{1-\tau^2}}\!\Big)^{\!\mathrm{H}}\right].
\nonumber
\end{eqnarray}

Distributing the terms in parenthesis and expending yields
\begin{eqnarray}
\bm{\mathit \Sigma}_{\tilde{\bm{n}}}
\hspace{-3.5ex}&&= \E{\hat{\bm{H}} \bm{w}\Big(\!\hat{\bm{H}} \bm{w} \!+\! \frac{\tau\bm{E}\bm{s}\!+\!\tau\bm{E}\bm{w}\! +\! \bm{n}}{\sqrt{1-\tau^2}}\!\Big)^{\!\mathrm{H}}}\nonumber\\
&&+\frac{\tau}{\sqrt{1-\tau^2}} \E{\bm{E}\bm{s}\Big(\!\hat{\bm{H}} \bm{w} \!+\! \frac{\tau\bm{E}\bm{s}\!+\!\tau\bm{E}\bm{w}\! +\! \bm{n}}{\sqrt{1-\tau^2}}\!\Big)^{\!\mathrm{H}}}\nonumber\\
&&+\frac{\tau}{\sqrt{1-\tau^2}} \E{\bm{E}\bm{w}\Big(\!\hat{\bm{H}} \bm{w} \!+\! \frac{\tau\bm{E}\bm{s}\!+\!\tau\bm{E}\bm{w}\! +\! \bm{n}}{\sqrt{1-\tau^2}}\!\Big)^{\!\mathrm{H}}}\nonumber\\
&&+\frac{1}{\sqrt{1-\tau^2}}\E{\bm{n}\Big(\!\hat{\bm{H}} \bm{w} \!+\! \frac{\tau\bm{E}\bm{s}\!+\!\tau\bm{E}\bm{w}\! +\! \bm{n}}{\sqrt{1-\tau^2}}\!\Big)^{\!\mathrm{H}}}\nonumber\\
&&\hspace{-2.5ex}= \hat{\bm{H}}\,\overbrace{\E{\bm{w}\bm{w}^{\mathrm{H}}}}^{= \eta\mathbf{I}_{N_t}}\!\hat{\bm{H}}^{\mathrm{H}}\! +\! \frac{\tau^2\big(\overbrace{\E{\bm{E}\bm{s}\bm{s}^\mathrm{H}\bm{E}^\mathrm{H}}\!\!+\!\E{\bm{E}\bm{w}\bm{w}^\mathrm{H}\bm{E}^\mathrm{H}}}^{=(1+\eta)\Tr{\bm{\Phi}_t}\bm{\Phi}_r}\big)}{1-\tau^2}\nonumber\\
&&+\frac{\sigma^2_{{\bm{n}}}}{1-\tau^2}\mathbf{I}_{N_r}\nonumber\\
&&\hspace{-2.5ex}= \eta\hat{\bm{H}}\hat{\bm{H}}^{\mathrm{H}}\! +\! \frac{\tau^2}{1\!-\!\tau^2}(1\!+\!\eta)\Tr{\bm{\Phi}_t}\bm{\Phi}_r + \frac{\sigma^2_{{\bm{n}}}}{1\!-\!\tau^2}\mathbf{I}_{N_r},
\label{eqn:intnoise_covariancefinal}
\end{eqnarray}
where, in the second-to-last equation we have substituted the expectations $\E{\bm{w}\bm{w}^{\mathrm{H}}}$, $\E{\bm{E}\bm{s}\bm{s}^\mathrm{H}\bm{E}^\mathrm{H}}$, $\E{\bm{E}\bm{w}\bm{w}^\mathrm{H}\bm{E}^\mathrm{H}}$ and $\E{\bm{n}\bm{n}^\mathrm{H}}$, taken over noise realizations and utilizing the identity $\E{\bm{X}\bm{A}\bm{X}^\mathrm{H}} = \sigma^2\Tr{\bm{A}}\mathbf{I}$, valid when the elements of $\bm{X}$ are zero-mean complex Gaussian variables with variance $\sigma^2$.

For the terms $\E{\bm{E}\bm{s}\bm{s}^\mathrm{H}\bm{E}^\mathrm{H}}$ and $\E{\bm{E}\bm{w}\bm{w}^\mathrm{H}\bm{E}^\mathrm{H}}$ in particular, which are subjected to correlation due to the relation $\bm{E}=\bm{\Phi}^{\frac{1}{2}}_r\bm{E}_\text{\ac{i.i.d.}}\bm{\Phi}^{\frac{1}{2}}_t$, we have used
\begin{subequations}
\label{eq:DistortedTXSignalCovariance}
\begin{eqnarray}
\E{\bm{E}\bm{s}\bm{s}^\mathrm{H}\bm{E}^\mathrm{H}} \hspace{-4ex}&&= \Ev{\bm{E}_\text{\ac{i.i.d.}}}{\Ev{\bm{s}}{\bm{E}\bm{s}\bm{s}^\mathrm{H}\bm{E}^\mathrm{H}\:|\:\bm{E}_\text{\ac{i.i.d.}}}}\\
&&=  \Ev{\bm{E}_\text{\ac{i.i.d.}}}{\bm{\Phi}^{\frac{1}{2}}_r\bm{E}_\text{\ac{i.i.d.}}\bm{\Phi}_t\bm{E}^\mathrm{H}_\text{\ac{i.i.d.}}{\bm{\Phi}^{\frac{1}{2}}}^\mathrm{H}_r} = \Tr{\bm{\Phi}_t}\bm{\Phi}_r,\nonumber
\end{eqnarray}
and 
\begin{eqnarray}
\E{\bm{E}\bm{w}\bm{w}^\mathrm{H}\bm{E}^\mathrm{H}} \hspace{-4ex}&&=  \Ev{\bm{E}_\text{\ac{i.i.d.}}}{\Ev{\bm{w}}{\bm{E}\bm{w}\bm{w}^\mathrm{H}\bm{E}^\mathrm{H}\:|\:\bm{E}_\text{\ac{i.i.d.}}}}\\
&&\hspace{-10ex}=\eta\!\cdot\! \mathbb{E}_{\bm{E}_\text{\ac{i.i.d.}}}\!\bigg[\bm{\Phi}^{\frac{1}{2}}_r\bm{E}_\text{\ac{i.i.d.}}\bm{\Phi}^{\frac{1}{2}}_t {\bm{\Phi}^{\frac{1}{2}}}^\mathrm{H}_t\bm{E}^\mathrm{H}_\text{\ac{i.i.d.}}{\bm{\Phi}^{\frac{1}{2}}}^\mathrm{H}_r\bigg]\! = \eta\Tr{\bm{\Phi}_t}\bm{\Phi}_r,\nonumber
\end{eqnarray}
\end{subequations}
where we assume that the correlation matrices are Hermitian and that $\mathbf{C}_{\bm{s}}=\mathbf{I}_{N_t}$.

We also remark that in case of $\bm{\Phi}_t=\bm{\Phi}_r=\mathbf{I}$, the covariance matrix of the effective noise becomes a function of the dimensionality of the transmit symbols, thus directly affected by the overloading factor $\gamma$.

\vspace{-2ex}
\bibliographystyle{IEEEtran}
\bibliography{IEEEabrv,\myreferences}

\begin{thebibliography}{10}
\providecommand{\url}[1]{#1}
\csname url@samestyle\endcsname
\providecommand{\newblock}{\relax}
\providecommand{\bibinfo}[2]{#2}
\providecommand{\BIBentrySTDinterwordspacing}{\spaceskip=0pt\relax}
\providecommand{\BIBentryALTinterwordstretchfactor}{4}
\providecommand{\BIBentryALTinterwordspacing}{\spaceskip=\fontdimen2\font plus
\BIBentryALTinterwordstretchfactor\fontdimen3\font minus
  \fontdimen4\font\relax}
\providecommand{\BIBforeignlanguage}[2]{{%
\expandafter\ifx\csname l@#1\endcsname\relax
\typeout{** WARNING: IEEEtran.bst: No hyphenation pattern has been}%
\typeout{** loaded for the language `#1'. Using the pattern for}%
\typeout{** the default language instead.}%
\else
\language=\csname l@#1\endcsname
\fi
#2}}
\providecommand{\BIBdecl}{\relax}
\BIBdecl

\bibitem{TasneemSys20}
T.~Assaf, A.~Al-Dweik, M.~S.~E. Moursi, H.~Zeineldin, and M.~Al-Jarrah,
  ``{NOMA} receiver design for delay-sensitive systems,'' \emph{{IEEE Systems
  J.}}, Early Access.

\bibitem{JeongTVT19}
J.~S. Yeom, H.~S. Jang, K.~S. Ko, and B.~C. Jung, ``{BER} performance of uplink
  {NOMA} with joint maximum-likelihood detector,'' \emph{{IEEE Trans. Veh.
  Technol.}}, vol.~68, no.~10, pp. 10\,295--10\,300, Oct. 2019.

\bibitem{NagaharaSPL15}
M.~Nagahara, ``Discrete signal reconstruction by sum of absolute values,''
  \emph{{IEEE Signal Process. Lett.}}, vol.~22, no.~10, pp. 1575--1579, Oct.
  2015.

\bibitem{MasoudIET20}
M.~Naderpour and H.~K. Bizaki, ``Low overhead {NOMA} receiver with automatic
  modulation classification techniques,'' \emph{{IET Commun.}}, vol.~14, no.~5,
  pp. 768--774, Feb. 2020.

\bibitem{HayakawaTWC2017}
R.~Hayakawa and K.~Hayashi, ``Convex optimization-based signal detection for
  massive overloaded {MIMO} systems,'' \emph{{IEEE Trans. Wireless Commun.}},
  vol.~16, no.~11, pp. 7080--7091, Nov. 2017.

\bibitem{HayakawaAccess2018}
------, ``Reconstruction of complex discrete-valued vector via convex
  optimization with sparse regularizers,'' \emph{{IEEE Access}}, vol.~6, pp.
  66\,499--66\,512, Oct. 2018.

\bibitem{MingICC20}
M.~Zeng, W.~Hao, A.~Yadav, N.-P. Nguyen, O.~A. Dobre, and H.~V. Poor,
  ``Energy-efficient joint power control and receiver design for uplink
  mmwave-{NOMA},'' in \emph{{Proc. IEEE ICC Wkshps}}, Dublin, Ireland, Jun.
  2020.

\bibitem{Popovski2018}
P.~Popovski, K.~F. Trillingsgaard, O.~Simeone, and G.~Durisi, ``{5G} wireless
  network slicing for {eMBB, URLLC, and mMTC}: {A} communication-theoretic
  view,'' \emph{{IEEE Access}}, vol.~6, pp. 55\,765--55\,779, Sept. 2018.

\bibitem{GiordaniComMag2020}
M.~{Giordani}, M.~{Polese}, M.~{Mezzavilla}, S.~{Rangan}, and M.~{Zorzi},
  ``Toward 6{G} networks: Use cases and technologies,'' \emph{{IEEE Commun.
  Mag.}}, vol.~58, no.~3, pp. 55--61, 2020.

\bibitem{AndreiAccessMCNOMA2019}
R.-A. Stoica, G.~Abreu, T.~Hara, and K.~Ishibashi, ``Massively concurrent
  non-orthogonal multiple access for {5G} networks and beyond,'' \emph{{IEEE
  Access}}, vol.~7, pp. 82\,080--82\,100, Jun. 2019.

\bibitem{CaireAsilomar2019}
A.~Fengler, S.~Haghighatshoar, P.~Jung, and G.~Caire, ``Grant-free massive
  random access with a massive {MIMO} receiver,'' in \emph{{P}roc. Asilomar
  {CSSC}}, Pacific Grove, USA, 2019.

\bibitem{GanesanSPAWC2020}
U.~K. {Ganesan}, E.~{Bj{\"o}rnson}, and E.~G. {Larsson}, ``An algorithm for
  grant-free random access in cell-free massive mimo,'' in \emph{{Proc. IEEE
  SPAWC}}, 2020, pp. 1--5.

\bibitem{MyListOfPapers:LiuTWC2018}
L.~Liu, C.~Yuen, Y.~L. Guan, Y.~Li, and C.~Huang, ``Gaussian message passing
  for overloaded massive {MIMO-NOMA},'' \emph{{IEEE Trans. Wireless Commun.}},
  vol.~18, no.~1, pp. 210--226, Jan. 2018.

\bibitem{DattaERTS12}
T.~Datta, N.~Srinidhi, A.~Chockalingam, and B.~S. Rajan, ``Low-complexity
  near-optimal signal detection in underdetermined large-{MIMO} systems,'' in
  \emph{{P}roc. {National Conf. Commun.}}, Kharagpur, India 2012.

\bibitem{ChenTWC2013}
C.~Qian, J.~Wu, Y.~R. Zheng, and Z.~Wang, ``Two-stage list sphere decoding for
  under-determined multiple-input multiple-output systems,'' \emph{{IEEE Trans.
  Wireless Commun.}}, vol.~12, no.~12, pp. 6476--6487, 2013.

\bibitem{RauhutBook:2013}
S.~Foucart and H.~Rauhut, \emph{{A Mathematical Introduction to Compressive
  Sensing}}.\hskip 1em plus 0.5em minus 0.4em\relax Springer Verlag, Aug. 2012.

\bibitem{das2013finite}
A.~K. Das and S.~Vishwanath, ``{On Finite Alphabet Compressive Sensing},'' in
  \emph{Proc. IEEE Intl. Conf. on Acoustics, Speech and Signal Process.
  (ICASSP)}.\hskip 1em plus 0.5em minus 0.4em\relax IEEE, 2013, pp. 5890--5894.

\bibitem{DonohoTIT06}
D.~Donoho, ``Compressed sensing,'' \emph{{IEEE Trans. Inf. Theory}}, vol.~52,
  no.~4, pp. 1289--1306, Apr. 2006.

\bibitem{ShenTSP2018}
K.~Shen and W.~Yu, ``Fractional programming for communication systems -- {P}art
  {I}: {P}ower control and beamforming,'' \emph{{IEEE Trans. Signal Process.}},
  vol.~66, no.~10, pp. 2616--2630, May 2018.

\bibitem{WenAccess2018}
F.~{Wen}, L.~{Chu}, P.~{Liu}, and R.~C. {Qiu}, ``A survey on nonconvex
  regularization-based sparse and low-rank recovery in signal processing,
  statistics, and machine learning,'' \emph{IEEE Access}, vol.~6, pp.
  69\,883--69\,906, 2018.

\bibitem{Bjornson2013}
E.~Bj\"{o}rnson and E.~Jorswieck, ``Optimal resource allocation in coordinated
  multi-cell systems,'' \emph{{Foundations and Trends in Communications and
  Information Theory}}, vol.~9, no. 2--3, pp. 113--381, Jan. 2013.

\bibitem{AreSP2007}
A.~Hj{\o}rungnes and D.~Gesbert, ``{C}omplex-valued matrix differentiation:
  {T}echniques and key results,'' \emph{{IEEE Trans. Signal Process.}},
  vol.~55, no.~6, pp. 2740--2746, Jun. 2007.

\bibitem{RosarioSPL2016}
F.~{Ros{\'a}rio}, F.~A. {Monteiro}, and A.~{Rodrigues}, ``Fast matrix inversion
  updates for massive {MIMO} detection and precoding,'' \emph{{IEEE Signal
  Process. Lett.}}, vol.~23, no.~1, pp. 75--79, 2016.

\bibitem{CandesMC09}
E.~J. Candes and B.~Recht, ``Exact matrix completion via convex optimization,''
  \emph{Found. Comput. Math.}, vol.~9, no.~6, pp. 717--772, Dec. 2009.

\bibitem{HassibiTIT2018}
C.~Thrampoulidis, E.~Abbasi, and B.~Hassibi, ``Precise error analysis of
  regularized $m$-estimators in high dimensions,'' \emph{{IEEE Trans. Inf.
  Theory}}, vol.~64, no.~8, pp. 5592--5628, Aug. 2018.

\bibitem{ItoTSP19}
D.~Ito, S.~Takabe, and T.~Wadayama, ``Trainable {ISTA} for sparse signal
  recovery,'' \emph{{IEEE Trans. Signal Process.}}, vol.~67, no.~12, pp.
  3113--3125, Jun. 2019.

\bibitem{MECKLENBRAUKER2017204}
C.~F. Mecklenbr{\"a}uker, P.~Gerstoft, and E.~Z{\"o}chmann, ``c--{LASSO} and
  its dual for sparse signal estimation from array data,'' \emph{Signal
  Processing}, vol. 130, pp. 204--216, 2017.

\bibitem{LuzziTIT2013}
L.~{Luzzi}, D.~{Stehl{\'e}}, and C.~{Ling}, ``Decoding by embedding: Correct
  decoding radius and {DMT} optimality,'' \emph{{IEEE Trans. Inf. Theory}},
  vol.~59, no.~5, pp. 2960--2973, May 2013.

\bibitem{5GZaidi}
A.~Zaidi, F.~Athley, J.~Medbo, U.~Gustavsson, G.~Durisi, and X.~Chen,
  \emph{5{G} Physical Layer: principles, models and technology
  components}.\hskip 1em plus 0.5em minus 0.4em\relax Academic Press, 2018.

\bibitem{MorePMS1993}
J.~J. Mor{\'{e}}, ``Generalizations of the trust region problem,'' \emph{Optim.
  Methods Softw.}, vol.~2, no. 3--4, pp. 189--209, 1993.

\bibitem{MyListOfPapers:Golub1996}
G.~H. Golub and C.~F. van Loan, \emph{Matrix Computations}, 3rd~ed.\hskip 1em
  plus 0.5em minus 0.4em\relax New York, NY: Johns Hopkins Univ. Press, Nov.
  1996.

\bibitem{AdachiMP2019}
S.~Adachi and Y.~Nakatsukasa, ``Eigenvalue-based algorithm and analysis for
  nonconvex {QCQP} with one constraint,'' \emph{Math. Program.}, vol. 173, no.
  1--2, pp. 79--116, Jan. 2019.

\bibitem{Golub1996}
G.~H. Golub and C.~F.~V. Loan, \emph{Matrix Computations (3rd ed.)}.\hskip 1em
  plus 0.5em minus 0.4em\relax Johns Hopkins University Press, 1996.

\bibitem{MyListOfPapers:CaiEUSIPCO2020}
H.~Cai, M.~F. Kaloorazi, J.~Chen, W.~Chen, and C.~Richard, ``Online dominant
  generalized eigenvectors extraction via a randomized method,'' in
  \emph{{Proc. EUSIPCO}}, 2020.

\bibitem{MyListOfPapers:ZhiqiangUAI2020}
Z.~Xu and P.~Li, ``A practical riemannian algorithm for computing dominant
  generalized eigenspace,'' in \emph{Proc. Conf. Uncertainty in Artificial
  Intell. (UAI)}, 2020.

\bibitem{MyListOfPapers:RommesMoC2007}
J.~Rommes, ``Arnoldi and {J}acobi-{D}avidson methods for generalized eigenvalue
  problems ${A}x = \lambda {B}x$ with singular ${B}$,'' \emph{Mathematics of
  Computation}, vol.~77, no. 262, pp. 995--1015, Apr. 2007.

\bibitem{FarahMScThesisUniBirm2012}
A.~M. Farah, ``Generalized and quadratic eigenvalue problems with {H}ermitian
  matrices,'' Master's thesis, University of Birmingham, 2012.

\bibitem{EmilTWC20}
E.~Bj\"{o}rnson and L.~Sanguinetti, ``Making cell-free massive {MIMO}
  competitive with {MMSE} processing and centralized implementation,''
  \emph{{IEEE Trans. Wireless Commun.}}, vol.~19, no.~1, pp. 77--90, Jan. 2020.

\bibitem{HyundongTWC03}
H.~Shin and J.~H. Lee, ``Capacity of multiple-antenna fading channels: spatial
  fading correlation, double scattering, and keyhole,'' \emph{{IEEE Trans. Inf.
  Theory}}, vol.~49, no.~10, pp. 2636--2647, Oct. 2003.

\bibitem{ShaoshiCST2015}
S.~Yang and L.~Hanzo, ``Fifty years of {MIMO} detection: {T}he road to
  large-scale {MIMO}s,'' \emph{{IEEE Commun. Surveys \& Tut.}}, vol.~17, no.~4,
  pp. 1941--1988, Fourthquarter 2015.

\bibitem{TeukolskyBook2007}
W.~H. Press, S.~A. Teukolsky, W.~T. Vetterling, and B.~P. Flannery,
  \emph{{Numerical Recipes: The Art of Scientific Computing}}, 3rd~ed.\hskip
  1em plus 0.5em minus 0.4em\relax USA: Cambridge University Press, 2007.

\bibitem{BehrangTSP2011}
B.~Nosrat-Makouei, J.~G. Andrews, and R.~W. Heath, ``{MIMO} interference
  alignment over correlated channels with imperfect {CSI},'' \emph{{IEEE Trans.
  Signal Process.}}, vol.~59, no.~6, pp. 2783--2794, Jun. 2011.

\bibitem{SuzukiTWC08}
H.~Suzuki, T.~V.~A. Tran, I.~B. Collings, G.~Daniels, and M.~Hedley,
  ``Transmitter noise effect on the performance of a {MIMO-OFDM} hardware
  implementation achieving improved coverage,'' \emph{{IEEE J. Sel. Areas
  Commun.}}, vol.~26, no.~6, pp. 867--876, Aug. 2008.

\bibitem{OmidTWC2018}
O.~Taghizadeh, A.~C. Cirik, and R.~Mathar, ``Hardware impairments aware
  transceiver design for full-duplex amplify-and-forward {MIMO} relaying,''
  \emph{{IEEE Trans. Wireless Commun.}}, vol.~17, no.~3, pp. 1644--1659, Mar.
  2018.

\bibitem{IimoriTWC19}
H.~Iimori, G.~T.~F. de~Abreu, and G.~C. Alexandropoulos, ``{MIMO} beamforming
  schemes for hybrid {SIC FD} radios with imperfect hardware and {CSI},''
  \emph{{IEEE Trans. Wireless Commun.}}, vol.~18, no.~10, pp. 4816--4830, Oct.
  2019.

\bibitem{Huffel1991}
S.~V. Huffel, \emph{Numerical Linear Algebra, Digital Signal Processing and
  Parallel Algorithms}, G.~H. Golub and P.~V. Dooren, Eds.\hskip 1em plus 0.5em
  minus 0.4em\relax Springer, 1991, vol.~70.

\bibitem{LloydSIAM1997}
L.~N. Trefethen and D.~B. I\hspace{-.1em}I\hspace{-.1em}I, \emph{Numerical
  linear algebra (1st ed.)}.\hskip 1em plus 0.5em minus 0.4em\relax SIAM, 1997.

\end{thebibliography}

\end{document}